\documentclass[AMA,Times1COL]{WileyNJDv5} 

\articletype{Article Type}%

\received{Date Month Year}
\revised{Date Month Year}
\accepted{Date Month Year}
\journal{Journal}
\volume{00}
\copyyear{2023}
\startpage{1}

\begin{document}

\title{Unified formulas for the effective conductivity of fibrous composites with circular inclusions and parallelogram periodicity and its influence on thermal gain in nanofluids}

\author[1]{Ra\'ul Guinovart-D\'iaz}
\author[2]{Juli\'an Bravo-Castillero}
\author[3]{Manuel E. Cruz}
\author[4]{Leslie D. P\'erez-Fern\'andez}
\author[5]{Federico J. Sabina}
\author[6]{David Guinovart}

\authormark{GUINOVART-DÍAZ \textsc{et al.}}
\titlemark{Unified formulas for the effective conductivity of fibrous
composites with circular inclusions and parallelogram periodicity
and its influence on the thermal gain in nanofluids}

\address[1]{\orgdiv{Facultad de Matem\'atica y Computaci\'on}, \orgname{Universidad de La Habana}, \orgaddress{\state{La Habana}, \country{Cuba}}}

\address[2]{\orgdiv{Unidad Acad\'emica del IIMAS en el Estado de Yucatán}, \orgname{Universidad Nacional Autónoma de México (UNAM)}, \orgaddress{\state{Yucatán}, \country{México}}}

\address[3]{\orgdiv{Departamento de Engenharia Mec\^anica, Polit\'ecnica/COPPE}, \orgname{Universidade Federal de Rio de Janeiro}, \orgaddress{\state{Rio de Janeiro}, \country{Brazil}}}

\address[4]{\orgdiv{Instituto de Física e Matemática}, \orgname{Universidade Federal de Pelotas}, \orgaddress{\state{Rio Grande do Sul}, \country{Brazil}}}

\address[5]{\orgdiv{Instituto de Investigaciones en Matemáticas Aplicadas (IIMAS) y en Sistemas}, \orgname{Universidad Nacional Autónoma de México}, \orgaddress{\state{Ciudad de México}, \country{México}}}

\address[6]{\orgdiv{The Hormel Institute}, \orgname{University of Minnesota}, \orgaddress{\state{Minnesota}, \country{USA}}}

\corres{Corresponding author Juli\'an Bravo-Castillero, Unidad Académica del IIMAS en el Estado de Yucatán, UNAM, Tablaje Catastral N° 6998, Carretera Mérida - Tetiz km. 4, Mérida, 97357, Mexico. \email{julian@mym.iimas.unam.mx}}



\abstract[Abstract]{A two-dimensional three-phase conducting composite with coated circular inclusions, periodically distributed in a parallelogram, is studied. The phases are assumed to be isotropic, and perfect contact conditions at the interfaces are considered. The effective behavior is determined by combining the asymptotic homogenization method with elements of the analytic function theory. The solution to local problems is sought as a series of Weierstrass elliptic functions and their derivatives with complex undetermined coefficients. The effective coefficients depend on the residue of such a solution, which in turn depends on products of vectors and matrices of infinite order. Systematic truncation of these vectors and matrices provides unified analytical formulas for the effective coefficients for any parallelogram periodic cell. The corresponding formulas for the particular cases of two-phase fibrous composites with perfect and imperfect contact at the interface are also explicitly provided. The results were applied to derive the critical normalized interfacial layer thickness and to analyze the enhancement of thermal conductivity in fibrous composites with annular cross sections. Furthermore, using a reiterated homogenization method, the analytical approach allows us to study the gains in the effective thermal conductivity tensor with thermal barriers and parallelogram cells. Numerical examples and comparisons validate the model. A simple and validated algorithm is provided that allows the calculation of effective coefficients for any parallelogram, any truncation order, and high fiber volume fractions very close to percolation. The programs created for validation are available in a freely accessible repository.}


\jnlcitation{\cname{%
\author{Guinovart-D\'iaz R.},
\author{Bravo-Castillero J.},
\author{Cruz M. E.},
\author{P\'erez-Fern\'andez L. D.},
\author{Sabina F. J.}, and
\author{Guinovart D.}}.
\ctitle{Simple formulas for the effective conductivity of three-phase fibrous composites with parallelogram periodicity and its influence on thermal gain.} 
\cjournal{\it Journal Name.} \cvol{2025;00(00):1--20}.}

\maketitle

\renewcommand\thefootnote{}
\renewcommand\thefootnote{\fnsymbol{footnote}}
\setcounter{footnote}{1}

\section{Introduction}\label{sec1}

Fiber-reinforced composite materials have gained increasing prominence in technological applications due to their outstanding mechanical, electrical, and thermal properties~\cite{zweben2015composite,chung2010functional}. For example, the epoxy/glass microbeads-based 1-3 piezoelectric composite described in~\cite{LiuQiyunZhou} is engineered to enhance electromechanical conversion efficiency. These materials typically comprise high-performance fibers---such as carbon or glass---embedded within a polymeric or ceramic matrix. This synergistic combination enables the development of composites with tailored, multifunctional properties that often surpass those of the individual constituents~\cite{agarwal2017analysis,mallick2007fiber}.

A crucial aspect in designing and applying composite materials is the ability to predict and control their effective electrical and thermal conductivity~\cite{progelhof1976methods}. These effective properties are influenced by various factors, including the orientation and volume fraction of the fibers, as well as the intrinsic characteristics of both the fibers and the matrix~\cite{progelhof1976methods,hashin1963variational}. A thorough understanding and accurate modeling of effective conductivity are essential for optimizing the performance of composites in applications requiring efficient heat or electricity transport~\cite{deng2014progress,nemat2013micromechanics,yariv2021conductivity,mityushev2001transport}. Analytical and numerical modeling approaches, particularly homogenization techniques, serve as foundational tools in the analysis and optimization of these materials~\cite{FERREIRA2024}.

Conductive fibrous composites, consisting of conductive fibers embedded in a polymer or ceramic matrix, have attracted considerable attention due to their unique combination of high electrical conductivity, mechanical strength, and low weight. These materials are widely utilized in aerospace, electronics, energy storage, and biomedical applications. This work employs the asymptotic homogenization method to investigate the effective conductivity of such fiber-reinforced composites. Previous studies~\cite{guinovart2011influence,JuanK2011,sabina2020,sabina2021} have derived closed-form expressions for the effective axial properties of fibrous composites with parallelogram-shaped unit cells, using infinite matrix formulations based on systems of equations. In the present study, these analytical formulations are applied to analyze the influence of microstructural parameters and to validate the results obtained through homogenization. Additionally, we introduce a truncation strategy to improve numerical implementation by considering different orders of the infinite matrices, enabling computational efficiency without sacrificing accuracy. The resulting formulation can be readily extended to validate other analytical or numerical approaches to study conductive composites.

The asymptotic homogenization method (AHM) is a powerful multiscale technique widely used to derive the effective macroscopic behavior of heterogeneous materials by rigorously accounting for their microstructural characteristics~\cite{bensoussan1978asymptotic,sanchezpalencia1980,bakhvalov1989}. Particularly effective for periodic or quasi-periodic composites, AHM enables accurate modeling of physical properties, such as stiffness, electrical and thermal conductivity, or piezoelectric response, by capturing the influence of fiber alignment, distribution, and phase interactions~\cite{Kalamkarov2009,sabina2020}. In contrast to full-scale numerical simulations, AHM offers a computationally efficient analytical framework incorporating detailed microstructural effects. In the present study, AHM is combined with the theory of complex functions to investigate the effective conductivity of conductive fibrous composites and to validate analytical formulations based on truncated matrix approximations. Its versatility is further demonstrated in recent developments, such as~\cite{Vignoli2024}, where AHM results have been combined with data-driven models to generate accurate predictions for elastic composite materials.

This paper presents a novel comprehensive framework that joins together analytical and numerical procedures to evaluate the effective thermal conductivity of fibrous composites, considering both perfect and imperfect interfacial contact scenarios. The theoretical development begins with closed-form approximations derived from the solution of the local problems using series expansions of elliptic functions, capturing the influence of microstructural geometry, fiber distribution, and interfacial resistance. Several interfacial configurations are explored, including spring-type contact resistance, coated fibers with mesophase layers, and dispersed inclusions modeled through reiterated homogenization. The methodology is extended to periodic cells with parallelogram symmetry to characterize anisotropic behavior. The proposed models are systematically compared in the numerical analysis to assess the influence of barrier thickness, material contrast, and microstructural parameters on overall conductivity. Key findings include the identification of critical interfacial thicknesses that govern conductivity gain or saturation, as well as the discovery of directional effects tied to unit cell geometry. In all cases, the inclusion of thermal barriers significantly alters effective conductivity, particularly at high inclusion volume fractions and thermal contrast. These results offer a unified analytical and numerical framework for designing fiber-reinforced composites with tailored thermal performance, especially in applications where interfacial engineering plays a pivotal role.


\section{Asymptotic homogenization, effective coefficients and local problems}\label{sec2}

In this section, we present the mathematical models for local problems and the related effective coefficients derived from applying the asymptotic homogenization method to a family of Dirichlet problems for the heat equation with periodic and rapidly oscillating coefficients.
The objective is to determine the effective (macroscopic) conductive behavior of the composite by rigorously linking it to the heterogeneous (microscopic) structure of the periodic unit cell.
\subsection{Statement of a family of problems for heat conduction}
\begin{figure*}[t]
    \centering
    \includegraphics[width=0.85\textwidth]{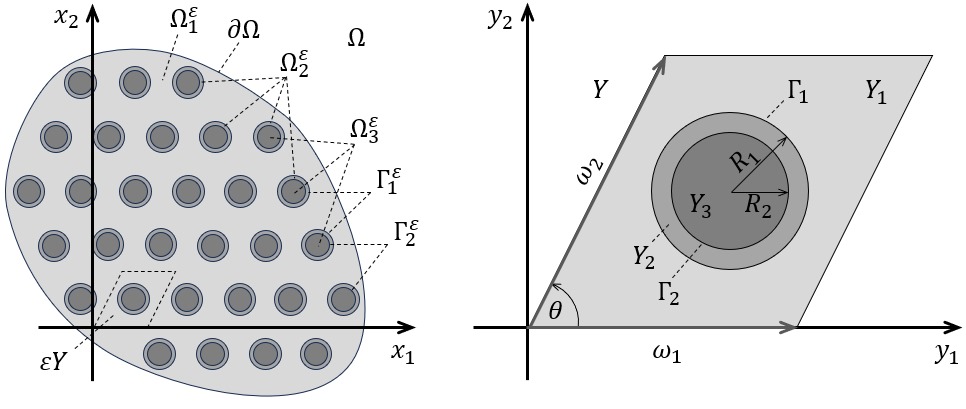} 
    \caption{Geometric representation of composite structures with interphases: (a) transversal section of the fiber composite with interphase; (b) periodic unit cell geometry with interphase thickness \( t = R_1-R_2 > 0 \) and orientation angle \( \theta \).}
    \label{fig:Fig_1}
\end{figure*}

Let $\varepsilon\in\mathbb{R}^{*}_{+}$ such that $0<\varepsilon\ll1$. Let $\Omega\subset\mathbb{R}^{2}$ be the domain occupied by a three-phase thermal conductive composite made of two-phase concentric circles periodically distributed in a matrix with $\varepsilon$-periodicity in the axes directions of the Cartesian Plane $Ox_{1}x_{2}$. The matrix, circular rings (interfacial region) and inner circles (inclusions) occupy the subdomains $\Omega^{\varepsilon}_{1}$, $\Omega^{\varepsilon}_{2}$, and $\Omega^{\varepsilon}_{3}$, respectively, with $\Omega=\Omega^{\varepsilon}_{1}\cup\Omega^{\varepsilon}_{2}\cup\Omega^{\varepsilon}_{3}\cup\Gamma^{\varepsilon}_{1}\cup\Gamma^{\varepsilon}_{2}$ where $\Gamma^{\varepsilon}_{s}$ is equal to the limiting curve between $\Omega^{\varepsilon}_{s}$ and $\Omega^{\varepsilon}_{s+1}$ ($s=1,2$). Let $Y\subset\mathbb{R}^{2}$ be the parallelogram periodic cell, which is described in terms of the local variables $y_{s}=\varepsilon^{-1}x_{s}$. Then, the matrix, the interphase and the inclusion occupy the subdomains $Y_{1}$, $Y_{2}$ and $Y_{3}$, respectively, such that $Y=Y_{1}\cup Y_{2}\cup Y_{3}\cup\Gamma_{1}\cup\Gamma_{2}$, being $\Gamma_{s}$ the limiting circumference between $Y_{s}$ and $Y_{s+1}$. A graphical representation of such a structure is shown in Figure~\ref{fig:Fig_1}.

For each $\varepsilon>0$, the temperature field $v^\varepsilon(\mathbf{x})$ is the solution of following boundary value problem for the heat equation:
\begin{equation}
	\label{eq:equilibrium}
	\nabla \cdot \left( \mathbf{\kappa}^\varepsilon(\mathbf{x}) \nabla v^\varepsilon(\mathbf{x}) \right) = f \quad \text{in } \Omega^{\varepsilon}_{1}\cup\Omega^{\varepsilon}_{2}\cup\Omega^{\varepsilon}_{3},
\end{equation}
\begin{equation}
            \label{eq:condition1}
		\left. v^\varepsilon \right|_{\Gamma_s^{\varepsilon}-} = \left. v^\varepsilon \right|_{\Gamma_s^{\varepsilon}+}, \quad s = 1,2,
	\end{equation}
\begin{equation}
             \label{eq:condition2}    
		\left. \mathbf{\kappa}^{\varepsilon}_{s} \frac{\partial v^\varepsilon}{\partial n}  \right|_{\Gamma_s^{\varepsilon}-} = \left. \mathbf{\kappa}^{\varepsilon}_{s+1} \frac{\partial v^\varepsilon}{\partial n} \right|_{\Gamma_s^{\varepsilon}+}, \quad s = 1,2,
	\end{equation}
\begin{equation}
	\label{eq:dirichlet}
	v^\varepsilon = 0 \quad \text{on } \partial \Omega,
\end{equation}
where $\nabla$ and $\nabla \cdot$  are the gradient and divergence operators, respectively, $f\in L^{2}(\Omega)$ represents the heat output source, $\kappa^\varepsilon(\mathbf{x})=\kappa(\mathbf{x}/\varepsilon)$ denotes the $\varepsilon{Y}$-periodic thermal conductivity tensor that varies with position according to the microscopic structure and the scale parameter $\varepsilon$ (so $\kappa(\mathbf{y})$ is $Y$-periodic in $\mathbf{y}$).  Furthermore, $\kappa^{\varepsilon}$ is a symmetric, bounded, positive-definite second-order tensor. The superscripts $^-$ and $^+$ denote the limits approaching the interface from inside region $s$ and region $s+1$, respectively. $\partial/\partial n$ is the normal derivative across the interface, and  $\kappa^{\varepsilon}_{i}$ be the value of the thermal conductivity tensor in the region $\Omega^{\varepsilon}_{i}$ $(i=1,2,3)$.
On the external boundary $\partial \Omega$ of the domain, we prescribe a Dirichlet boundary condition. The mathematical aspects related to the well-posedness of this family of problems and the application of the asymptotic homogenization method can be found, for example, in ~\cite{bensoussan1978asymptotic,sanchezpalencia1980,bakhvalov1989}.

\subsection{Asymptotic homogenization}
In this subsection, we briefly review the basic steps of the asymptotic homogenization method applied to problem \eqref{eq:equilibrium}-\eqref{eq:dirichlet} up to the presentation of the mathematical models corresponding to the homogenized problem, effective coefficients and local problems. The details of the homogenization procedure can be found in Chapter 4 of ~\cite{bakhvalov1989}.

A formal asymptotic solution of \eqref{eq:equilibrium}-\eqref{eq:dirichlet} is sought as an expansion of the thermal field \( v^\varepsilon(\mathbf{x}) \) in powers of the small parameter \( \varepsilon \), as follows:
\begin{equation}
	\label{eq:expansion_u}
	v^\varepsilon(\mathbf{x}) = v_0(\mathbf{x}, \mathbf{y}) + \varepsilon\, v_1(\mathbf{x}, \mathbf{y}) + \varepsilon^2\, v_2(\mathbf{x}, \mathbf{y}) + \cdots,
\end{equation}
where each term \( v_k(\mathbf{x}, \mathbf{y}) \) is assumed to be ${Y}$-periodic in \( \mathbf{y} \).

By substituting the asymptotic expansion \eqref{eq:expansion_u} into the equation \eqref{eq:equilibrium}, and applying the chain rule to rewrite derivatives as
\[
\nabla \to \nabla_{\mathbf{x}} + \frac{1}{\varepsilon} \nabla_{\mathbf{y}}
\]
we collect terms of equal powers of $\varepsilon$, leading to a hierarchy of local problems. These problems enable the separation of microscopic and macroscopic behaviors and ultimately allow us to characterize the composite's effective properties.

In the general case of anisotropic conductivities, this procedure yields the following homogenized problem:
\begin{align}
	\nabla\cdot\left( \hat {\kappa}\nabla v_0 \right) &= f && \text{in } \Omega, \label{eq:homog_problem}\\
	v_0 &= 0 && \text{on } \partial \Omega, \label{eq:homog_bc}
\end{align}
where $v_0=v_0(\mathbf{x})$ is the non-perturbed part of the asymptotic \eqref{eq:expansion_u}, and \( \hat {\kappa} \) is the effective conductivity tensor which represents the macroscopic response of the heterogeneous medium and is computed from the solution of periodic cell problems.

The effective conductivity tensor is defined by
\begin{equation}
	\label{eq:keff_def}
	\hat {\kappa} = \left\langle \kappa(\mathbf{y})  - \kappa(\mathbf{y}) \cdot \nabla_y u  \right\rangle_Y,
\end{equation}
where \( \langle * \rangle_Y \) denotes the average over the periodic unit cell \( Y \), and \( u=u(\mathbf{y}) \) is the vector field of the local functions, which are zero-average periodic solutions of the so-called local problems, governed by the following equation on $Y$:

\begin{equation}
	\label{eq:localproblem_eq}
	\nabla_{\mathbf{y}}\cdot\left(\kappa(\mathbf{y})  - \kappa(\mathbf{y}) \cdot \nabla_y u(\mathbf{y})\right)=0,  
\end{equation}
which must be completed with the contact conditions at the interfaces that are derived after inserting \eqref{eq:expansion_u} in \eqref{eq:condition1} and \eqref{eq:condition2}. 
This formulation follows the classical asymptotic homogenization framework for anisotropic media and has been previously applied in the context of thermal gain in nanofluids, treated as fibrous composites with circular inclusions and square periodic cell, in\cite{iglesias2023conductivity}. In the following section, we restrict our attention to a particular case where all phases are isotropic but the periodic cell is a parallelogram (Figure 1(b)), and we solve the associated local problems explicitly using complex variable techniques.

\subsection{Local problems}
In this subsection, we derive the local problems that arise from the homogenization method applied to a composite medium. As shown in \cite{sabina2020}, the local behavior is governed by a set of Laplace equations defined within each of the three distinct regions: matrix, interface, and fiber, respectively. For each subdomain indexed by \( s = 1, 2, 3 \), the local field \( u^{(s)} \) satisfies:
\begin{equation}
	\nabla^2 u^{(s)} = 0,
	\label{ec_LP}
\end{equation}
subject to interface conditions derived from the continuity of temperatures:
\begin{align}
	u^{(1)}|_{\Gamma_1} &= u^{(2)}|_{\Gamma_1}, \label{ec_CDa} \\
	u^{(2)}|_{\Gamma_2} &= u^{(3)}|_{\Gamma_2}, \label{ec_CDb}
\end{align}
and of the continuity of heat flux across interfaces:
\begin{align}
	\left[
	\kappa_1 (u_{,1}^{(1)} + \delta_{1\alpha}) n_1^{(1)} +
	\kappa_1 (u_{,2}^{(1)} + \delta_{2\alpha}) n_2^{(1)}
	\right] \Big|_{\Gamma_1}
	&=
	-\left[
	\kappa_2 (u_{,1}^{(2)} + \delta_{1\alpha}) n_1^{(2)} +
	\kappa_2 (u_{,2}^{(2)} + \delta_{2\alpha}) n_2^{(2)}
	\right] \Big|_{\Gamma_1}, \label{ec_CT1} \\
	\left[
	\kappa_2 (u_{,1}^{(2)} + \delta_{1\alpha}) n_1^{(2)} +
	\kappa_2 (u_{,2}^{(2)} + \delta_{2\alpha}) n_2^{(2)}
	\right] \Big|_{\Gamma_2}
	&=
	-\left[
	\kappa_3 (u_{,1}^{(3)} + \delta_{1\alpha}) n_1^{(3)} +
	\kappa_3 (u_{,2}^{(3)} + \delta_{2\alpha}) n_2^{(3)}
	\right] \Big|_{\Gamma_2}, \label{ec_CT2}
\end{align}
where $\alpha=1,2$ corresponds to the components of the vector field $u$. 
The solutions of the Laplace equations \eqref{ec_LP} in each region \( s = 1, 2, 3 \) are obtained by expressing \( u^{(s)} \) as the real part of a holomorphic function of a complex variable \( z \), denoted by \( \phi_s(z) \). Since the real part of any holomorphic function automatically satisfies Laplace’s equation, we can write:
\begin{equation}
	u^{(s)} = \operatorname{Re}[\phi_s(z)].
	\label{Sol_Gn}
\end{equation}
The expressions for the holomorphic functions \( \phi_s(z) \) in each region are defined as follows \cite{Pobedrya}:
\paragraph*{Matrix Region (\( s = 1 \)):}
\begin{equation}
	\phi_1(z) = \frac{a_0 z}{R_1} + \sum_{p = 1}^{\infty} {}^o a_p \frac{\zeta^{(p-1)}(z / R_1)}{(p - 1)!},
	\label{zetaweirtrass}
\end{equation}
where \( \zeta(z) \) denotes the Weierstrass quasi-periodic function, given by:
\begin{equation}
	\zeta(z) = \frac{1}{z} + \sum_{(m,n) \ne (0,0)} \left( \frac{1}{z - \beta_{mn}} + \frac{1}{\beta_{mn}} + \frac{z}{\beta_{mn}^2} \right).
    \label{zetaWeier}
\end{equation}
and $\zeta^{(p-1)}$ the derivative of order (p-1).\\
\( \beta_{mn} = m \omega_1 + n \omega_2 \), for \( m, n \in \mathbb{Z} \), and \( \omega_1, \omega_2 \) are the fundamental periods of the periodic cell. The symbol \( \sum^o \) indicates that the summation is taken over odd values of the index only.

\paragraph*{Interfacial Region (\( s = 2 \)):}
\begin{equation}
	\phi_2(z) = \sum_{p = 1}^{\infty} {}^o b_p \left( \frac{R_1}{z} \right)^p + \sum_{p = 1}^{\infty} {}^o b_{-p} \left( \frac{z}{R_2} \right)^p.
	\label{Sol_2}
\end{equation}

\paragraph*{Fiber Region (\( s = 3 \)):}
\begin{equation}
	\phi_3(z) = \sum_{p = 1}^{\infty} {}^o c_p \left( \frac{z}{R_2} \right)^p.
	\label{Sol_3}
\end{equation}

The Laurent expansion of the function $\varphi_1(z)$ is given by:
\begin{equation}
	\varphi_1(z) = \frac{z}{R}a_0 + \sum_{p=3}^{\infty}\!\!{\vphantom{\sum}}^{o}\left(\frac{R}{z}\right)^p a_p - \sum_{p=1}^{\infty}\!\!{\vphantom{\sum}}^{o}\sum_{k=1}^{\infty}\!\!{\vphantom{\sum}}^{o} \left(\frac{z}{R}\right)^p \sqrt{\frac{k}{p}} W_{kp} a_k,
	\label{Sol_1}
\end{equation}
where:
\begin{align}
	W_{kp} &= \frac{(k+p-1)! \sqrt{kp}}{k!p!}{R^{k+p}}S_{k+p}, \\
	S_{k+p} &= \sum_{m,n} (m\omega_1 + n\omega_2)^{-(k+p)}, \quad m^2+n^2 \neq 0, \quad k+p > 2,
\end{align}
and by definition $S_2 = 0$. The constants $a_0$, $a_p$, $b_p$, and $z$ are complex numbers. The undetermined coefficients $a_0$, $a_p$, $b_p$  will be identified by applying the interface (\ref{ec_CDa}) - (\ref{ec_CT2}) and periodicity conditions.

We impose periodicity of the solution \( u^{(1)} \) with respect to both lattice directions by requiring that:
\begin{equation}
	u^{(1)} (z + \omega_\alpha) - u^{(1)} (z) = \operatorname{Re} \left( \frac{\omega_\alpha}{R_1} a_0 + R_1\, \delta_\alpha\, a_1 \right) = 0,
\end{equation}
where \( \omega_\alpha \in \{\omega_1, \omega_2\} \) are the fundamental periods of the unit cell, and the quantity \( \delta_\alpha \) is defined by the jump of the Weierstrass zeta function:
\begin{equation}
	\delta_\alpha = \zeta(z + \omega_\alpha) - \zeta(z) = 2\, \zeta\left( \frac{\omega_\alpha}{2} \right).
\end{equation}

From the periodicity condition, the coefficient \( a_0 \) can be explicitly determined in terms of \( a_1 \) as:
\begin{equation}
	a_0 = - R_1^2 \frac{\delta_1\, \overline{\omega_2} - \delta_2\, \overline{\omega_1}}{\omega_1\, \overline{\omega_2} - \omega_2\, \overline{\omega_1}}\, a_1 
	- R_1^2 \frac{\overline{\delta_1}\, \overline{\omega_2} - \overline{\delta_2}\, \overline{\omega_1}}{\omega_1\, \overline{\omega_2} - \omega_2\, \overline{\omega_1}}\, \overline{a_1}.
	\label{eq:a0}
\end{equation}
Here, the overline symbol \( \overline{\,\cdot\,} \) denotes complex conjugation.

By applying the temperature continuity conditions at the interfaces \( \Gamma_\gamma \) (\( \gamma=1,2 \)) as expressed in equations~\eqref{ec_CDa} and~\eqref{ec_CDb}, and substituting the functional forms from  (\ref{Sol_2})-(\ref{Sol_1}), we obtain algebraic relations between the undetermined coefficients appearing in the holomorphic functions \( \phi_1(z), \phi_2(z) \), and \( \phi_3(z) \). The displacement continuity condition takes the form:
\begin{equation}
	\operatorname{Re} \left[ \phi_\gamma(z) \right]_{\Gamma_\gamma} = \operatorname{Re} \left[ \phi_{\gamma + 1}(z) \right]_{\Gamma_\gamma},
	\label{eq:cont_phi}
\end{equation}
which yields the following matching conditions between coefficients:
\begin{align}
	b_p \left( \frac{R_1}{R_2} \right)^p + \overline{b_{-p}} &= \overline{a_p} + \delta_{1p} a_0 + \sum_{k = 1}^\infty {}^o \sqrt{\frac{k}{p}} W_{kp} a_k, \label{eq:match1} \\
	b_p + \overline{b_{-p}} \left( \frac{R_1}{R_2} \right)^p &= c_p,\label{eq:match2}
\end{align}

for $p=1,3,5, ...$, while the Kronecker delta \( \delta_{1p} \) selects the term involving \( a_0 \) when \( p = 1 \), and the summation is restricted to odd values of \( k \), as previously defined.

On the other hand, the continuity of normal flux at the interfaces, originally given by equations ~\eqref{ec_CT1} and \eqref{ec_CT2} can be reformulated using the Cauchy-Riemann equations relations for holomorphic functions, obtaining
\begin{equation}
	\left. \left\{ \left( \phi_1(z) - \overline{\phi}_1(z) \right) + (1 - \rho_1) \left[ (z - \overline{z}) \delta_{1\alpha} - i(z + \overline{z}) \delta_{2\alpha} \right] \right\} \right|_{\Gamma_1} = 
	\rho_1 \left. \left[ \phi_2(z) - \overline{\phi}_2(z) \right] \right|_{\Gamma_1}, \label{eq:stress2}
\end{equation}
and
\begin{equation}
	\left. \left\{ \rho_1 \left( \phi_2(z) - \overline{\phi}_2(z) \right) + (\rho_1 - \rho_2) \left[ (z - \overline{z}) \delta_{1\alpha} - i(z + \overline{z}) \delta_{2\alpha} \right] \right\} \right|_{\Gamma_2} = 
	\rho_2 \left. \left[ \phi_3(z) - \overline{\phi}_3(z) \right] \right|_{\Gamma_2}, \label{eq:stress3}
\end{equation}
where the parameters \( \rho_1 = \kappa_2 / \kappa_1 \) and \( \rho_2 = \kappa_3 / \kappa_1 \) represent the ratios of thermal conductivities (or analogous physical properties, depending on the context).

By substituting the expressions for the functions \eqref{Sol_3} - \eqref{Sol_1} into the interfacial condition \eqref{eq:stress2}, and performing algebraic manipulations by comparing terms with the same powers of \( z \), we obtain the following new set of identities relating the undetermined coefficients
\begin{equation}
	\delta_{1p} a_0 - \overline{a_p} + \sum_{k = 1}^\infty {}^o \sqrt{\frac{k}{p}} W_{kp} a_k 
	+ (1 - \rho_1)\, \delta_{1p} (\delta_{1\alpha} - i \delta_{2\alpha}) R_1 
	= \rho_1 \left[ \left( \frac{R_1}{R_2} \right)^p b_p - \overline{b_{-p}} \right].
	\label{eq:stress_cont1}
\end{equation}

Applying the same procedure to the second continuity condition at the interface \( \Gamma_2 \), from equation~\eqref{eq:stress3}, we obtain:
\begin{equation}
	\rho_1 \left[ b_p - \left( \frac{R_1}{R_2} \right)^p \overline{b_{-p}} \right] 
	+ (\rho_1 - \rho_2)\, \delta_{1p} (\delta_{1\alpha} - i \delta_{2\alpha}) R_2 
	= \rho_2\, c_p.
	\label{eq:stress_cont2}
\end{equation}

The grouping of the resulting expressions leads to a system of equations for the undetermined coefficients. In particular, we derive two equivalent representations for \( c_p \), expressed entirely in terms of the coefficients \( a_p \), \( a_0 \), and the known material and geometric parameters:
\begin{equation}
	c_p = \frac{\rho_1}{A^{+}} \left( \overline{a_p} + \delta_{1p} a_0 + \sum_{k = 1}^\infty {}^o \sqrt{\frac{k}{p}} W_{kp} a_k 
	+ \left( \frac{R_1^{2p} - R_2^{2p}}{R_1^p R_2^p} \right) \frac{\rho_1 - \rho_2}{2 \rho_1}\,
	\delta_{1p} (\delta_{1\alpha} - i \delta_{2\alpha}) R_2 \right),
	\label{eq:cp1}
\end{equation}
\begin{equation}
	c_p = \frac{1}{A^{-}} \left( \delta_{1p} a_0 - \overline{a_p} + \sum_{k = 1}^\infty {}^o \sqrt{\frac{k}{p}} W_{kp} a_k \right) 
	+ \frac{1}{A^{-}} \left( (1 - \rho_1) R_1 
	+ \left( \frac{R_1^{2p} + R_2^{2p}}{R_1^p R_2^p} \right) \frac{\rho_1 - \rho_2}{2} R_2 \right) 
	\delta_{1p} (\delta_{1\alpha} - i \delta_{2\alpha}),
	\label{eq:cp2}
\end{equation}

where 
\[
	A^{\pm}=\frac{R_1^{2p}(\kappa_1+\kappa_2)\pm R_2^{2p}(\kappa_1-\kappa_2)}{2R_1^{p}R_2^{p}}.
\]

By equating the two representations \eqref{eq:cp1} and \eqref{eq:cp2}, we arrive at an infinite linear system for the unknown coefficients \( a_k \), given by:
\begin{equation}
	\overline{a_p} - \chi_p \left( \delta_{1p} a_0 + \sum_{k = 1}^\infty {}^o \sqrt{\frac{k}{p}} W_{kp} a_k \right) 
	= R_1 \chi_1\, \delta_{1p} (\delta_{1\alpha} - i \delta_{2\alpha}),
	\label{eq:system}
\end{equation}
where the scalar coefficient \( \chi_p \) is defined as:
\begin{equation}
	\chi_p = 
	\frac{
	(1 - \rho_1)(\rho_1 + \rho_2)(V_2 + V_3)^p + (1 + \rho_1)(\rho_1 - \rho_2) V_3^p
	}{
	(1 + \rho_1)(\rho_1 + \rho_2)(V_2 + V_3)^p + (1 - \rho_1)(\rho_1 - \rho_2) V_3^p
	}.
	\label{eq:chip}
\end{equation}

By enforcing the continuity of both temperature and flux across the interfaces and applying the functional representations of the local solutions, we have derived an infinite linear system governing the undetermined coefficients \( a_k \). This system captures the influence of material contrast, interfacial geometry, and lattice structure through the terms \( \rho_{\alpha} \), \( V_{i} \), \( W_{kp} \), and \( \chi_p \), respectively. In the next section, we use the solution of this system to obtain closed-form formulas for the effective conductivity tensor.

\subsection{Analytical Derivation of Effective Conductivity Tensor}
The effective thermal conductivity coefficients are calculated using the solutions of the local problems described in equations~\eqref{ec_CDa}–\eqref{Sol_Gn}. Based on the asymptotic homogenization framework presented in~\cite{sabina2020}, the components of the effective conductivity tensor of the fibrous composite with isotropic constitutes take the following form:
\begin{equation}
\begin{aligned}
	\hat{\kappa}_{11} &= \kappa_1 V_1 + \kappa_2 V_2 + \kappa_3 V_3 + \left\langle \kappa\, u_{,1}(x) \right\rangle, \\
	\hat{\kappa}_{12} &= \hat{\kappa}_{21} = \left\langle \kappa\, u_{,2}(x) \right\rangle = \left\langle \kappa\, u_{,1}(x) \right\rangle, \\
	\hat{\kappa}_{22} &= \kappa_1 V_1 + \kappa_2 V_2 + \kappa_3 V_3 + \left\langle \kappa\, u_{,2}(x) \right\rangle,
\end{aligned}
\label{EFFec}
\end{equation}
where \( \langle \cdot \rangle \) denotes the average over the periodic cell. These expressions correspond to double integrals on the periodic unit cell shown in Figure~\ref{fig:Fig_1}. Applying Green's theorem together with the periodicity conditions on opposite sides of the periodic cell, and the continuity conditions at the interfaces, we obtain, for the first pair of coefficients, the following complex-valued expression:
\begin{equation}
	\hat{\kappa}_{11} - i \hat{\kappa}_{21} = \left\langle \kappa \right\rangle  
	- \frac{\kappa_1 - \kappa_2}{V} \left( \int_{\Gamma_1} u^{(1)}\, dx_2 + i \int_{\Gamma_1} u^{(1)}\, dx_1 \right)
	- \frac{\kappa_2 - \kappa_3}{V} \left( \int_{\Gamma_2} u^{(3)}\, dx_2 + i \int_{\Gamma_2} u^{(3)}\, dx_1 \right).
\end{equation}
By evaluating the local fields on the interfaces and expressing the result in terms of the residue \( a_1 \) of the holomorphic function \( \phi_1(z) \), the expression simplifies to:
\begin{equation}
\begin{aligned}
	\hat{\kappa}_{11} - i \hat{\kappa}_{21} = \left\langle \kappa \right\rangle 
	&- \kappa_1 (V_2 + V_3) \left[ 1 - \rho_1 
	+ \frac{2 V_3 \rho_1 (\rho_1 - \rho_2)}{V_2 (\rho_1 + \rho_2) + 2 V_3 \rho_1} \right] 
	\cdot \frac{(\chi_1 + 1)\, \overline{a_1} - R_1 \chi_1}{R_1 \chi_1} \\
	&- \frac{\kappa_1 (\rho_1 - \rho_2)^2 V_2 V_3}{V_2 (\rho_1 + \rho_2) + 2 V_3 \rho_1}.
\end{aligned}
\label{c1313masC2313}
\end{equation}

Similarly, for the remaining pair of effective coefficients, we obtain:
\begin{equation}
\begin{aligned}
	\hat{\kappa}_{12} - i \hat{\kappa}_{22} = -i \left\langle \kappa \right\rangle 
	&- \kappa_1 (V_2 + V_3) \left[ 1 - \rho_1 
	+ \frac{2 V_3 \rho_1 (\rho_1 - \rho_2)}{V_2 (\rho_1 + \rho_2) + 2 V_3 \rho_1} \right] 
	\cdot \frac{(\chi_1 + 1)\, \overline{a_1} + i R_1 \chi_1}{R_1 \chi_1} \\
	&+ \frac{i\, \kappa_1 (\rho_1 - \rho_2)^2 V_2 V_3}{V_2 (\rho_1 + \rho_2) + 2 V_3 \rho_1}.
\end{aligned}
\label{c1323masC2323}
\end{equation}

By evaluating the interfacial integrals and expressing the local solutions in terms of the residue \( a_1 \), we have obtained closed-form expressions for the effective conductivity tensor components. In the following subsection, we focus on the solution of the infinite linear systems governing the coefficients \( a_k \), and explore how these solutions lead to simplified analytical approximations of the effective thermal properties.

\subsection{Solution of the Infinite Systems and Analytical Formulas for the Effective Coefficients}
To derive closed-form expressions for the effective thermal conductivity coefficients, it is necessary to solve the infinite linear system ~\eqref{eq:system}, in order to obtain the undetermined constant \( a_1 \) which plays a central role in the residue-based formulation.

The system can be rewritten in the following compact form:
\begin{equation}
	\overline{a}_p 
	+ \chi_1 R_1^2 H_1\, \delta_{1p} \overline{a}_1 
	+ \chi_1 R_1^2 H_2\, \delta_{1p} a_1 
	+ \chi_p \sum_{k = 1}^{\infty} {}^o W_{kp} a_k 
	= R_1 \chi_1 \delta_{1p} (\delta_{1\alpha} - i \delta_{2\alpha}),
	\label{sistemafinal}
\end{equation}
where the auxiliary constants and matrix elements are defined as:
\begin{equation}
	H_1 = \frac{\pi} {Img(\overline{\omega_1}\omega_2)}, \quad
	H_2 = \frac{\delta_1 \, \overline{\omega_2} - \delta_2 \, \overline{\omega_1}}{-2iImg(\overline{\omega_1}\omega_2)}, \quad
	W_{kp} = \frac{(k + p - 1)!}{(k - 1)! (p - 1)!} \cdot \frac{S_{k + p} R^{k + p}}{\sqrt{kp}},
    \label{H1H2}
\end{equation}
\begin{equation}
	\delta_\alpha = \zeta(z + \omega_\alpha) - \zeta(z) = 2\,\zeta(\omega_\alpha / 2),
    \label{deltaalp}
\end{equation}
and $\zeta(z)$ denotes the Weierstrass zeta function (\ref{zetaWeier})
and the interfacial contrast function \( \chi_p \) defined in  ~\eqref{eq:chip}.

To solve the infinite system~\eqref{sistemafinal}, we express it in matrix form by separating real and imaginary parts. Let
\[
a_p = x_p + i y_p, \quad 
W_{kp} = w_{1kp} + i w_{2kp}, \quad 
H_1=h ,\quad 
H_2 = h_{12} + i h_{22},
\]
where \( x_p, y_p, w_{1kp}, w_{2kp}, h_{1\alpha}, h_{2\alpha} \in \mathbb{R} \) represent the real and imaginary parts of the corresponding complex quantities.

The system is then written in the following real-valued matrix form:
\begin{equation}
\begin{aligned}
	\left( \begin{array}{cc}
		1 & 0 \\
		0 & 1
	\end{array} \right)
	\left( \begin{array}{c}
		x_p \\
		y_p
	\end{array} \right)
	&+ \chi_1 R_1^2 
	\left( \begin{array}{cc}
		h + h_{12} &  - h_{22} \\
		 - h_{22} & h - h_{12}
	\end{array} \right)
	\left( \begin{array}{c}
		x_1 \\
		y_1
	\end{array} \right) \delta_{1p} \\
	&+ \chi_p 
	\left( \begin{array}{cc}
		w_{1kp} & -w_{2kp} \\
		- w_{2kp} & -w_{1kp}
	\end{array} \right)
	\left( \begin{array}{c}
		x_k \\
		y_k
	\end{array} \right)
	= R_1 \chi_1 
	\left( \begin{array}{c}
		\delta_{1\alpha} \\
		\delta_{2\alpha}
	\end{array} \right) \delta_{1p}.
\end{aligned}
\label{Sistema_Matricial2}
\end{equation}

Here, the summation implied in the term with \( W_{kp} \) is taken over odd indices \( k = 1, 3, 5, \dots \) for each fixed \( p = 1, 3, 5, \dots \). The index \( \alpha = 1, 2 \) selects the applied direction in the local problem.
 
The system~\eqref{Sistema_Matricial2} can be solved explicitly. To do so, we proceed by first isolating the equation corresponding to \( p = 1 \), which yields:
\[
\left[
	\left( \begin{array}{cc}
		1 & 0 \\
		0 & 1
	\end{array} \right)
	+ \chi_1 R_1^2
	\left( \begin{array}{cc}
		h + h_{12} &  - h_{22} \\
		 - h_{22} & h - h_{12}
	\end{array} \right)
\right]
\begin{pmatrix}
	x_1 \\
	y_1
\end{pmatrix}
+ \chi_1
\left( \begin{array}{cc}
	w_{1k1} & -w_{2k1} \\
	- w_{2k1} & -w_{1k1}
\end{array} \right)
\begin{pmatrix}
	x_k \\
	y_k
\end{pmatrix}
= R_1 \chi_1
\begin{pmatrix}
	\delta_{1\alpha} \\
	\delta_{2\alpha}
\end{pmatrix}.
\]

This can be compactly written as:
\begin{equation}
	\left( I + \chi_1 R_1^2 J_1 \right) X + \chi_1 N_1 X_1 = R_1 \chi_1 B,
	\label{Sistama_p1}
\end{equation}
where:
\[
J_1 =\left( \begin{array}{cc}
		h + h_{12} &  - h_{22} \\
		 - h_{22} & h - h_{12}
	\end{array} \right)
 \quad
X =
\begin{pmatrix}
	x_1 \\
	y_1
\end{pmatrix}, \quad
B =
\begin{pmatrix}
	\delta_{1\alpha} \\
	\delta_{2\alpha}
\end{pmatrix}.
\]

The matrix \( N_1 \equiv N_1(n_{k1}) \) is composed of two rows of square \( 2 \times 2 \) blocks of the form:
\[
n_{k1} =
\begin{pmatrix}
	w_{1k1} & -w_{2k1} \\
	- w_{2k1} & -w_{1k1}
\end{pmatrix}, \quad \text{for } k = 2t + 1.
\]

The corresponding vector \( X_1 \) stacks the unknowns for higher-order indices and is defined by:
\[
X_1^T = (x_3, y_3, x_5, y_5, \dots).
\]

Now, considering the case \( p > 1 \) in system~\eqref{Sistema_Matricial2}, and summing over \( k = 1, 3, 5, \dots \), we obtain—for each value of \( p \)—a pair of homogeneous equations of the form:
\begin{equation}
	\left( \begin{array}{cc}
		1 & 0 \\
		0 & 1
	\end{array} \right)
	\begin{pmatrix}
		x_p \\
		y_p
	\end{pmatrix}
	+ \chi_p
	\left( \begin{array}{cc}
		w_{1kp} & -w_{2kp} \\
		- w_{2kp} & -w_{1kp}
	\end{array} \right)
	\begin{pmatrix}
		x_k \\
		y_k
	\end{pmatrix}
	= 
	\begin{pmatrix}
		0 \\
		0
	\end{pmatrix}.
	\label{Sistema_Matricial3}
\end{equation}

To obtain an analytical solution to the infinite system, we truncate it by considering that the indices \( k \) and \( p \) take values in a finite set of odd integers
\[
k = 1, 3, 5, \dots, 2n + 1, \qquad p = 3, 5, \dots, 2n + 1.
\]
Separating the components \( x_1, y_1 \) in the \( p > 1 \) equations of~\eqref{Sistema_Matricial3} and moving them to the right-hand side yields the system:
\begin{equation}
	(I + W) X_1 = - N_2 X,
	\label{ImasW}
\end{equation}
where \( X_1 \) is the column vector of unknowns \((x_k, y_k)^T\) for \( k > 1 \). The matrix \( N_2 \equiv N_2(n_{1p}) \) consists of two columns of \( 2 \times 2 \) blocks defined by:
\[
n_{1p} = \chi_p 
\begin{pmatrix}
	w_{11p} & -w_{21p} \\
	- w_{21p} & -w_{11p}
\end{pmatrix}.
\]

Solving~\eqref{ImasW} gives:
\begin{equation}
	X_1 = - (I + W)^{-1} N_2 X,
	\label{X1}
\end{equation}
which expresses the higher-order unknowns in terms of \( x_1 \) and \( y_1 \). The matrix \( W \equiv W(w_{kp}) \) is composed of \( 2 \times 2 \) blocks given by:
\[
w_{kp} = \chi_p 
\begin{pmatrix}
	w_{1kp} & -w_{2kp} \\
	- w_{2kp} & -w_{1kp}
\end{pmatrix}, \quad k = 2t + 1, \quad p = 2s + 1.
\]

Substituting~\eqref{X1} into~\eqref{Sistama_p1}, and defining the matrix \( Z \) as:
\begin{equation}
	Z = I + \chi_1 R_1^2 J_1 - \chi_1 N_1 Y^{-1} N_2,
	\label{matrizZ}
\end{equation}
where
\begin{equation}
	Y = I + W,
	\label{Y}
\end{equation}
the solution for \( a_1 \) becomes:
\begin{equation}
	a_1 = \begin{pmatrix} 1 & i \end{pmatrix}
	\begin{pmatrix} x_1 \\ y_1 \end{pmatrix}
	= R_1 \chi_1 
	\begin{pmatrix} 1 & i \end{pmatrix}
	Z^{-1} 
	\begin{pmatrix} \delta_{1\alpha} \\ \delta_{2\alpha} \end{pmatrix}.
\end{equation}

Substituting this expression for \( a_1 \) into equations~\eqref{c1313masC2313}–\eqref{c1323masC2323}, and simplifying, we obtain the following closed-form expressions for the components of the effective conductivity tensor:

\begin{equation}
	\hat{\kappa}_{11}= \kappa_1\left(1 - 2(V_2+ V_3)\Delta\frac{z_{22}}{|Z|} \right),
	\label{cc1313}
\end{equation}

\begin{equation}
	\hat{\kappa}_{21} = 2\kappa_1(V_2 + V_3)\Delta\frac{ z_{12}}{|Z|},
	\label{cc2313}
\end{equation}

\begin{equation}
	\hat{\kappa}_{12} = 2\kappa_1(V_2 + V_3)\Delta\frac{z_{21}}{|Z|},
	\label{cc1323}
\end{equation}

\begin{equation}
	\hat{\kappa}_{22}= \kappa_1 \left(1 - 2(V_2 + V_3) \Delta\frac{z_{11}}{|Z|} \right),
	\label{cc2323}
\end{equation}
where the scalar functions \( \Delta \) is:
\begin{equation}
\Delta  = \frac{V_2(\rho_1 + \rho_2)(1 - \rho_1) + 2V_3\rho_1(1 - \rho_2)}{(1+\rho_1)(\rho_1+\rho_2)(V_2+V_3) + (1-\rho_1)(\rho_1-\rho_2)V_3}.
	\label{Fchi_nueva}
\end{equation}

Here, \( z_{ij} \) are the components of the matrix \( Z \), and \( |Z| \) denotes its determinant. The parameters \( V_2 \) and \( V_3 \) represent the volume fractions of the interphase and central fiber, respectively. The contrast parameters are given by \( \rho_1 = \kappa_2 / \kappa_1 \) and \( \rho_2 = \kappa_3 / \kappa_1 \).

The expressions~\eqref{cc1313}–\eqref{cc2323} define the effective thermal conductivity tensor \( \hat{\kappa} \) of the analyzed  three-phase fibrous composite 
\begin{equation}
	\hat{\kappa} = 
	\begin{pmatrix}
		\hat{\kappa}_{11} & \hat{\kappa}_{12} \\
		\hat{\kappa}_{21} & \hat{\kappa}_{22}
	\end{pmatrix}.
	\label{matriz_efectiva}
\end{equation}

This tensor is symmetric, i.e., \( \hat{\kappa}_{12} = \hat{\kappa}_{21} \). Furthermore, for composites with hexagonal (\( \theta = 60^\circ \)) or square (\( \theta = 90^\circ \)) periodic cells and a configuration such that \( |\omega_1| = |\omega_2| = 1 \), the symmetry of the structure implies that \( \hat{\kappa}_{11} = \hat{\kappa}_{22} \) and \( \hat{\kappa}_{12} = 0 \).

The area of the periodic unit cell is given by:
\[
V = |\omega_1|\, |\omega_2|\, \sin \theta,
\]
and the maximum fiber area fraction (percolation limit) is:
\[
V_{f,\text{max}} = V_2 + V_3 = \frac{\pi r_m^2}{V}, \quad 
r_m =\frac{1}{2} \min \left\{ {|\omega_1|},{|\omega_2|},d \right\},\quad d=\sqrt{{|\omega_1|}^2+{|\omega_2|}^2-2|\omega_1||\omega_2|\cos\theta}.
\]

\subsection{Critical Relative Interfacial Layer Thickness}
In composite materials with coated or debonded inclusions, the interfacial layer plays a crucial role in heat transport performance. As discussed in~\cite{Lu1995}, two contrasting scenarios arise depending on the thermal conductivity of the inclusion and interfacial layer. For composites with low-conductivity interphases (e.g., debonded inclusions), the effective thermal conductivity decreases as the interfacial volume fraction \( V_2/V_3 \) increases, eventually reaching a critical value where the inclusion's beneficial effects are fully suppressed. Conversely, when a poorly conducting inclusion is coated with a high-conductivity interphase, the effective conductivity increases with \( V_2/V_3 \). In this case, a critical thickness also exists, above which the coating compensates for the inclusion’s poor conduction.

This critical interfacial thickness ratio \( \lambda = V_2/V_3 \) marks the point where the composite’s effective conductivity equals that of the matrix, i.e., \( \hat{\kappa} = \kappa_1 I \), regardless of fiber volume fraction \( V_f = V_2 + V_3 \). Based on the expressions derived in equations~\eqref{cc1313}–\eqref{cc2323}, we now define and derive an analytical expression for \( \lambda \) under the assumption of isotropic matrix behavior, i.e.
\begin{equation}
	\hat \kappa = \kappa_1 
	\begin{pmatrix}
		1 & 0 \\
		0 & 1
	\end{pmatrix},
\end{equation}
which implies that \( \hat{\kappa}_{11} = \hat{\kappa}_{22} = \kappa_1 \) and \( \hat{\kappa}_{12} = \hat{\kappa}_{21} = 0 \). Substituting this condition into equations~\eqref{cc1313}–\eqref{cc2323}, we require the parameter \( \Delta = 0 \) for the matrix-dominated limit.

Taking the relation
\[
V_2(\rho_1 + \rho_2)(1 - \rho_1) + 2 V_3 \rho_1 (1 - \rho_2) = 0.
\]
and solving for the volume ratio \( \lambda = V_2/V_3 \), we obtain the expression for the critical relative interfacial layer thickness:
\begin{equation}
	\lambda = \frac{V_2}{V_3} = \frac{2 \rho_1 (\rho_2 - 1)}{(1 - \rho_1)(\rho_1 + \rho_2)}.
	\label{lam}
\end{equation}

Substituting \( \Delta = 0 \) into the effective conductivity expressions~\eqref{cc1313}–\eqref{cc2323}, we find that \( \hat{\kappa}_{11} = \hat{\kappa}_{22} = \kappa_1 \) and \( \hat{\kappa}_{12} = \hat{\kappa}_{21} = 0 \), confirming that the effective response of the composite reduces to that of the isotropic matrix.\\

To ensure that the expression in~\eqref{lam} is physically meaningful, the right-hand side must be positive. This occurs only when \( (\rho_2 - 1) \) and \( (1 - \rho_1) \) have the same sign. Therefore, a critical interfacial thickness can be defined under the following two conditions:
(a) when \( \rho_1 < 1 \) and \( \rho_2 > 1 \), or 
(b) when \( \rho_1 > 1 \) and \( \rho_2 < 1 \). In both cases, the critical ratio exists only when the value 1 lies strictly between the two conductivity contrasts \( \rho_1 \) and \( \rho_2 \). Since \( V_2/V_3 > 0 \), this ensures that the right-hand side of~\eqref{lam} remains positive and physically consistent.

Notably, the critical interfacial thickness \( \lambda \) depends solely on the contrast parameters \( \rho_1 \) and \( \rho_2 \), and is entirely independent of the inclusion volume fraction or the periodic geometry of the composite. It thus provides an intrinsic material-based threshold beyond which the interfacial layer either shields or enhances the inclusion’s contribution to the composite’s effective thermal conductivity.

An expression for the critical relative interfacial layer thickness was previously given as equation (9) in~\cite{Lu1995}. The critical condition derived in~\eqref{lam} coincides with Lu's result when translated into our notation for the interphase-to-inclusion geometry. From Figure~\ref{fig:Fig_1}, we have:
\[
V_3 = \frac{\pi R_2^2}{V}, \qquad V_2 = \frac{\pi(R_2 + t)^2}{V} - \frac{\pi R_2^2}{V},
\]
so the volume ratio becomes:
\[
\frac{V_2}{V_3} = \left( \frac{t}{R_2} + 1 \right)^2 - 1 = \frac{2\rho_1(\rho_2 - 1)}{(1 - \rho_1)(\rho_1 + \rho_2)}.
\]
This leads to the following relation for the critical normalized interfacial thickness $t$:
\begin{equation}
	\left( \frac{t}{R_2} + 1 \right)^2 = \frac{(1 + \rho_1)(\rho_2 - \rho_1)}{(1 - \rho_1)(\rho_1 + \rho_2)}.
	\label{lam1}
\end{equation}




\section{Simple Analytical Formulas for Effective Thermal Conductivity of Three-phase Fibrous Composites}\label{section3}

This section presents simplified analytical approximations for the effective thermal conductivity of fibrous composites by truncating the infinite system to low-order terms to offer a balance between accuracy and mathematical tractability.

To obtain manageable expressions, we evaluate the systems defined in equations~\eqref{Sistama_p1} and~\eqref{Sistema_Matricial3} for specific orders \(n = 0, 1, 2, 3, 4, 5\). The effective conductivity coefficients given in equations~\eqref{cc1313}–\eqref{cc2323} depend on the components of the matrix \(Z\) (equation~\eqref{matrizZ}), which in turn depends on the matrices \(N_1\) (of size \(2 \times 2n\)), \(N_2\) (\(2n \times 2\)), and \(Y\) (\(2n \times 2n\)) defined in equation~\eqref{Y}. By choosing specific values of \(n\), we construct increasingly refined approximations of effective coefficients.

Below we present explicit expressions for these matrices and corresponding conductivity approximations for selected values of \(n\).

\paragraph*{Case \( n = 0 \) \textnormal{($O_0$)}:}

In this approximation, we consider a truncated form of the matrix \( Z \), defined as:
\begin{equation}
	Z = I + \chi_1 R_1^2 J_1,
	\label{Zcorta}
\end{equation}
where
\begin{equation}
    I = \begin{pmatrix}
        1 & 0 \\
        0 & 1
    \end{pmatrix}, \quad
    J_1 = \begin{pmatrix}
        h + h_{12} &  - h_{22} \\
         - h_{22} & h - h_{12}
    \end{pmatrix},
    \label{Unit_J1}
\end{equation}
with
\begin{equation}
h = H_1, \quad h_{12} = \text{Re}(H_2), \quad h_{22} = \text{Img}(H_2). 
\label{h12h22}
\end{equation}

The interfacial parameter \( \chi_1 \) is given by:
\[
\chi_1 = \frac{(1 - \rho_1)(\rho_1 + \rho_2)(V_2 + V_3) + (1 + \rho_1)(\rho_1 - \rho_2)V_3}
{(1 + \rho_1)(\rho_1 + \rho_2)(V_2 + V_3) + (1 - \rho_1)(\rho_1 - \rho_2)V_3}.
\]

This approximate expression for the effective thermal conductivity coincides with formula (9a) in \cite{Lu1995}, p. 2617, for \(\theta = 60^\circ\) (hexagonal configuration) and \(\theta = 90^\circ\) (square configuration).

In higher-order approximations (\( n \geq 1 \)), the matrix \( Z \) is defined as:
\begin{equation}
	Z = I + \chi_1 R_1^2 J_1 - \chi_1 N_1 Y^{-1} N_2,
	\label{matrizZn0}
\end{equation}
with
\[
Y = I + W,
\]
where \( I \) is the identity matrix of size \( 2n \times 2n \).

The following matrix \( L_k \) plays a fundamental role in simplifying the analytical expressions:
\begin{equation}
	L_k = \begin{pmatrix}
		\text{Re}(S_k) & -\text{Im}(S_k) \\
		-\text{Im}(S_k) & -\text{Re}(S_k)
	\end{pmatrix}.
    \label{Lk}
\end{equation}

The higher-order interfacial parameters \( \chi_p \) used in subsequent approximations are defined as:
\begin{equation}
	\chi_p = \frac{(1 - \rho_1)(\rho_1 + \rho_2)(V_2 + V_3)^p + (1 + \rho_1)(\rho_1 - \rho_2)V_3^p}
	{(1 + \rho_1)(\rho_1 + \rho_2)(V_2 + V_3)^p + (1 - \rho_1)(\rho_1 - \rho_2)V_3^p}, \quad p = 1, 3, 5, \dots
	\label{chi_p}
\end{equation}

Note that as the truncation order \( n \) increases in system~\eqref{sistemafinal}, the only changes in the matrix \( Z \) occur through the matrices \( N_1 \), \( N_2 \), and \( Y \). Once the components of these matrices are specified, the effective thermal conductivity can be computed. Since \( Y = I + W \), and \( I \) is known, it is sufficient to determine the components of matrix \( W \) for each case.

\paragraph*{Case \( n = 1 \) \textnormal{($O_1$)}:}

In this first-order approximation, the matrices \( N_1 \), \( N_2 \), and \( W \) take the following simplified forms:
\begin{equation}
	N_1 = \sqrt{3}\, R_1^4 L_4, \quad 
	N_2 = \chi_3 N_1, \quad 
	W = 10\, R_1^6 \chi_3 L_6.
    \label{Orden1}
\end{equation}

\paragraph*{Case \( n = 2 \) \textnormal{($O_2$)}:}
The matrix \( N_1 \) is structured as a row of two \( 2 \times 2 \) blocks:
\begin{equation}
	N_1 = \begin{pmatrix}
		N_{11} & N_{12}
	\end{pmatrix}, \quad \text{where} \quad
	N_{11} = \sqrt{3}\, R_1^4 L_4, \quad
	N_{12} = \sqrt{5}\, R_1^6 L_6.
\end{equation}

The transpose matrix \( N_2 \) is given by:
\begin{equation}
	N_2 = \begin{pmatrix}
		\chi_3 N_{11} & \chi_5 N_{12}
	\end{pmatrix}^T.
\end{equation}

The matrix \( W \) is a \( 2 \times 2 \) block matrix, symmetric by blocks:
\begin{equation}
	W = \begin{pmatrix}
		W_{11} & W_{12} \\
		W_{21} & W_{22}
	\end{pmatrix},
\end{equation}
with block components:
\begin{align*}
	W_{11} &= 10\, R_1^6 \chi_3 L_6, &
	W_{12} &= 7\sqrt{15}\, R_1^8 \chi_3 L_8, \\
	W_{21} &= 7\sqrt{15}\, R_1^8 \chi_5 L_8, &
	W_{22} &= 126\, R_1^{10} \chi_5 L_{10}.
\end{align*}

\paragraph*{Case \( n = 3 \) \textnormal{($O_3$)}:}
The matrix \( N_1 \) is structured as a row of three \( 2 \times 2 \) blocks:
\begin{equation}
	N_1 = \begin{pmatrix}
		N_{11} & N_{12} & N_{13}
	\end{pmatrix}, \quad \text{where} \quad
	N_{11} = \sqrt{3}\, R_1^4 L_4,\quad
	N_{12} = \sqrt{5}\, R_1^6 L_6,\quad
	N_{13} = \sqrt{7}\, R_1^8 L_8.
\end{equation}

The transpose matrix \( N_2 \) is given by:
\begin{equation}
	N_2 = \begin{pmatrix}
		\chi_3 N_{11} & \chi_5 N_{12} & \chi_7 N_{13}
	\end{pmatrix}^T.
\end{equation}

The matrix \( W \) is a \( 3 \times 3 \) block matrix:
\begin{equation}
	W = \begin{pmatrix}
		W_{11} & W_{12} & W_{13} \\
		W_{21} & W_{22} & W_{23} \\
		W_{31} & W_{32} & W_{33}
	\end{pmatrix},
\end{equation}
with block components:
\begin{align*}
	W_{11} &= 10\, \chi_3 R_1^6 L_6, &
	W_{12} &= 7\sqrt{15}\, \chi_3 R_1^8 L_8, &
	W_{13} &= 12\sqrt{21}\, \chi_3 R_1^{10} L_{10}, \\
	W_{21} &= 7\sqrt{15}\, \chi_5 R_1^8 L_8, &
	W_{22} &= 126\, \chi_5 R_1^{10} L_{10}, &
	W_{23} &= 66\sqrt{35}\, \chi_5 R_1^{12} L_{12}, \\
	W_{31} &= 12\sqrt{21}\, \chi_7 R_1^{10} L_{10}, &
	W_{32} &= 66\sqrt{35}\, \chi_7 R_1^{12} L_{12}, &
	W_{33} &= 1716\, \chi_7 R_1^{14} L_{14}.
\end{align*}

\paragraph*{Case \( n = 4 \) \textnormal{($O_4$)}:}

The matrix \( N_1 \) is structured as a row of four \( 2 \times 2 \) blocks:
\begin{equation}
	N_1 = \begin{pmatrix}
		N_{11} & N_{12} & N_{13} & N_{14}
	\end{pmatrix},
\end{equation}
where
\[
N_{11} = \sqrt{3}\, R_1^4 L_4, \quad
N_{12} = \sqrt{5}\, R_1^6 L_6, \quad
N_{13} = \sqrt{7}\, R_1^8 L_8, \quad
N_{14} = 3\, R_1^{10} L_{10}.
\]

\begin{equation}
	N_2 = \begin{pmatrix}
		\chi_3 N_{11} & \chi_5 N_{12} & \chi_7 N_{13} & \chi_9 N_{14}
	\end{pmatrix}^T,
\end{equation}

\begin{equation}
	W = \begin{pmatrix}
		W_{11} & W_{12} & W_{13} & W_{14} \\
		W_{21} & W_{22} & W_{23} & W_{24} \\
		W_{31} & W_{32} & W_{33} & W_{34} \\
		W_{41} & W_{42} & W_{43} & W_{44}
	\end{pmatrix},
\end{equation}

with components:
\begin{align*}
W_{11} &= 10\, \chi_3 R_1^6 L_6, &
W_{12} &= 7\sqrt{15}\, \chi_3 R_1^8 L_8, &
W_{13} &= 12\sqrt{21}\, \chi_3 R_1^{10} L_{10}, &
W_{14} &= 55\sqrt{3}\, \chi_3 R_1^{12} L_{12}, \\
W_{21} &= 7\sqrt{15}\, \chi_5 R_1^8 L_8, &
W_{22} &= 126\, \chi_5 R_1^{10} L_{10}, &
W_{23} &= 66\sqrt{35}\, \chi_5 R_1^{12} L_{12}, &
W_{24} &= 429\sqrt{5}\, \chi_5 R_1^{14} L_{14}, \\
W_{31} &= 12\sqrt{21}\, \chi_7 R_1^{10} L_{10}, &
W_{32} &= 66\sqrt{35}\, \chi_7 R_1^{12} L_{12}, &
W_{33} &= 1716\, \chi_7 R_1^{14} L_{14}, &
W_{34} &= 2145\sqrt{7}\, \chi_7 R_1^{16} L_{16}, \\
W_{41} &= 55\sqrt{3}\, \chi_9 R_1^{12} L_{12}, &
W_{42} &= 429\sqrt{5}\, \chi_9 R_1^{14} L_{14}, &
W_{43} &= 2145\sqrt{7}\, \chi_9 R_1^{16} L_{16}, &
W_{44} &= 24310\, \chi_9 R_1^{18} L_{18}.
\end{align*}

\paragraph*{Case \( n = 5 \) \textnormal{($O_5$)}:}

The matrix \( N_1 \) is structured as a row of five \( 2 \times 2 \) blocks:
\begin{equation}
	N_1 = \begin{pmatrix}
		N_{11} & N_{12} & N_{13} & N_{14} & N_{15}
	\end{pmatrix},
\end{equation}
where
\[
N_{11} = \sqrt{3}\, R_1^4 L_4, \quad
N_{12} = \sqrt{5}\, R_1^6 L_6, \quad
N_{13} = \sqrt{7}\, R_1^8 L_8, \quad
N_{14} = 3\, R_1^{10} L_{10}, \quad
N_{15} = \sqrt{11}\, R_1^{12} L_{12}.
\]

\begin{equation}
	N_2 = \begin{pmatrix}
		\chi_3 N_{11} & \chi_5 N_{12} & \chi_7 N_{13} & \chi_9 N_{14} & \chi_{11} N_{15}
	\end{pmatrix}^T,
\end{equation}

\begin{equation}
	W = \begin{pmatrix}
		W_{11} & W_{12} & W_{13} & W_{14} & W_{15} \\
		W_{21} & W_{22} & W_{23} & W_{24} & W_{25} \\
		W_{31} & W_{32} & W_{33} & W_{34} & W_{35} \\
		W_{41} & W_{42} & W_{43} & W_{44} & W_{45} \\
		W_{51} & W_{52} & W_{53} & W_{54} & W_{55}
	\end{pmatrix},
	\label{W5}
\end{equation}
with components:
\begin{align*}
W_{11} &= 10\, \chi_3 R_1^6 L_6, &
W_{12} &= 7\sqrt{15}\, \chi_3 R_1^8 L_8, &
W_{13} &= 12\sqrt{21}\, \chi_3 R_1^{10} L_{10}, &
W_{14} &= 55\sqrt{3}\, \chi_3 R_1^{12} L_{12}, \\
W_{15} &= 26\sqrt{33}\, \chi_3 R_1^{14} L_{14}, &
W_{21} &= 7\sqrt{15}\, \chi_5 R_1^8 L_8, &
W_{22} &= 126\, \chi_5 R_1^{10} L_{10}, &
W_{23} &= 66\sqrt{35}\, \chi_5 R_1^{12} L_{12}, \\
W_{24} &= 429\sqrt{5}\, \chi_5 R_1^{14} L_{14}, &
W_{25} &= 273\sqrt{55}\, \chi_5 R_1^{16} L_{16}, &
W_{31} &= 12\sqrt{21}\, \chi_7 R_1^{10} L_{10}, &
W_{32} &= 66\sqrt{35}\, \chi_7 R_1^{12} L_{12}, \\
W_{33} &= 1716\, \chi_7 R_1^{14} L_{14}, &
W_{34} &= 2145\sqrt{7}\, \chi_7 R_1^{16} L_{16}, &
W_{35} &= 1768\sqrt{77}\, \chi_7 R_1^{18} L_{18}, &
W_{41} &= 55\sqrt{3}\, \chi_9 R_1^{12} L_{12}, \\
W_{42} &= 429\sqrt{5}\, \chi_9 R_1^{14} L_{14}, &
W_{43} &= 2145\sqrt{7}\, \chi_9 R_1^{16} L_{16}, &
W_{44} &= 24310\, \chi_9 R_1^{18} L_{18}, &
W_{45} &= 25194\sqrt{11}\, \chi_9 R_1^{20} L_{20}, \\
W_{51} &= 26\sqrt{33}\, \chi_{11} R_1^{14} L_{14}, &
W_{52} &= 273\sqrt{55}\, \chi_{11} R_1^{16} L_{16}, &
W_{53} &= 1768\sqrt{77}\, \chi_{11} R_1^{18} L_{18}, &
W_{54} &= 25194\sqrt{11}\, \chi_{11} R_1^{20} L_{20}, \\
W_{55} &= 352716\, \chi_{11} R_1^{22} L_{22}.
\end{align*}

\paragraph*{General Case \( n = m \) \textnormal{($O_m$)}:}
The matrix \( N_1 \) is structured as a row of \( m \) aligned \( 2 \times 2 \) blocks:
\begin{equation}
	N_1 = \begin{pmatrix}
		N_{11} & N_{12} & \cdots & N_{1m}
	\end{pmatrix}, \qquad
	N_2 = \begin{pmatrix}
		\chi_3 N_{11} & \chi_5 N_{12} & \cdots & \chi_{2m+1} N_{1m}
	\end{pmatrix}^T,
\end{equation}
with block components
\begin{equation}
	N_{1j} = \sqrt{2j + 1}\, R_1^{2j + 2} L_{2j + 2}, \quad j = 1, \dots, m.
\end{equation}

The matrix \( W \) is structured as a block matrix of size \( m \times m \):
\begin{equation}
	W = \begin{pmatrix}
		W_{11} & W_{12} & \cdots & W_{1m} \\
		W_{21} & W_{22} & \cdots & W_{2m} \\
		\vdots & \vdots & \ddots & \vdots \\
		W_{m1} & W_{m2} & \cdots & W_{mm}
	\end{pmatrix},
	\label{Wm1}
\end{equation}
with each block defined as:
\begin{equation}
	W_{st} = \frac{(k + p - 1)! \sqrt{kp}}{k! \, p!} \chi_k R^{k + p} L_{k + p}, \quad
	k = 2s + 1, \quad p = 2t + 1, \quad s,t = 1, \dots, m.
\label{Wst}
\end{equation}

The computation of $L_{k+p}$ in (\ref{Lk}) requires first determining the lattice sums $S_{k+p}$, which are obtained through the recursive formulas presented in Appendix A of \cite{Yan2016}. The $S_4$ and $S_6$  sums are given by:
\begin{equation}
S_{4}=\frac{1}{60}\left(\frac{\pi}{\omega_1}\right)^4\left(\frac{4}{3}+320\sum_{m=1}^{\infty} \frac{m^3 \xi^{2m}}{1-\xi^{2m}}\right),
\end{equation}

\begin{equation}
    S_{6} = \frac{1}{140}\left(\frac{\pi}{\omega_1}\right)^6 \left(\frac{8}{27} - \frac{448}{3}\sum_{m=1}^{\infty} \frac{m^3 \xi^{2m}}{1-\xi^{2m}}\right),
    \label{eq:S6}
\end{equation}
where the geometric parameter $\xi = e^{\pi i \omega_2/\omega_1}$ characterizes the periodicity structure.

For all higher-order sums with $k \geqslant 4$, the recurrence relation takes the form:

\begin{equation}
    S_{2k} = \frac{3}{(4k^2-1)(k-3)}\sum_{r=2}^{k-2}(2r-1)(2k-2r-1)S_{2r}S_{2k-2r},
    \label{eq:recurrence}
\end{equation}
providing an efficient method for calculating higher-order terms.

The fundamental parameters $\delta_1$ and $\delta_2$ in (\ref{H1H2}), necessary to calculate (\ref{Unit_J1}), are determined by the expressions:

\begin{align}
    \delta_1 &= \omega_1 \left(\frac{\pi}{\omega_1}\right)^2 \left[ \frac{1}{3} - 8 \sum_{m=1}^\infty \frac{m \xi^{2m}}{1 - \xi^{2m}} \right], 
    \label{eq:delta1} \\
    \delta_2 &= \frac{\delta_1 \omega_2 - 2\pi i}{\omega_1},
    \label{eq:delta2}
\end{align}
here, $\omega_1$ and $\omega_2$ represent the fundamental periodicity vectors of the lattice structure, and $i$ is the imaginary unit satisfying $i^2 = -1$.

By systematically following this procedure we obtain a hierarchy of approximate formulas for the effective coefficients for any finite truncation order \(m\), These approximations highlight the influence of the inclusions periodic net through compact matrix forms, providing a balance between computational efficiency and accuracy, as we will illustrate in the numerical examples. 



\section{Effective Thermal Conductivity of Fibrous Composites with Interfacial Spring Barriers}

In practical applications, the interface between fiber and matrix in a composite material often presents imperfect thermal contact, which leads to additional resistance to heat flow. This interfacial resistance can be modeled using a thermal spring barrier, where the heat flux is proportional to the temperature jump across the interface.

The asymptotic homogenization method has previously been applied to calculate the effective elastic properties of angular fibrous composites with imperfect bonding, as shown in \cite{JuanK2011}. In that work, both spring-type and three-phase models were used to derive closed-form expressions for the effective shear moduli of composites with anisotropic constituents.

The same modeling strategy has been extended to the thermal conductivity problem. In particular, the analytical expressions obtained in \cite{JuanK2011} (Equations (36)–(39)) and further adapted for thermal analysis in \cite{iglesias2023conductivity} describe the effective conductivity tensor components for two-phase composites with interfacial spring barriers. In this section, we describe the derivation of these expressions from formulas (\ref{cc1313})-(\ref{cc2323}) but specify the particularities for modeling the existence of thermal barriers of interfacial spring type.

Let us show that this model can be derived from (\ref{cc1313})-(\ref{cc2323})   by taking the limit as the thickness of the interface approaches zero (\( t \rightarrow 0 \)) and assuming \( \rho_1 = (t/R_2)K \), where \( K \) is the imperfection parameter.  

We know that \( V_3 = \pi R_2^2/V \) and \( V_2 + V_3 = \pi (R_2 + t)^2/V \), where \( t \) is the thickness of the interface. From this, we obtain:  
\begin{equation}   
		\frac{V_2}{V_2 + V_3} = \frac{t}{R_2} \left( \frac{2 + t/R_2}{1 + 2t/R_2 + (t/R_2)^2} \right),  
\end{equation}

\begin{equation}  
	\frac{V_3}{V_2 + V_3} = \frac{1}{1 + 2t/R_2 + (t/R_2)^2}.  
\end{equation}  

Now, we rewrite \( \Delta \) in terms of \( t/R_2 \). Additionally, we keep the terms containing factors of the form \( \rho_1(1 - \rho_2) \) and \( \rho_1(1 + \rho_2) \) in the numerator and denominator of \( \Delta \), respectively, and try to group similar terms. With these considerations, \( \Delta \) can be rewritten as follows:  

\begin{equation}  
	\Delta = \frac{V_2 \rho_1 (1 - \rho_2) + V_2 (\rho_2 - \rho_1^2) + 2V_3 \rho_1 (1 - \rho_2)}{(V_2 + V_3) \rho_1 (1 + \rho_2) + V_2 (\rho_2 + \rho_1^2) + V_3 \rho_1 (1 + \rho_2)}.  
	\label{Fchi_simp2a}  
\end{equation} 

By appropriately grouping similar terms and dividing the numerator and denominator of \( \Delta \) by \( V_2 + V_3 \), we obtain the following.  

\begin{equation}  
	\Delta = \frac{\rho_1 (1 - \rho_2) + \frac{V_2}{V_2 + V_3} (\rho_2 - \rho_1^2) + \frac{V_3}{V_2 + V_3} \rho_1 (1 - \rho_2)}{\rho_1 (1 + \rho_2) + \frac{V_2}{V_2 + V_3} (\rho_2 + \rho_1^2) + \frac{V_3}{V_2 + V_3} \rho_1 (1 + \rho_2)}.  
	\label{Fchi_simp2b}  
\end{equation}  

Substituting the expressions for \( \frac{V_2}{V_2 + V_3} \), \( \frac{V_3}{V_2 + V_3} \), and \( \rho_1 = (t/R_2)K \), we get the following:  

\begin{equation}
	\Delta  = \frac{K(1-\rho_2) +\left(\frac{2+t/R_2}{1+2t/R_2+ (t/R_2)^2}\right)(\rho_2-(t/R_2)^2K^2)  + \frac{1}{1+2t/R_2+ (t/R_2)^2}K(1 - \rho_2)}{K(1+\rho_2) + \left(\frac{2+t/R_2}{1+2t/R_2+ (t/R_2)^2}\right)(\rho_2+(t/R_2)^2K^2)+ \frac{1}{1+2t/R_2+ (t/R_2)^2}K(1+\rho_2)}.
	\label{Fchi_simp2c}
\end{equation}

Taking the limit as \( t \) approaches zero, we obtain:  

\begin{equation}  
	\lim_{t \to 0} \Delta(t) = \frac{(1 - \rho_2) K + \rho_2}{(1 + \rho_2) K + \rho_2} = \beta_1.  
	\label{Fchi_simp2d}  
\end{equation}
With slight modifications and following the same reasoning, it can be shown that \( \chi_p \) from formula (\ref{chi_p}) tends to 

\begin{equation}
	\beta_p=\frac{(1 - \rho_2)K+p\rho_2}{(1 + \rho_2)K +p\rho_2},
	\label{beta_p1}
\end{equation}
as \( t \rightarrow 0 \), for all admissible \( p \).

As a result of the limit processes, the analytical expressions (\ref{cc1313})-(\ref{cc2323}) are transformed into the corresponding formulae for the effective thermal conductivity tensor components for a two-phase fibrous composite with interfacial spring barriers, which are given by: 

\begin{align}
	\hat{\kappa}_{11} &= \kappa_1 \left(1 - \frac{2 V_2 \beta_1 Z_{22}}{|Z|} \right), &
	\hat{\kappa}_{12} &= \frac{2 \kappa_1 V_2 \beta_1 Z_{21}}{|Z|},\label{kappa11spring} \\
	\hat{\kappa}_{21} &= \frac{2 \kappa_1 V_2 \beta_1 Z_{12}}{|Z|}, &
	\hat{\kappa}_{22} &= \kappa_1 \left(1 - \frac{2 V_2 \beta_1 Z_{11}}{|Z|} \right),
	\label{kappa22spring}
\end{align}
where \( Z \) is a \( 2 \times 2 \) matrix defined in equation~\eqref{matrizZ}, and \( |Z| \) denotes its determinant.

The parameter \( \beta_p \), which characterizes the thermal transmission across the interface, is defined by:
\begin{equation}
	\beta_p = \frac{(1 - \rho)K + p \rho}{(1 + \rho)K + p \rho}, \quad \text{for } p = 1, 3, 5, \dots,
	\label{beta_p}
\end{equation}
where \( \rho = \kappa_2 / \kappa_1 \) is the contrast between the thermal conductivities of the fiber and matrix phases.

The expressions for the effective coefficients in equations~\eqref{kappa11spring}–\eqref{kappa22spring} are derived using the same mathematical framework developed earlier, including the matrices \( Z \), \( N_1 \), \( N_2 \), and \( W \), as defined in equations~\eqref{Zcorta}–\eqref{W5}. However, in this case, the interfacial parameter \( \chi_p \) must be replaced by \( \beta_p \) as defined in equation~\eqref{beta_p}.

The spring barrier model provides an effective approach to account for imperfect thermal contact at the fiber–matrix interface. By introducing the interfacial parameter \( \beta_p \), the model captures the influence of the interface resistance on the effective thermal conductivity. The resulting expressions generalize the perfect contact case and could be useful for analyzing composites with more complex configurations.



\section{Effective Thermal Conductivity of Fibrous Composites Without Interfacial Barriers}

In this section, we focus on the evaluation of the effective thermal conductivity of two-phase fibrous composites assuming perfect interfacial contact between the constituents. This serves as a baseline model, where no thermal resistance is present at the fiber-matrix interfaces. The goal is to establish analytical expressions for the effective conductivity tensor under ideal conditions, which will later be contrasted with models incorporating thermal barriers.

\begin{figure}
	\centering
	\includegraphics[width=0.5\textwidth]{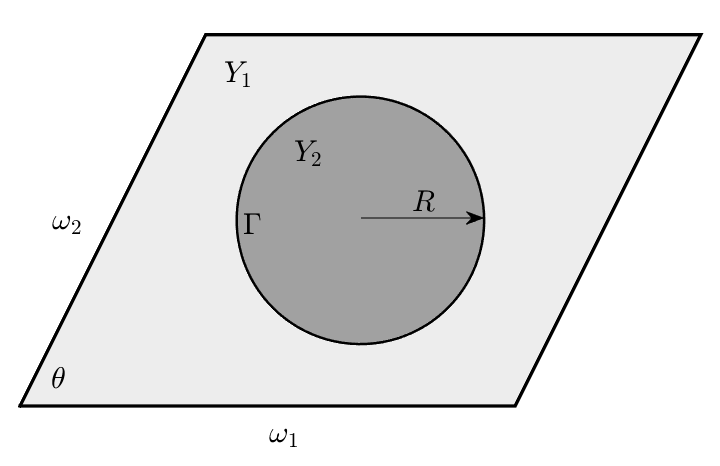} 
	\caption{Periodic unit cell without thermal barrier ($t = 0$) and orientation angle $\theta$. The regions $Y_1$ and $Y_2$ represent the matrix and fiber domains, respectively. $R$ denotes the fiber radius, and $\Gamma$ represents the interface between matrix and fiber.}
	\label{Celda_rombica_2f}
\end{figure}

The effective longitudinal thermal conductivity of two-phase fibrous composites with perfect contact (see Figure~\ref{Celda_rombica_2f}) was derived using the asymptotic homogenization method. In particular, we consider a periodic array of circular fibers arranged in a parallelogram (rhombic) lattice, as analyzed in \cite{guinovart2011influence}.

For this configuration, the effective thermal conductivity tensor components can be obtained by appropriately adapting the expressions provided in equations (\ref{cc1313})-(\ref{cc2323}). To obtain the analytical expressions for the perfect case, it suffices to set \( V_2 = 0 \) and \( \rho_1 = 1 \). 

The effective conductivity tensor components of the composite without interfacial barriers take the following form:
\begin{align}
	\hat{\kappa}_{11} &= \kappa_1 \left(1 - \frac{2 V_2 \chi Z_{22}}{|Z|} \right), 
	&\quad \hat{\kappa}_{12} &= \frac{2 \kappa_1 V_2 \chi Z_{21}}{|Z|}, \label{k11_2f}\\
	\hat{\kappa}_{21} &= \frac{2 \kappa_1 V_2 \chi Z_{12}}{|Z|}, 
	&\quad \hat{\kappa}_{22} &= \kappa_1 \left(1 - \frac{2 V_2 \chi Z_{11}}{|Z|} \right),
	\label{k22_2f}
\end{align}
where $|Z|$ denotes the determinant of the $2 \times 2$ matrix $Z$, defined as:
\begin{equation}
	Z = I + \chi J_1 - \chi^2 N_1 Y^{-1} N_2,
	\label{Z_2f}
\end{equation}
with
\begin{equation}
	\chi = \frac{1 - \rho}{1 + \rho}, \quad \rho = \frac{\kappa_2}{\kappa_1}, \qquad
	Y = I + \chi W.
\end{equation}
Here, $I$ and $J_1$ are defined in equation~\eqref{Unit_J1}. The remaining matrices $N_1$, $N_2$, and $W$ are analogous in structure and notation to those introduced in \eqref{Orden1}-\eqref{Wst}. For this case $N_2=N_1^T$ and $W$ is obtained replacing $\chi_p$ (for $p>1$) by $1$.  We consider the expressions of these matrices for  arbitrary order \( m \), the matrix \( N_1 \) is composed of \(m\) aligned \(2 \times 2\) blocks:
\begin{equation}
	N_1 = \begin{pmatrix}
		N_{11} & N_{12} & \cdots & N_{1m}
	\end{pmatrix}, \qquad
	N_2 = N_1^T,
\end{equation}
with
\begin{equation}
	N_{1j} = \sqrt{2j + 1}\, R^{2j + 2} L_{2j + 2}, \quad j = 1, \dots, m.
\end{equation}

The matrix \( W \) is a symmetric block matrix of size \( m \times m \), composed of \(m^2\) blocks:
\begin{equation}
	W = \begin{pmatrix}
		W_{11} & W_{12} & \cdots & W_{1m} \\
		W_{21} & W_{22} & \cdots & W_{2m} \\
		\vdots & \vdots & \ddots & \vdots \\
		W_{m1} & W_{m2} & \cdots & W_{mm}
	\end{pmatrix},
	\label{Wm+1}
\end{equation}
where each block \( W_{st} \) is given by:
\begin{equation}
	W_{st} = \frac{(k + p - 1)! \sqrt{kp}}{k!\, p!} \, R^{k + p} L_{k + p},
	\quad \text{with} \quad
	k = 2s + 1,\; p = 2t + 1,\quad s,t = 1, \dots, m.\label{W5_2f}
\end{equation}

The analytical expressions derived in this section characterize the effective thermal conductivity of fibrous composites under ideal interfacial conditions. By progressively increasing the harmonic order, we obtain increasingly accurate approximations that capture the influence of microstructural geometry.
Special cases occur when the period-parallelogram takes one of these forms:
\begin{enumerate}
    \item[(i)] Rectangular cells with primitive periods $\omega_1 = 1$, $\omega_2 = ci$ ($c > 0$)
    \item[(ii)] Parallelogram cells with primitive periods $\omega_1 = 1$, $\omega_2 = 1/2 + ci$ ($c > 0$)
    \item[(iii)] Rhombic cells with primitive periods $\omega_1 = e^{ih/2}$, $\omega_2 = e^{-ih/2}$
\end{enumerate}

These three configurations are particularly significant because, as shown in \cite{Chih-Bing1965}, they represent the only cases where the lattices $S_{2k}$ yield real numbers. This special property leads to a diagonal matrix $\mathbf{Z}$, which in turn causes the off-diagonal effective coefficients to vanish analytically ($\hat{\kappa}_{12} = \hat{\kappa}_{21} = 0$). 

Furthermore, the effective conductivities along the principal directions ($\hat{\kappa}_{11}$ and $\hat{\kappa}_{22}$) satisfy Keller's reciprocal relations \cite{Keller1964,Perrins1979}. This mathematical symmetry reflects the physical behavior of the composite material when the microstructure follows these particular geometric arrangements. These results serve as a reference for comparison with more complex configurations involving interfacial thermal barriers, which are addressed in the following sections.




\section{Results and discussion}

This section presents a detailed comparative analysis of the effective behavior of thermal conductive composite materials featuring various thermal barrier configurations.The study evaluates the influence on the effective coefficients of volume fractions, thermal conductivities, and the geometrical angle of the rhombic periodic unit cell. Five configurations are examined to investigate the impact of thermal barriers on the composite performance. . First, a reference model without thermal barriers is analyzed to establish a baseline. This is followed by four structurally-enhanced, conduction-impaired configurations: a composite with a ring-shaped thermal barrier; a spring-based thermal barrier model, which accounts for compliant interfacial behavior; a coated thermal barrier configuration that simulates protective surface layers; and a dispersed fiber composite incorporating embedded thermal barriers. Each case provides insights into how specific micro-structural designs can be leveraged to tailor and improve the effective properties of composite materials.

\subsection {Composite without thermal barrier} 

In this subsection, we analyze a two-phase fibrous composite without a thermal barrier, arranged in a rhombic periodic structure. The goal is to evaluate the effective thermomechanical coefficients of the system using different approximation strategies. Tables~\ref{Tabla_1}--\ref{Tabla_2} present a comparison of the results obtained through various truncation levels applied to the infinite system~(\ref{sistemafinal}). The fiber arrangement is illustrated in Figure~\ref{Celda_rombica_2f}, and the effective coefficients are calculated using analytical expressions from equations~(\ref{k11_2f})--(\ref{k22_2f}). The approximations are based on progressively increasing levels of precision, defined by the sets~(\ref{Z_2f})--(\ref{W5_2f}) and corresponding to truncation orders $O_k$, with $k = 0, \ldots, 5, 30$.

The composite material exhibits anisotropic behavior for a general array of fibers, as illustrated in the following subsections. The dimensionless effective longitudinal conductivity moduli $\hat{\kappa}_{11}/\kappa_1$, $\hat{\kappa}_{22}/\kappa_1$ and $\hat{\kappa}_{12}/\kappa_1$ are reported in Tables~\ref{Tabla_1} and~\ref{Tabla_2} for two rhombic configurations with angles $\theta = 45^\circ$ and $\theta = 75^\circ$, respectively, and density contrast $\rho=120$. These numerical results are compared with those obtained in \cite{guinovart2011influence} for the $O_{0}$-$O_{30}$ approximations, as well as the results from \cite{JIANG2004225} with good coincidence. The calculations are extended up to near the percolation threshold to capture the maximum error behavior.

As shown in Tables~\ref{Tabla_1}--\ref{Tabla_2}, the discrepancies between different approximation orders are minimal for volume fractions below $V_2 = 0.5$. However, the absolute error increases significantly for higher fiber volume fractions approaching the percolation limit, specifically at $V_2 = 0.65$ for $\theta=45^0$ and $V_2 = 0.81$ for $\theta=75^0$ . Notably, truncating the infinite system at order $O_5$ already yields accurate results in all volume fractions examined, making it a reliable analytical approximation for the effective coefficients; in fact, these are given in Tables 1 and 2 for $V_2$ values beyond the ranges given in \cite{JIANG2004225}.
\begin{center}
\begin{table*}
\caption{Results obtained by different approximations for the effective coefficients of a rhombic arrangement, \( \theta = 45^\circ \), $\rho=120$, and perfect contact (\( t = 0 \)).\label{Tabla_1}}
\footnotesize
\begin{tabular*}{\textwidth}{@{\extracolsep\fill}rrrrrrrrrr@{}}
\toprule

&\(\mathbf{V_2}\) & \(\mathbf{O_0}\) & \(\mathbf{O_1}\) & \(\mathbf{O_2}\) & \(\mathbf{O_3}\) & \(\mathbf{O_4}\) & \(\mathbf{O_5}\) & \(\mathbf{O_{30}}\) & \textbf{Ref.~\cite{JIANG2004225}} \\
\midrule
$\hat{\kappa}_{11}/\kappa_1 $
&0.10 & 1.21364 & 1.21365 & 1.21365 & 1.21365 & 1.21365 & 1.21365 & 1.21365 & 1.21365 \\
&0.20 & 1.46821 & 1.46841 & 1.46841 & 1.46841 & 1.46841 & 1.46841 & 1.46841 & 1.46841 \\
&0.30 & 1.77847 & 1.78035 & 1.78042 & 1.78042 & 1.78042 & 1.78042 & 1.78042 & 1.78042 \\
&0.40 & 2.16919 & 2.17931 & 2.18001 & 2.18002 & 2.18002 & 2.18002 & 2.18002 & 2.18002 \\
&0.50 & 2.68845 & 2.73247 & 2.73761 & 2.73779 & 2.73786 & 2.73786 & 2.73782 & 2.73782 \\
&0.60 & 3.45782 & 3.67903 & 3.74042 & 3.75253 & 3.76211 & 3.76330 & 3.75734 & -- \\
&0.65 & 4.05140 & 4.76606 & 5.31578 & 5.71560 & 6.65585 & 7.29076 & 7.30173 & -- \\

\midrule
$\hat{\kappa}_{22}/\kappa_1 $
&0.10     & 1.22306      &1.22306     & 1.22306     & 1.22306     & 1.22306      &1.22306      &1.22306 &  1.22306 \\
&0.20     & 1.51582      &1.51607     &1.51608      &1.51608      &1.51608       &1.51608      &1.51608 & 1.51608  \\
&0.30     & 1.91885      &1.92161     & 1.92172     & 1.92172     & 1.92172      &1.92172      &1.92172 & 1.92172  \\
&0.40     & 2.51292      &2.5319      & 2.5334      &2.53343      &2.53343       &2.53343      &2.53343 &  2.53343 \\
&0.50     & 3.48671      &3.60232     &3.61951      &3.62055      &3.62068       &3.62069      &3.62069 &  3.62069 \\
&0.60     & 5.41217      & 6.2942     & 6.6064      &6.67673      &6.69705       &6.70253      &6.70467 & -- \\
&0.65    & 7.34028      &10.8466     & 13.9724     &16.3006      &18.1535       &19.61010      &25.9050& -- \\
\midrule
$\hat{\kappa}_{12}/\kappa_1 $
&0.10   &-0.0047066  &-0.0047068  &-0.0047068 & -0.0047068 &  -0.0047068 &  -0.0047068&  -0.0047068& -0.00471 \\
&0.20   &-0.023808   &-0.023831  & -0.023832  & -0.023832 &  -0.023832  & -0.023832 &  -0.023832 & -0.02383\\
&0.30   &-0.070191   &-0.07063  & -0.07065  & -0.07065 &  -0.07065  & -0.07065 &  -0.07065 &  -0.07065\\
&0.40   & -0.17186   & -0.17629   & -0.17669  &  -0.17671 &   -0.17671  &  -0.17671 &   -0.17671 & -0.17671 \\
&0.50   &  -0.39913  &   -0.43493   & -0.440947  &  -0.441377 &    -0.44143  &  -0.441435 &   -0.441436 & -044144\\
&0.60   & -0.977175  &   -1.30758   &  -1.43299  &    -1.4621 &    -1.47051  &   -1.47278 &   -1.47366  & -- \\
&0.65   & -1.64444   &  -3.04027    &  -4.3283   &  -5.29252  &   -6.05998   &  -6.6633   &  -9.27074   & --\\
\bottomrule
\end{tabular*}
\end{table*}
\end{center}

\begin{center}
\begin{table*}
\caption{Results obtained by different approximations for the effective coefficients of a rhombic arrangement, \( \theta = 75^\circ \), $\rho=120$, and perfect contact (\( t = 0 \)).\label{Tabla_2}}
\footnotesize
\begin{tabular*}{\textwidth}{@{\extracolsep\fill}rrrrrrrrrr@{}}
\toprule
&\(\mathbf{V_2}\) & \(\mathbf{O_0}\) & \(\mathbf{O_1}\) & \(\mathbf{O_2}\) & \(\mathbf{O_3}\) & \(\mathbf{O_4}\) & \(\mathbf{O_5}\) & \(\mathbf{O_{30}}\) & \textbf{Ref.~\cite{JIANG2004225}} \\
\midrule
$\hat{\kappa}_{11}/\kappa_1 $
 &0.10     & 1.21851     & 1.21852     & 1.21852     & 1.21852    &  1.21852    &  1.21852    &  1.21852 & 1.21852 \\
&0.20     & 1.49161     & 1.49176      &1.49176     & 1.49176     & 1.49176     & 1.49176     & 1.49176& 1.49176 \\
&0.30      & 1.8428     & 1.84424      &1.84428     & 1.84428     & 1.84428     & 1.84428     & 1.84428& 1.84428\\
&0.40      &2.31149     & 2.31973      & 2.32016    & 2.32016     & 2.32016     & 2.32016     &2.32016 & 2.32016\\
&0.50      &2.96926     &3.00559       &3.00879      &3.00884     &3.00884      & 3.00884     & 3.00884 & 3.00884\\
&0.60      & 3.9619     & 4.10657      & 4.12815     & 4.12874     & 4.12883    &  4.12884    &  4.12884 & 4.12884\\
&0.70      &5.64193     & 6.25001      &6.42292     &6.43319      &6.43686     & 6.43748      &6.43758 & --\\
&0.80     & 9.16257     & 12.9716      & 16.5079    &  17.2459    & 18.1823    & 18.5996     &19.1531 & \\
&0.81     & 9.74537     & 14.6412      &20.3884     & 21.931      & 24.4194    & 25.8781     &30.1607 & --\\
\midrule
$\hat{\kappa}_{22}/\kappa_1 $
&0.10 & 1.2178  & 1.2178  & 1.2178  & 1.2178  & 1.2178  & 1.2178  & 1.2178  & 1.2178 \\
&0.20 & 1.48801 & 1.48815 & 1.48816 & 1.48816 & 1.48816 & 1.48816 & 1.48816 & 1.48816 \\
&0.30 & 1.83228 & 1.83370 & 1.83374 & 1.83374 & 1.83374 & 1.83374 & 1.83374 & 1.83374 \\
&0.40 & 2.28620 & 2.29438 & 2.29476 & 2.29477 & 2.29477 & 2.29477 & 2.29477 & 2.29477 \\
&0.50 & 2.91288 & 2.94935 & 2.95195 & 2.95202 & 2.95203 & 2.95203 & 2.95203 & 2.95203 \\
&0.60 & 3.83659 & 3.98480 & 4.00019 & 4.00116 & 4.00122 & 4.00123 & 4.00123 & 4.00126 \\
&0.70 & 5.34386 & 5.98713 & 6.09497 & 6.11238 & 6.11467 & 6.11530 & 6.11543 & - \\
&0.80 & 8.30570 & 12.6525 & 14.7357 & 15.9550 & 16.5716 & 16.9854 & 17.4845 & - \\
&0.81 & 8.77404 & 14.4624 & 17.8549 & 20.3976 & 22.0588 & 23.5037 & 27.3349 & - \\
\midrule
$\hat{\kappa}_{12}/\kappa_1 $
&0.10   &0.0013332  & 0.0013332  & 0.0013332  & 0.0013332  & 0.0013332   &0.0013332  & 0.0013332& 0.00133 \\
&0.20   &0.0067195  & 0.0067225  & 0.0067225  & 0.0067225  & 0.0067225  & 0.0067225  & 0.0067225& 0.00672\\
&0.30    &0.019639  &  0.019671  &  0.019675  &  0.019675  &  0.019675  &  0.019675  & 0.019675 &0.01968\\
&0.40    &0.047192  & 0.047306   & 0.047392   & 0.047391  & 0.047391   & 0.047391    & 0.047391 &0.04739\\
&0.50    & 0.10522  &   0.10494  &   0.10606  &   0.10602  &   0.10602  &   0.10602  & 0.10602 & 0.10602\\
&0.60    & 0.23383   &  0.22724  &   0.23878  &   0.23806  &   0.23812  &   0.23812  & 0.23812 & 0.23812\\
&0.70    &  0.5562   &  0.49054  &   0.61197  &   0.59863  &   0.60111  &   0.60112  & 0.60113 &-- \\
&0.80    &  1.5989   &  0.59546  &     3.307  &    2.4089  &     2.997  &    3.0155  & 3.1133  & --\\
&0.81    &  1.8125   &  0.33361   &   4.7275  &    2.8614  &    4.3839  &    4.4596  & 5.1845  &--\\
\bottomrule
\end{tabular*}
\end{table*}
\end{center}

\begin{center}
\begin{table*}
\caption{Effective conductivity properties of periodic fibrous composites with perfect contact. Periodic cells defined by \( \omega_1 = 1 \) and \( \omega_2 = re^{i\theta} \) and $\rho = 50$.\label{tabla1Yang}}
\footnotesize
\begin{tabular*}{\textwidth}{@{\extracolsep\fill}lcccccccccc@{}}
\toprule
& \multicolumn{4}{c}{\textbf{Rectangular} ($r=0.5$, $\theta=\pi/2$, $V_2=0.36$)} & \multicolumn{6}{c}{\textbf{Rhombic} ($r=\sqrt{3}/3$, $\theta=\pi/6$, $V_2=0.48$)} \\
\cmidrule(r){2-5} \cmidrule(r){6-11}
& 
\multicolumn{2}{c}{$\kappa_{11}/\kappa_1$} & 
\multicolumn{2}{c}{$\kappa_{22}/\kappa_1$} & 
\multicolumn{2}{c}{$\kappa_{11}/\kappa_1$} & 
\multicolumn{2}{c}{$\kappa_{22}/\kappa_1$} & 
\multicolumn{2}{c}{$\kappa_{12}/\kappa_1$} \\
\cmidrule(r){2-3} \cmidrule(r){4-5} \cmidrule(r){6-7} \cmidrule(r){8-9} \cmidrule(r){10-11}
\textbf{$k$} & \cite{Yan2016} & AHM $O_{k-1}$  & \cite{Yan2016} & AHM $O_{k-1}$ & \cite{Yan2016} & AHM $O_{k-1}$  & \cite{Yan2016} & AHM $O_{k-1}$ & \cite{Yan2016} & AHM $O_{k-1} $ \\
\midrule
1 & 1.66993 & 1.66993 & 3.50961 & 3.50961 & 2.69362 & 2.69362 & 3.7933 & 3.7933 & 0.943634 & 0.943634 \\
2 & 1.68932 & 1.68932 & 4.02376 & 4.02376 & 2.90209 & 2.90209 & 4.32367 & 4.32367 & 1.22573 & 1.22573 \\
3 & 1.68987 & 1.68987 & 4.15497 & 4.15497 & 2.94546 & 2.94546 & 4.45468 & 4.45468 & 1.29986 & 1.29986 \\
4 & 1.68988 & 1.68988 & 4.18703 & 4.18703 & 2.95635 & 2.95635 & 4.48683 & 4.48683 & 1.31857 & 1.31857 \\
6 & 1.68988 & 1.68988 & 4.19710 & 4.19710 & 2.95970 & 2.95970 & 4.49690 & 4.49690 & 1.32437 & 1.32437 \\
8 & 1.68988 & 1.68988 & 4.19779 & 4.19779 & 2.95993 & 2.95993 & 4.49759 & 4.49759 & 1.32476 & 1.32476 \\
10 & 1.68988 & 1.68988 & 4.19784 & 4.19784 & 2.95995 & 2.95995 & 4.49764 & 4.49764 & 1.32479 & 1.32479 \\
\bottomrule
\end{tabular*}
\end{table*}
\end{center}

\begin{table*}[ht]
	\caption{Effective conductivity properties of periodic fibrous composites with perfect contact. Periodic cells defined by $\omega_1 = 1$ and $\omega_2 = re^{i\theta}$, $\rho = 50$ and high fiber volume fraction.\label{tabla4higthV2}}
	\footnotesize
	\begin{tabular*}{\textwidth}{@{\extracolsep\fill}lccccccc@{}}
		\toprule
		\textbf{Model} & 
		\multicolumn{2}{c}{Hex./Sqr ($r = 1$)} & 
		\multicolumn{2}{c}{Rect. ($r = 0.5, V_2=0.392699$)} & 
		\multicolumn{3}{c}{Rhom. ($r = \frac{\sqrt{3}}{3},V_2=0.523598$)} \\
		\cmidrule(lr){2-3} \cmidrule(lr){4-5} \cmidrule(lr){6-8}
		& $\theta = \pi/3, V_2 = 0.9068993$ & $\theta = \pi/2, V_2 = 0.785398$ &
		$\hat{\kappa}_{11}/\kappa_1$ & $\hat{\kappa}_{22}/\kappa_1$ &
		$\hat{\kappa}_{11}/\kappa_1$ & $\hat{\kappa}_{22}/\kappa_1$ & $\hat{\kappa}_{12}/\kappa_1$ \\
		\midrule
		$O_{12}$  &   29.6869 &     20.3868&      1.75784 &     10.7503&      5.26261&      11.0898 &     5.03885\\
		$O_{32}$  &  30.3462 &     20.9468 &     1.75784 &      11.028 &     5.35517&      11.3674  &    5.19912\\
		$O_{42}$  & 30.5737  &    21.1024  &    1.75784 &     11.1053 &     5.38091&      11.4445 &     5.24369\\
		\cite{Perrins1979}&30.59   &    21.1   &  -   &   -&  - &  -&  - \\
		\bottomrule
	\end{tabular*}
    \label{table1YanghighVFperfect}
\end{table*}

Table \ref{tabla1Yang} presents the effective coefficients of two-phase fibrous composites for two distinct periodic unit cells, rectangular and rhombic. These configurations were previously analyzed in class form \cite{Yan2016} and \cite{ZEMLYANOVA2023}. Here, we adopt representative cells defined by the periodic vectors $\omega_1 = 1$ and $\omega_2 = re^{i\theta}$, where $r$ and $\theta$ are specified in the table captions. The material parameters were taken from Table 1 in \cite{Yan2016}, with fixed contrast $\rho = 50$. The results provide significant information on the convergence as the truncation order increases, as well as its agreement with previous models \cite{Yan2016}\cite{ZEMLYANOVA2023}\cite{Perrins1979}.

The table \ref{tabla1Yang} presents a remarkable agreement between the effective conductivity properties calculated by \cite{Yan2016} and the Asymptotic Homogenization Method (AHM) for rectangular and rhombic periodic fibrous composites. For all approximation orders ($O_k$) from 1 to 10, both models yield identical numerical results for all conductivity components ($\kappa_{11}$, $\kappa_{22}$, and $\kappa_{12}$), with perfect matching up to the fifth decimal place. This exact correspondence validates both methodologies and suggests their mathematical equivalence for these configurations. The rectangular case shows faster convergence in the $\kappa_{11}$ component (stabilizing by $O_4$) compared to the $\kappa_{22}$ component, while the rhombic configuration demonstrates more gradual convergence across all components, particularly for the off-diagonal term $\kappa_{12}$. The identical results persist even at higher approximation orders, indicating numerical stability in both methods. This perfect agreement provides strong evidence that both approaches correctly capture the effective conductive behavior of these periodic microstructures, regardless of the geometric configuration (rectangular or rhombic) or the specific conductivity component being calculated.

The effective conductivity properties of periodic fibrous composites with perfect thermal contact between phases have been analyzed in Table \ref{table1YanghighVFperfect} for different fiber arrangements and high volume fractions. The computed effective conductivity components $\hat{\kappa}_{11}/\kappa_1$, $\hat{\kappa}_{22}/\kappa_1$, and $\hat{\kappa}_{12}/\kappa_1$ show excellent agreement with established theoretical results. For the Hex./Sqr arrangement, our high-order approximation $O_{42}$ yields $\hat{\kappa}_{11}/\kappa_1 = 30.5737$ ($\theta = \pi/3$) and $\hat{\kappa}_{22}/\kappa_1 = 21.1024$ ($\theta = \pi/2$), matching the reference values of 30.59 and 21.1 from \cite{Perrins1979} with remarkable precision (differences < 0.1\%). This agreement serves as strong validation of our computational approach. The rectangular configuration exhibits significant anisotropy, with $\hat{\kappa}_{22}/\kappa_1$ (ranging from 10.7503 to 11.1053) being substantially larger than $\hat{\kappa}_{11}/\kappa_1$ (constant at 1.75784 across all orders). The rhombic arrangement demonstrates intermediate conductivity values with notable off-diagonal components ($\hat{\kappa}_{12}/\kappa_1$ increasing from 5.03885 to 5.24369), reflecting coupling effects due to non-orthogonal fiber orientation. The convergence of results through successive approximation orders ($O_{12}$ to $O_{42}$) confirms the robustness of the method, particularly for high volume fractions where analytical solutions are limited.

\begin{figure}
	\centering
	\includegraphics[width=1.\textwidth] {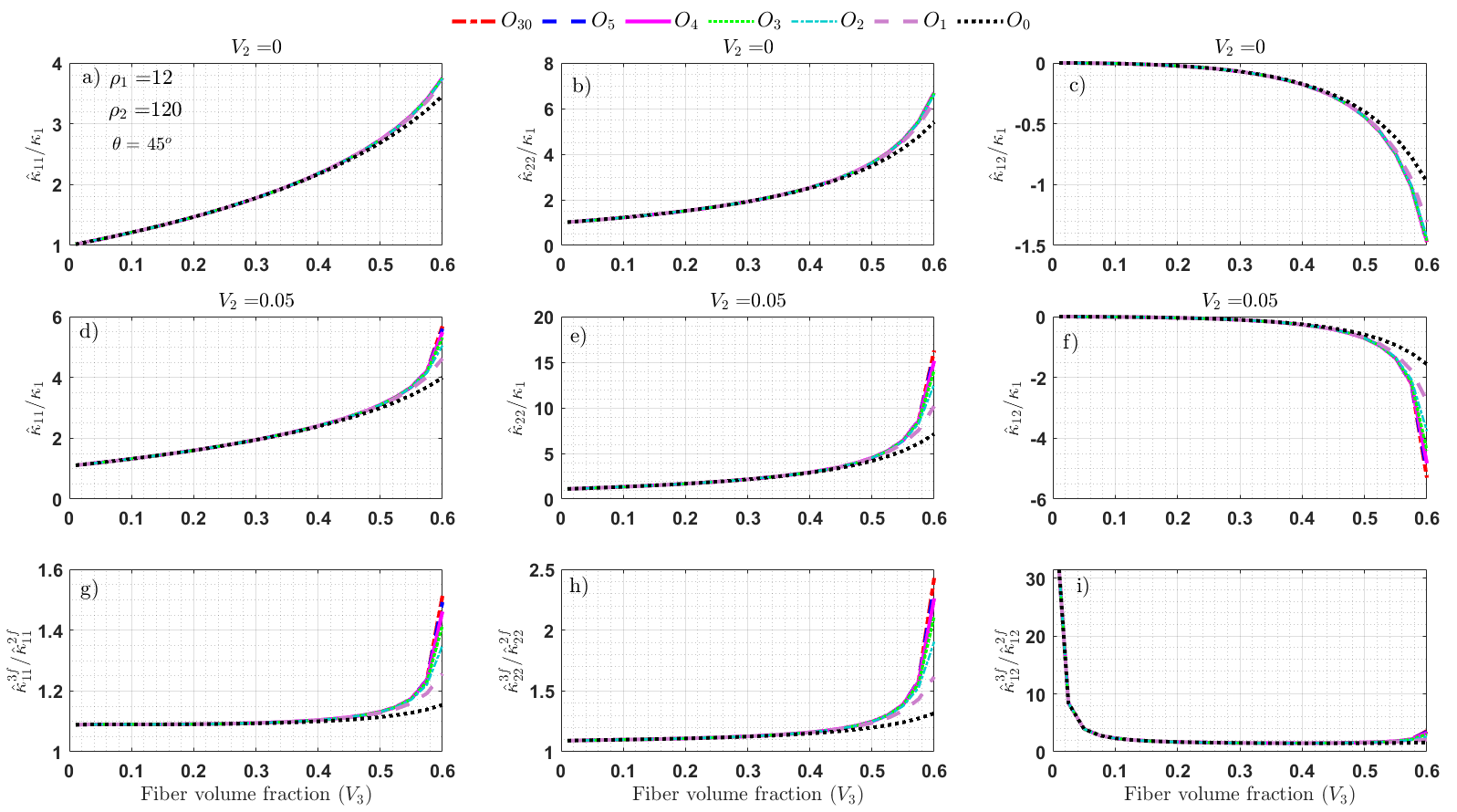} %
	\caption{ Effective conductive properties of a fibrous with rhombic array. Figures a)-c) corresponding to  composite without thermal barrier, figures d)-f) corresponding to  composite with thermal barrier and figures g)-i) are displaying the corresponding ratio of the effective properties. } 
	\label{3f_VS_2f_angulo_45}
\end{figure}

Figure~\ref{3f_VS_2f_angulo_45} presents the effective conductive coefficients of a fibrous composite with a rhombic distribution of fibers, both without and with a ring-shaped thermal barrier. The calculations correspond to the three-phase model defined by equations~(\ref{cc1313})--(\ref{cc2323}), and are performed for varying interface volume fractions \( V_2 \) and inner fiber volume fractions \( V_3 \). When \( V_2 = 0 \), the structure reduces to a two-phase composite without a thermal barrier, and the behavior of the effective properties is depicted in panels (a), (b), and (c), which serve as the reference baseline. The influence of the thermal barrier becomes evident when \( V_2 > 0 \), corresponding to the three-phase composite configuration. This effect is illustrated in panels (d), (e), and (f) of Figure~\ref{3f_VS_2f_angulo_45}. Notably, even the introduction of a thin thermal barrier layer (\( V_2 = 0.05 \)) leads to a measurable increase in the effective conductivity coefficients when compared to the two-phase baseline. All results are obtained using successive approximations \( O_k \), with \( k = 0, \ldots, 5, 30\), ensuring consistent comparison across the range of microstructural parameters.

Figure~\ref{3f_VS_2f_angulo_45} g)-i) presents the ratio of the effective conductivity coefficients between the three-phase and two-phase composites, defined as \( \kappa_{\text{gain}} = \hat{\kappa}_{3f} / \hat{\kappa}_{2f} \). The results confirm that the presence of the ring-shaped conductive layer leads to significant enhancements in the composite’s conductive performance. The most pronounced gains are observed for interface volume fractions \( V_2 \) near the percolation threshold of approximately 0.6, particularly for the principal conductivity components \( \kappa_{11} \) and \( \kappa_{22} \). In contrast, the off-diagonal coefficient \( \kappa_{12} \) shows consistent improvement across the entire range of \( V_2 \), with additional enhancement as \( V_2 \) approaches the percolation limit.

\subsection{Composite with ring-shaped conductive layer} 

\begin{figure}
	\centering
	\includegraphics[width=.5\textwidth] {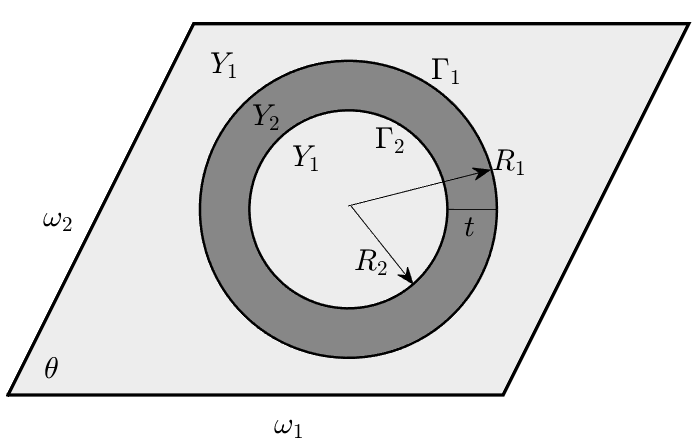} %
	\caption{ Periodic cell of a composite with tubular fibers. $Y_1, Y_2$, regions occupied by matrix, ring, and matrix, respectively.  $R_1$ outer radius, fiber radius plus coated thickness $t$, $R_2$ fiber radius. $\Gamma_1$ interphase between matriz and coated, $\Gamma_2$ interphase between coated an fiber.}
	\label{Celda_rombica_fibraAnillo}
\end{figure}

This subsection examines a three-phase fibrous composite system in which a central fiber core is surrounded by a concentric ring-shaped thermal barrier embedded within a matrix. The geometry of the periodic unit cell is illustrated in Figure~\ref{Celda_rombica_fibraAnillo}, where the fiber of radius \( R_2 \) is coated with a thermal barrier of thickness \( t \), yielding an outer radius \( R_1 \). The regions are separated by interfaces \( \Gamma_1 \) and \( \Gamma_2 \), distinguishing the matrix, coating, and fiber phases. The ring-shaped region defines a volume fraction \( V_2 \), while the inner fiber core occupies \( V_3 \). To isolate the effect of the thermal barrier, the effective conductive coefficients of this three-phase configuration are compared to those of a two-phase composite with solid fibers, as described in Figure~\ref{Celda_rombica_2f}. In both models, the total fiber volume fraction is kept constant by setting \( V_2 + V_3 \) in the three-phase case equal to the fiber volume fraction in the two-phase case. 

The goal is to evaluate how the introduction and variation of the thermal barrier influence the composite's effective conductive coefficients. All calculations are performed using approximations \( O_k \), with \( k = 0, \ldots, 5 \text{ and } 30\), applied to the analytical expressions given by equations~(\ref{cc1313})--(\ref{cc2323}) for the three-phase model (Figure~\ref{Celda_rombica_fibraAnillo}) and equations~(\ref{k11_2f})--(\ref{k22_2f}) for the two-phase counterpart (Figure~\ref{Celda_rombica_2f}). For the numerical analysis, the matrix and core fiber conductivities are set as \( \kappa_1 = \kappa_3 = 1 \), while the coating layer is assigned a high conductivity \( \kappa_2 = 120 \). In the two-phase model, \( \kappa_1 = 1 \) and \( \kappa_2 = 120 \) represent the matrix and fiber, respectively.

Additionally, we introduce the ratio  
\[
\alpha = \frac{V_2}{V_2 + V_3}, \quad 0 \leq \alpha \leq 1,
\]  
which represents the relative contribution of the ring-shaped interface to the total fiber volume fraction (\( V_2 + V_3 \)) in the three-phase composite.

\begin{figure}
	\centering
	\includegraphics[width=1\textwidth] {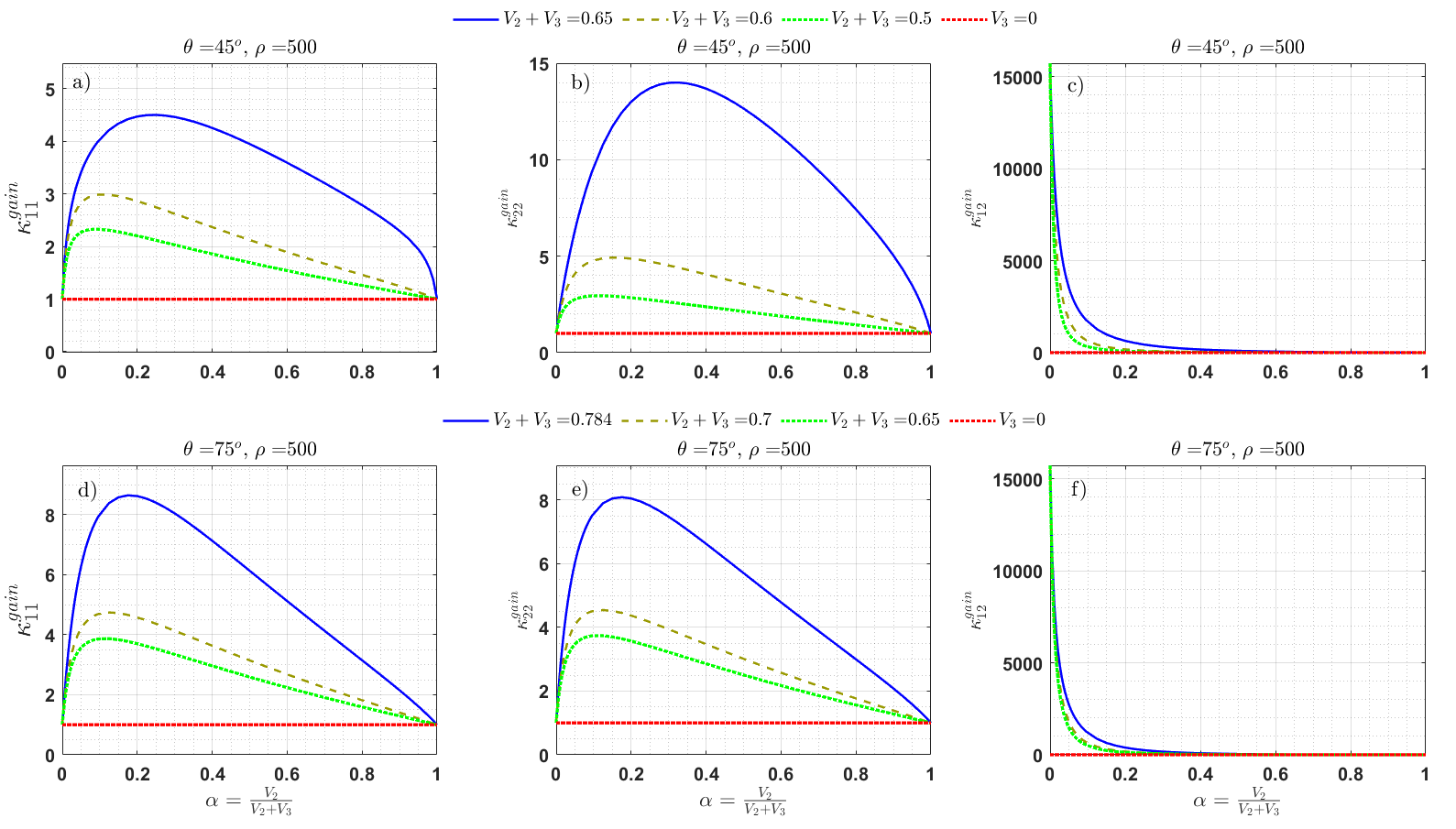} %
	\caption{ The gain in effective conductivity between a composite of
		tubular fibers vs a composite with solid fibers.} 
	\label{Kgain_Anillo_varia_angulo 45_75}
\end{figure}

Figure~\ref{Kgain_Anillo_varia_angulo 45_75} presents the results of this analysis. Panels (a)–(c) display the gain parameter \( \kappa_{\text{gain}} \) for the \( \theta = 45^\circ \) configuration, while panels (d)–(f) correspond to the \( \theta = 75^\circ \) case. In both geometries, the introduction of a tubular thermal barrier leads to enhanced effective conductivity compared to the solid fiber configuration. For \( \kappa_{\text{gain},11} \) and \( \kappa_{\text{gain},22} \), the maximum gain occurs in the range \( 0.2 \leq \alpha \leq 0.4 \) for the \( 45^\circ \) cell, and between \( 0.1 \leq \alpha \leq 0.2 \) for the \( 75^\circ \) cell. The contrast in conductivity becomes more pronounced as the total volume fraction \( V_2 + V_3 \) approaches the percolation threshold.

These findings indicate that incorporating a thin, highly conductive tubular interface can significantly enhance the overall conductivity of the composite, especially when the barrier exhibits a strong contrast with the matrix phase.

To quantify the improvement in effective conductivity resulting from the use of tubular fibers instead of solid fibers, we define a gain parameter:  
\[
\kappa_{\text{gain}} = \frac{\hat{\kappa}_{3f}}{\hat{\kappa}_{2f}},
\]  
where \( \hat{\kappa}_{3f} \) and \( \hat{\kappa}_{2f} \) denote the effective properties obtained using the three-phase and two-phase models, respectively. This metric allows direct comparison between the two configurations. The analysis is conducted for rhombic periodic cells with angles \( \theta = 45^\circ \) and \( \theta = 75^\circ \).

\subsection{Spring thermal barrier model} 

In this subsection, we analyze the influence of interfacial thermal resistance on the effective thermal conductivity of fibrous composites using a spring thermal barrier model. This configuration simulates imperfect thermal contact between fibers and matrix through a temperature discontinuity at the interface, characterized by the Biot number (\( Bi \)). The effective properties are examined as functions of three key parameters: the Biot number, the fiber volume fraction (\( V_2 \)), and the conductivity contrast (\( \rho = \kappa_2 / \kappa_1 \)). We compare predictions from the Asymptotic Homogenization Method (AHM), specifically the sixth-order truncation (AHM-\( O_6 \)), against the model proposed by Lu \cite{Lu1995}, which assumes hexagonal symmetry.

The results presented in Tables~\ref{Tabla3}--\ref{Tabla6} reveal an excellent agreement between the AHM results -\( O_6 \) and Lu’s model for all Biot numbers and for all conductivity contrasts when the fiber volume fraction is low to moderate (\( V_2 \leq 0.8 \)). 
However, as \( V_2 \) increases toward the percolation threshold (\( V_2 \geq 0.88 \)), differences become increasingly pronounced—particularly for moderate to high conductivity contrasts (e.g., \( \rho = 101 \) and \( \rho = 1001 \)). In these regimes, Lu’s model tends to predict higher effective conductivities, potentially reflecting its ability to account for intensified fiber interactions. Both models converge to the matrix conductivity value (\( \kappa = 1.0 \)) at the critical Biot number, which in the AHM model corresponds to \( K = \rho / (\rho - 1) \), indicating total interfacial thermal resistance. Lu’s model, however, retains slightly elevated values in this limit, possibly due to residual thermal effects. At lower contrast values (e.g., \( \rho = 11 \)), both models remain in close agreement throughout the tested range, with discrepancies appearing only at the highest values of \( V_2 \) and \( Bi \), where Lu’s model does not report results for some combinations.

\begin{center}
\begin{table*}
\caption{Comparison of AHM-\(O_6\) with results from \cite{Lu1995} for \( \rho = 1001 \).\label{Tabla3}}
\footnotesize
\begin{tabular*}{\textwidth}{@{\extracolsep\fill}lrrrrrr@{}}
\toprule
& \multicolumn{2}{c}{\( Bi = 10^{12} \)} 
& \multicolumn{2}{c}{\( Bi = 1 \)} 
& \multicolumn{2}{c}{\( Bi = 10^{-1} \)} \\
\cmidrule(lr){2-3} \cmidrule(lr){4-5} \cmidrule(lr){6-7}
\textbf{\( V_2 \)} & \textbf{AHM} & \textbf{\cite{Lu1995}} & \textbf{AHM} & \textbf{\cite{Lu1995}} & \textbf{AHM} & \textbf{\cite{Lu1995}} \\
\midrule
0.200000  & 1.498756 & 1.49876  & 1.497513 & 1.497510 & 1.486489 & 1.486490 \\
0.500000  & 2.996711 & 2.99671  & 2.988707 & 2.988730 & 2.918805 & 2.918960 \\
0.700000  & 5.777989 & 5.77799  & 5.744044 & 5.744630 & 5.456978 & 5.461800 \\
0.800000  & 9.856891 & 9.85697  & 9.747525 & 9.752310 & 8.872281 & 8.906640 \\
0.880000  & 24.64129 & 24.73310 & 23.737639 & 23.916400 & 18.137860 & 18.607400 \\
0.900000  & 48.51346 & 52.45030 & 44.106085 & 47.526900 & 25.889222 & 27.682900 \\
0.905000  & 69.45809 & 96.96000 & 59.749882 & 75.614400 & 29.333050 & 32.287700 \\
0.906899  & 84.75738 & --       & 70.032774 & 113.20000 & 30.956052 & 34.648000 \\
\midrule
& \multicolumn{2}{c}{\( Bi = 10^{-2} \)} 
& \multicolumn{2}{c}{\( Bi = 10^{-3} \)} 
& \multicolumn{2}{c}{\( Bi = 10^{-4} \)} \\
\cmidrule(lr){2-3} \cmidrule(lr){4-5} \cmidrule(lr){6-7}
\textbf{\( V_2 \)} & \textbf{AHM} & \textbf{\cite{Lu1995}} & \textbf{AHM} & \textbf{\cite{Lu1995}} & \textbf{AHM} & \textbf{\cite{Lu1995}} \\
\midrule
0.200000  & 1.390456 & 1.390460 & 1.000000 & 1.000000 & 0.718793 & 0.718793 \\
0.500000  & 2.381116 & 2.381580 & 1.000000 & 1.000000 & 0.419048 & 0.419058 \\
0.700000  & 3.684625 & 3.694060 & 1.000000 & 1.000000 & 0.268471 & 0.268559 \\
0.800000  & 4.829263 & 4.867800 & 1.000000 & 1.000000 & 0.200798 & 0.201035 \\
0.880000  & 6.291171 & 6.418830 & 1.000000 & 1.000000 & 0.144686 & 0.145317 \\
0.900000  & 6.790226 & 6.967270 & 1.000000 & 1.000000 & 0.127563 & 0.127947 \\
0.905000  & 6.926787 & 7.119430 & 1.000000 & 1.000000 & 0.122523 & 0.121990 \\
0.906899  & 6.980022 & 7.179000 & 1.000000 & 1.000000 & 0.120467 & 0.119000 \\
\bottomrule
\end{tabular*}
\end{table*}
\end{center}

\begin{center}
\begin{table*}
\caption{Comparison of AHM-\(O_6\) with results from \cite{Lu1995} for \( \rho = 101 \).\label{Tabla4}}
\footnotesize
\begin{tabular*}{\textwidth}{@{\extracolsep\fill}lrrrrrr@{}}
\toprule
& \multicolumn{2}{c}{\( Bi = 10^{12} \)} 
& \multicolumn{2}{c}{\( Bi = 1 \)} 
& \multicolumn{2}{c}{\( Bi = 10^{-1} \)} \\
\cmidrule(lr){2-3} \cmidrule(lr){4-5} \cmidrule(lr){6-7}
\textbf{\( V_2 \)} & \textbf{AHM} & \textbf{\cite{Lu1995}} & \textbf{AHM} & \textbf{\cite{Lu1995}} & \textbf{AHM} & \textbf{\cite{Lu1995}} \\
\midrule
0.200000  & 1.487808 & 1.48781  & 1.475964 & 1.47596  & 1.382979 & 1.38298  \\
0.500000  & 2.927360 & 2.92736  & 2.854058 & 2.85420  & 2.343983 & 2.34441  \\
0.700000  & 5.499097 & 5.49910  & 5.211344 & 5.21554  & 3.585300 & 3.59391  \\
0.800000  & 9.036660 & 9.03672  & 8.214202 & 8.24214  & 4.653686 & 4.68814  \\
0.880000  & 19.737296 & 19.78700 & 15.440260 & 15.75370 & 5.989045 & 6.10023  \\
0.900000  & 31.726534 & 32.99270 & 20.561879 & 21.56270 & 6.437497 & 6.59024  \\
0.905000  & 38.897667 & 43.95000 & 22.603374 & 24.11690 & 6.559560 & 6.72533  \\
0.906899  & 42.903800 & 55.00000 & 23.520393 & 25.33380 & 6.607069 & 6.77812  \\
\midrule
& \multicolumn{2}{c}{\( Bi = 10^{-2} \)} 
& \multicolumn{2}{c}{\( Bi = 10^{-3} \)} 
& \multicolumn{2}{c}{\( Bi = 10^{-4} \)} \\
\cmidrule(lr){2-3} \cmidrule(lr){4-5} \cmidrule(lr){6-7}
\textbf{\( V_2 \)} & \textbf{AHM} & \textbf{\cite{Lu1995}} & \textbf{AHM} & \textbf{\cite{Lu1995}} & \textbf{AHM} & \textbf{\cite{Lu1995}} \\
\midrule
0.200000  & 1.000000 & 1.00000  & 0.719188 & 0.719188 & 0.672238 & 0.672238 \\
0.500000  & 1.000000 & 1.00000  & 0.419722 & 0.419732 & 0.341772 & 0.341773 \\
0.700000  & 1.000000 & 1.00000  & 0.269235 & 0.269325 & 0.181972 & 0.181984 \\
0.800000  & 1.000000 & 1.00000  & 0.201599 & 0.201837 & 0.110668 & 0.110699 \\
0.880000  & 1.000000 & 1.00000  & 0.145527 & 0.146162 & 0.050187 & 0.050142 \\
0.900000  & 1.000000 & 1.00000  & 0.128426 & 0.128820 & 0.030449 & 0.029139 \\
0.905000  & 1.000000 & 1.00000  & 0.123394 & 0.122880 & 0.024326 & 0.020600 \\
0.906899  & 1.000000 & 1.00000  & 0.121342 & 0.120000 & 0.021771 & -- \\
\bottomrule
\end{tabular*}
\end{table*}
\end{center}

\begin{center}
\begin{table*}
\caption{Comparison of AHM-\(O_6\) with results from \cite{Lu1995} for \( \rho = 11 \).\label{Tabla5}}
\footnotesize
\begin{tabular*}{\textwidth}{@{\extracolsep\fill}lrrrrrr@{}}
\toprule
& \multicolumn{2}{c}{\( Bi = 10^{12} \)} 
& \multicolumn{2}{c}{\( Bi = 1 \)} 
& \multicolumn{2}{c}{\( Bi = 10^{-1} \)} \\
\cmidrule(lr){2-3} \cmidrule(lr){4-5} \cmidrule(lr){6-7}
\textbf{\( V_2 \)} & \textbf{AHM} & \textbf{\cite{Lu1995}} & \textbf{AHM} & \textbf{\cite{Lu1995}} & \textbf{AHM} & \textbf{\cite{Lu1995}} \\
\midrule
0.200000  & 1.400002 & 1.40000  & 1.321429 & 1.321430 & 1.000000 & 1.000000 \\
0.500000  & 2.430579 & 2.43058  & 2.059212 & 2.059440 & 1.000000 & 1.000000 \\
0.700000  & 3.842480 & 3.84248  & 2.887217 & 2.891140 & 1.000000 & 1.000000 \\
0.800000  & 5.181830 & 5.18184  & 3.505442 & 3.519200 & 1.000000 & 1.000000 \\
0.880000  & 7.187101 & 7.18870  & 4.176696 & 4.214270 & 1.000000 & 1.000000 \\
0.900000  & 8.040418 & 8.05223  & 4.380322 & 4.429100 & 1.000000 & 1.000000 \\
0.905000  & 8.304141 & 8.32640  & 4.434026 & 4.486150 & 1.000000 & 1.000000 \\
0.906899  & 8.411798 & 8.44102  & 4.454735 & 4.508200 & 1.000000 & 1.000000 \\
\midrule
& \multicolumn{2}{c}{\( Bi = 10^{-2} \)} 
& \multicolumn{2}{c}{\( Bi = 10^{-3} \)} 
& \multicolumn{2}{c}{\( Bi = 10^{-4} \)} \\
\cmidrule(lr){2-3} \cmidrule(lr){4-5} \cmidrule(lr){6-7}
\textbf{\( V_2 \)} & \textbf{AHM} & \textbf{\cite{Lu1995}} & \textbf{AHM} & \textbf{\cite{Lu1995}} & \textbf{AHM} & \textbf{\cite{Lu1995}} \\
\midrule
0.200000  & 0.723076 & 0.723076 & 0.672726 & 0.672726 & 0.667276 & 0.667276 \\
0.500000  & 0.426381 & 0.426391 & 0.342559 & 0.342560 & 0.333788 & 0.333788 \\
0.700000  & 0.276797 & 0.276891 & 0.182842 & 0.182855 & 0.173162 & 0.173164 \\
0.800000  & 0.209520 & 0.209772 & 0.111570 & 0.111604 & 0.101534 & 0.101537 \\
0.880000  & 0.153846 & 0.154523 & 0.051132 & 0.051097 & 0.040616 & 0.040475 \\
0.900000  & 0.136959 & 0.137454 & 0.031423 & 0.030137 & 0.020584 & 0.019024 \\
0.905000  & 0.132013 & 0.131690 & 0.025313 & 0.021660 & 0.014334 & 0.010150 \\
0.906899  & 0.129999 & 0.129000 & 0.022763 & --       & 0.011718 & --       \\
\bottomrule
\end{tabular*}
\end{table*}
\end{center}

The most significant deviations between models occur for high conductivity contrast (\( \rho = 1001 \)), as shown in Table~\ref{Tabla3}, where Lu’s model predicts substantially higher effective conductivity than AHM-\( O_6 \) for fiber volume fractions \( V_2 \geq 0.9 \). Moreover, Lu’s model does not provide results for \( V_2 = 0.906899 \) under certain Biot number conditions, suggesting possible numerical instabilities or theoretical limitations when applied to extreme parameter regimes. In contrast, AHM-\( O_6 \) yields more conservative predictions at high \( V_2 \) and \( \rho \), which may be attributed to its underlying homogenization framework that could underrepresent localized interfacial effects.

To assess the accuracy of the AHM-\( O_6 \) approximation in this high-volume, high-contrast regime, additional calculations were performed using the higher-order AHM-\( O_{30} \) model from \cite{JuanK2011}. As presented in Table~\ref{Tabla6}, the results show excellent agreement between AHM-\( O_{30} \) and Lu’s model wherever data is available. This close correspondence suggests that when extended to higher truncation orders, the AHM approach provides reliable and accurate estimates of the effective thermal conductivity, even under challenging conditions involving strong interfacial resistance and densely packed fibers.

\begin{center}
\begin{table*}
\caption{Comparison of \cite{JuanK2011} (\( O_{35} \)) vs \cite{Lu1995} for higher values of \( V_2 \).\label{Tabla6}}
\small
\begin{tabular*}{\textwidth}{@{\extracolsep\fill}lrrrrrr@{}}
\toprule
{\textbf{\( \rho = 1001 \)}}
& \multicolumn{2}{c}{\( Bi = 10^{12} \)} 
& \multicolumn{2}{c}{\( Bi = 1 \)} 
& \multicolumn{2}{c}{\( Bi = 10^{-1} \)} \\
\cmidrule(lr){2-3} \cmidrule(lr){4-5} \cmidrule(lr){6-7}
\textbf{\( V_2 \)} & \( O_{35} \) & \textbf{\cite{Lu1995}} & \( O_{35} \) & \textbf{\cite{Lu1995}} & \( O_{35} \) & \textbf{\cite{Lu1995}} \\
\midrule
0.900000  & 52.450156 & 52.450300 & 46.524881 & 47.526900 & 26.012919 & 27.682900 \\
0.905000  & 96.842677 & 96.960000 & 71.325711 & 75.614400 & 29.594309 & 32.287700 \\
0.906899  & 243.443097 & --        & 98.874361 & 113.200000 & 31.312325 & 34.648000 \\
\midrule

{\textbf{\( \rho = 1001 \)}} 
& \multicolumn{2}{c}{\( Bi = 10^{-2} \)} 
& \multicolumn{2}{c}{\( Bi = 10^{-3} \)}
& \multicolumn{2}{c}{\( Bi = 10^{-4} \)} \\
\cmidrule(lr){2-3} \cmidrule(lr){4-5} \cmidrule(lr){6-7}
\textbf{\( V_2 \)} & \( O_{35} \) & \textbf{\cite{Lu1995}} & \( O_{35} \) & \textbf{\cite{Lu1995}} & \( O_{35} \) & \textbf{\cite{Lu1995}} \\
\midrule
0.900000 &      6.790222 &      6.967270 &      1.000000&       1.000000&      0.126628 &      0.127947\\
0.905000 &      6.926783 &      7.119430 &      1.000000&       1.000000&       0.120200&       0.121990\\
0.906899 &      6.980017 &      7.179000 &      1.000000 &      1.000000&      0.116355 &      0.119000\\
\midrule
{\textbf{\( \rho = 101 \)}} 
& \multicolumn{2}{c}{\( Bi = 10^{12} \)} 
& \multicolumn{2}{c}{\( Bi = 1 \)} 
& \multicolumn{2}{c}{\( Bi = 10^{-1} \)} \\
\cmidrule(lr){2-3} \cmidrule(lr){4-5} \cmidrule(lr){6-7}
\textbf{\( V_2 \)} & \( O_{35} \) & \textbf{\cite{Lu1995}} & \( O_{35} \) & \textbf{\cite{Lu1995}} & \( O_{35} \) & \textbf{\cite{Lu1995}} \\
\midrule
0.900000  & 32.992679 & 32.992700 & 20.628281 & 21.562700 & 6.437494 & 6.590240 \\
0.905000  & 43.939153 & 43.950000 & 22.733744 & 24.116900 & 6.559557 & 6.725330 \\
0.906899  & 54.558947 & 55.000000 & 23.692205 & 25.333800 & 6.607065 & 6.778120 \\
\midrule
{\textbf{\( \rho = 101 \)}} 
& \multicolumn{2}{c}{\( Bi = 10^{-2} \)} 
& \multicolumn{2}{c}{\( Bi = 10^{-3} \)}
& \multicolumn{2}{c}{\( Bi = 10^{-4} \)} \\
\cmidrule(lr){2-3} \cmidrule(lr){4-5} \cmidrule(lr){6-7}
\textbf{\( V_2 \)} & \( O_{35} \) & \textbf{\cite{Lu1995}} & \( O_{35} \) & \textbf{\cite{Lu1995}} & \( O_{35} \) & \textbf{\cite{Lu1995}} \\
\midrule
0.900000  & 1.000000 & 1.000000 & 0.127495 & 0.128820 & 0.028940 & 0.029139 \\
0.905000  & 1.000000 & 1.000000 & 0.121084 & 0.122880 & 0.020348 & 0.020600 \\
0.906899  & 1.000000 & 1.000000 & 0.117254 & 0.120000 & 0.014202 & 0.000000 \\
\midrule
{\textbf{\( \rho = 11 \)}} 
& \multicolumn{2}{c}{\( Bi = 10^{12} \)} 
& \multicolumn{2}{c}{\( Bi = 1 \)} 
& \multicolumn{2}{c}{\( Bi = 10^{-1} \)} \\
\cmidrule(lr){2-3} \cmidrule(lr){4-5} \cmidrule(lr){6-7}
\textbf{\( V_2 \)} & \( O_{35} \) & \textbf{\cite{Lu1995}} & \( O_{35} \) & \textbf{\cite{Lu1995}} & \( O_{35} \) & \textbf{\cite{Lu1995}} \\
\midrule
0.900000  & 8.052233 & 8.052230 & 4.380321 & 4.429100 & 1.000000 & 1.000000 \\
0.905000  & 8.326403 & 8.326400 & 4.434025 & 4.486150 & 1.000000 & 1.000000 \\
0.906899  & 8.441013 & 8.441020 & 4.454734 & 4.508200 & 1.000000 & 1.000000 \\
\midrule
{\textbf{\( \rho = 11 \)}} 
& \multicolumn{2}{c}{\( Bi = 10^{-2} \)} 
& \multicolumn{2}{c}{\( Bi = 10^{-3} \)}
& \multicolumn{2}{c}{\( Bi = 10^{-4} \)} \\
\cmidrule(lr){2-3} \cmidrule(lr){4-5} \cmidrule(lr){6-7}
\textbf{\( V_2 \)} & \( O_{35} \) & \textbf{\cite{Lu1995}} & \( O_{35} \) & \textbf{\cite{Lu1995}} & \( O_{35} \) & \textbf{\cite{Lu1995}} \\
\midrule
0.900000  & 0.136069 & 0.137454 & 0.029921 & 0.030137 & 0.019001 & 0.019024 \\
0.905000  & 0.129816 & 0.131690 & 0.021356 & 0.021660 & 0.010133 & 0.010150 \\
0.906899  & 0.126136 & 0.129000 & 0.015241 & --       & 0.003658 & --       \\
\bottomrule
\end{tabular*}
\end{table*}
\end{center}

In summary, both AHM-\( O_6 \) and Lu’s model demonstrate strong performance under moderate conditions. However, their reliability decreases in extreme regimes characterized by high fiber volume fractions (\( V_2 \)) and high conductivity contrasts (\( \rho \)). Both models offer stable predictions across the full range of Biot numbers for practical applications. Nonetheless, experimental validation is recommended in these critical regimes to determine which model best reflects real-world thermal behavior in fibrous composites.

\begin{figure}
	\centering
	\includegraphics[width=1.\linewidth]{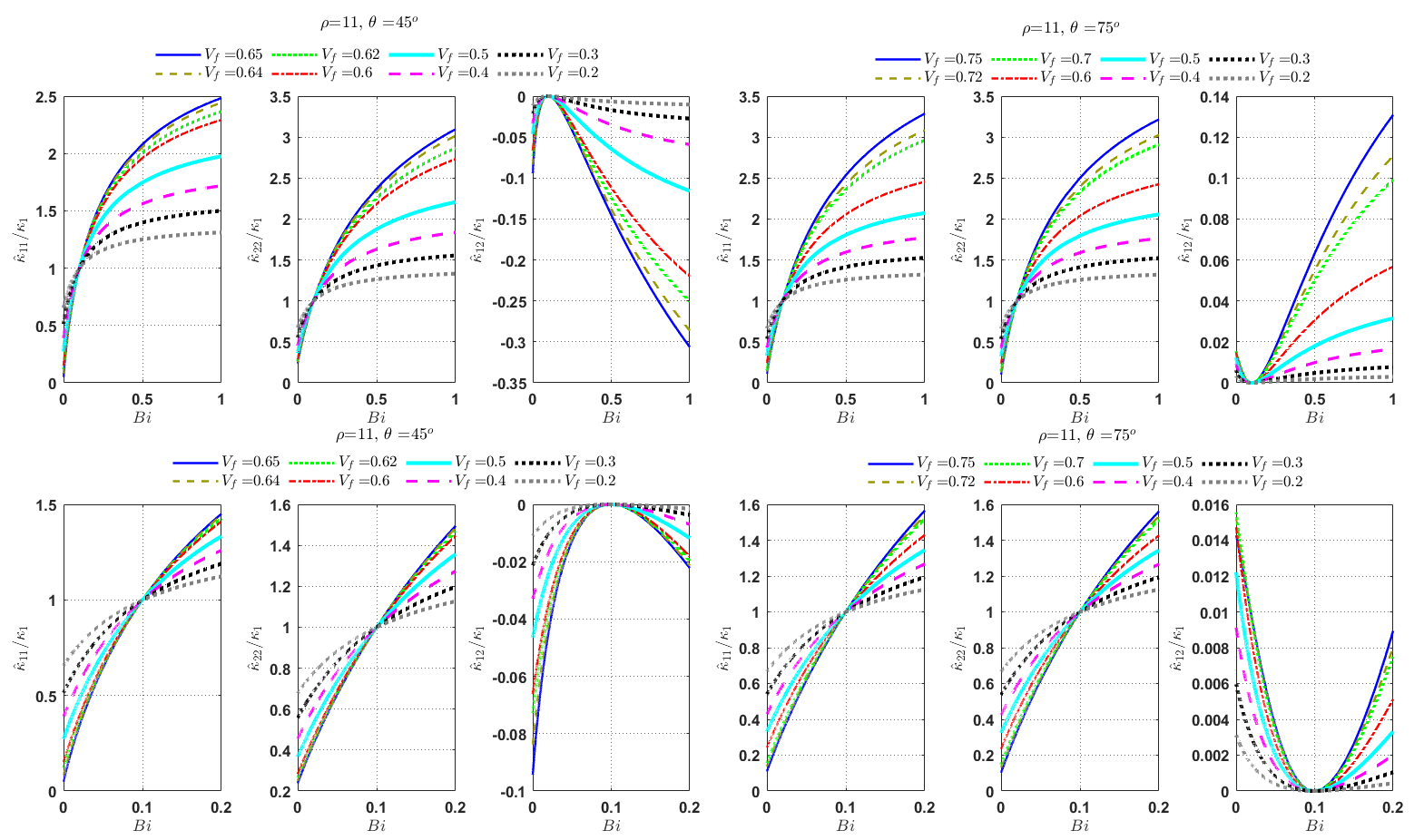}
	\caption{Effective coefficients for two types of periodic symmetry and for three different values of $\rho$. In the lower row,  the initial interval near of the critical point are shown. }
	\label{fig:k11k22k12vsbi4575rho11}
\end{figure}

\begin{figure}
	\centering
	\includegraphics[width=1\linewidth]{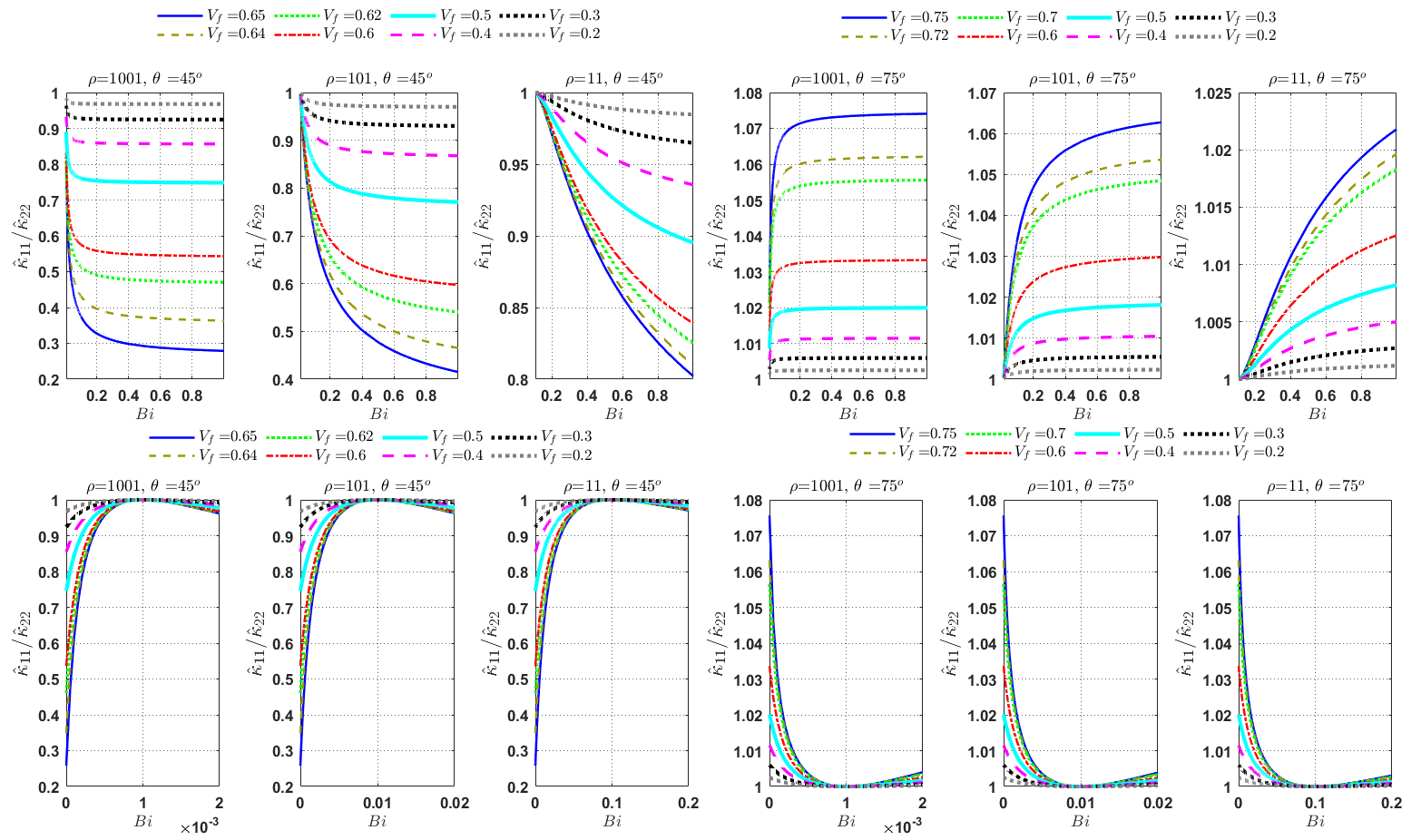}
	\caption{The contrast $\hat{\kappa}_{11}/\hat{\kappa}_{22}$ between the effective coefficients for two types of periodic symmetry and for three different values of $\rho$ are displayed.}
	\label{fig:k11overk22spring}
\end{figure}

We now extend the application of the AHM model to additional configurations of rhombic periodic cells, analyzing the effects of the Biot number (\( Bi \)), fiber volume fraction (\( V_2 \)), and conductivity contrast (\( \rho \)) on the effective thermal properties. Figure~\ref{fig:k11k22k12vsbi4575rho11} illustrates the effective thermal conductivity coefficients as functions of the Biot number for two types of periodic cells with angles \( \theta = 45^\circ \) and \( \theta = 75^\circ \). Each row of the figure corresponds to one of these geometries. The top row shows the complete conductivity variation as \( Bi \) increases from 0 to 1. In contrast, the bottom row presents a zoomed-in view near the critical Biot number (\( Bi = 0.1 \), for \( \rho = 11 \)), where significant changes in thermal behavior are observed.

The critical point marks a transition in effective conductivity behavior, indicating whether the thermal barrier enhances or inhibits heat conduction relative to the matrix. Interestingly, this transition occurs at the same \( Bi \) value for both geometries, confirming that the periodicity angle does not influence the critical Biot number and validating the robustness of the numerical implementation. The top row shows that the effective thermal properties stabilize for \( Bi > 1 \). At the same time, the most significant variations are observed in the interval \( Bi \in [0, 1] \), especially as \( Bi \) decreases and \( V_2 \) increases.

Figure~\ref{fig:k11overk22spring} further examines the thermal anisotropy of the composite through the ratio \( \hat{\kappa}_{11}/\hat{\kappa}_{22} \), indicating whether heat flows more easily along the fiber axis (\( \hat{\kappa}_{11} \)) or transversely (\( \hat{\kappa}_{22} \)). The figure is organized into two rows: the top row presents this ratio beyond the critical Biot point, while the bottom row focuses on values near it. The results show that for \( \theta = 45^\circ \), the transverse conductivity \( \hat{\kappa}_{22} \) dominates, whereas for \( \theta = 75^\circ \), the longitudinal conductivity \( \hat{\kappa}_{11} \) is more prominent. As the conductivity contrast \( \rho \) decreases, the anisotropy ratio \( \hat{\kappa}_{11}/\hat{\kappa}_{22} \) also decreases, reflecting reduced directional dependence in heat flow.

\begin{table*}
\caption{Effective conductivity properties of periodic fibrous composites with imperfect thermal barrier and periodic cell defined by \( \omega_1 = 1 \) and \( \omega_2 = re^{i\theta} \). Variations in effective conductivity with different approximation orders \( O_k \) and fiber arrangements. Comparison of AHM results with numerical data from Table 3 in \cite{Yan2016}. Material parameters: \( \rho = 1001 \), \( K = 1/2\beta \), \( \beta = 4.995 \times 10^{-3} \).\label{tabla2Yang}}
\footnotesize
\begin{tabular*}{\textwidth}{@{\extracolsep{\fill}}lcccccccccc@{}}
\toprule
& \multicolumn{4}{c}{\textbf{Rectangular} ($r=\frac{\sqrt{3}}{2}$, $\theta=\frac{\pi}{2}$, $V_2=0.62$)} & \multicolumn{6}{c}{\textbf{Rhombic} ($r=\frac{\sqrt{6}}{1+\sqrt{3}}$, $\theta=\frac{5\pi}{12}$, $V_2=0.66$)} \\
\cmidrule(r){2-5} \cmidrule(r){6-11}
& 
\multicolumn{2}{c}{$\hat{\kappa}_{11}/\kappa_1$} & 
\multicolumn{2}{c}{$\hat{\kappa}_{22}/\kappa_1$} & 
\multicolumn{2}{c}{$\hat{\kappa}_{11}/\kappa_1$} & 
\multicolumn{2}{c}{$\hat{\kappa}_{22}/\kappa_1$} & 
\multicolumn{2}{c}{$\hat{\kappa}_{12}/\kappa_1$} \\
\cmidrule(r){2-3} \cmidrule(r){4-5} \cmidrule(r){6-7} \cmidrule(r){8-9} \cmidrule(r){10-11}
\textbf{$k$} & \cite{Yan2016} & AHM $O_{k-1}$ & \cite{Yan2016} & AHM $O_{k-1}$ & \cite{Yan2016} & AHM $O_{k-1}$ & \cite{Yan2016} & AHM $O_{k-1}$ & \cite{Yan2016} & AHM $O_{k-1}$\\
\midrule
1 & 3.38835 & 3.38835 & 5.34695 & 5.34695 & 4.16683 & 4.16683 & 5.62170 & 5.62170 & 0.666908 & 0.666908 \\
2 & 3.60232 & 3.60232 & 6.43325 & 6.43325 & 4.38226 & 4.38226 & 6.45305 & 6.45305 & 0.775892 & 0.775892 \\
3 & 3.60237 & 3.60237 & 6.61906 & 6.61906 & 4.43359 & 4.43360 & 6.64675 & 6.64675 & 0.862522 & 0.862522 \\
4 & 3.60286 & 3.60286 & 6.66858 & 6.66858 & 4.43406 & 4.43406 & 6.68391 & 6.68391 & 0.866679 & 0.866679 \\
6 & 3.60292 & 3.60292 & 6.68037 & 6.68037 & 4.43496 & 4.43496 & 6.69323 & 6.69323 & 0.869527 & 0.869527 \\
8 & 3.60293 & 3.60293 & 6.68096 & 6.68099 & 4.43499 & 4.43499 & 6.69365 & 6.69367 & 0.869641 & 0.869645 \\
10 & 3.60293 & 3.60293 & 6.68099 & 6.68099 & 4.43499 & 4.43499 & 6.69367 & 6.69367 & 0.869646 & 0.869647 \\
\bottomrule
\end{tabular*}
\end{table*}
\begin{table*}[ht]
	\caption{Effective conductivity properties of periodic fibrous composites with imperfect thermal barrier and periodic cell defined by \( \omega_1 = 1 \) and \( \omega_2 = re^{i\theta} \). Variations in effective conductivity with different approximation orders \( O_k \), fiber arrangements and high fiber volume fraction. Material parameters: \( \rho = 1001 \), \( K = 1/2\beta \), \( \beta = 4.995 \times 10^{-3} \).\label{tabla2YanghighVF}}
	\footnotesize
	\begin{tabular*}{\textwidth}{@{\extracolsep\fill}lccccccc@{}}
		\toprule
		\textbf{Model} & 
		\multicolumn{2}{c}{Hex/Sq (\( r = 1 \))} & 
		\multicolumn{2}{c}{Rect. (\(\theta=\pi/2, r = \sqrt{3}/2, V_2=0.680174 \))} & 
		\multicolumn{3}{c}{Rhom. \(\left(\theta=\frac{5\pi}{12}, r = \frac{\sqrt{6}}{1+\sqrt{3}}, V_2=0.729009\right)\)} \\
		\cmidrule(lr){2-3} \cmidrule(lr){4-5} \cmidrule(lr){6-8}
		& \( \theta = \pi/3,\ V_2 = 0.9068993\) & \( \theta = \pi/2,\ V_2 = 0.785398 \) &
		\( \hat{\kappa}_{11}/\kappa_1 \) & \( \hat{\kappa}_{22}/\kappa_1 \) &
		\( \hat{\kappa}_{11}/\kappa_1 \) & \( \hat{\kappa}_{22}/\kappa_1 \) & \( \hat{\kappa}_{12}/\kappa_1 \) \\
		\midrule
		\( O_{10} \)      & 31.2597  &    19.6887 &     4.31421 &     17.2852 &     6.31149 &     17.5963&      3.64049\\
		\( O_{21} \)      &   31.3126 &     19.7137 &     4.31421&      17.3066 &     6.31302&      17.6177 &     3.64623\\
		\( O_{25} \)      & 31.3126   &   19.7137  &    4.31421&      17.3067&      6.31302  &    17.6177 &     3.64623\\
		\bottomrule
	\end{tabular*}
    \label{YangHighVFimperf}
\end{table*}
The effective conductivity properties of periodic fibrous composites with a thermal barrier interface are analyzed in Table~\ref{tabla2Yang}, comparing results from the Asymptotic Homogenization Method (AHM) at approximation orders $O_0$ through $O_{10}$ against numerical benchmarks from \cite{Yan2016}. Using the material parameters from Table~2 in \cite{Yan2016} with $\rho = 1001$ and $t/R_1 = 0.1$, we examine configurations with complex cell periods $\omega_1 = 1$ and $\omega_2 = re^{i\theta}$, where all relevant parameters are specified in Table~\ref{tabla2Yang}.
The numerical results demonstrate an exceptional agreement between the Asymptotic Homogenization Method (AHM) and the reference data from \cite{Yan2016}, validating the computational approach for rectangular and rhombic periodic cell configurations. The rectangular configuration ($r=\sqrt{3}/2$, $\theta=\pi/2$, $V_2=0.62$) shows rapid convergence in the $\hat{\kappa}_{11}/\kappa_1$ components, stabilizing at $k=2$ with a final value of 3.60293, while the $\hat{\kappa}_{22}/\kappa_1$ components require higher values $k$ to reach their convergent state of 6.68099 at $k=6$. Compared to the corollary configuration ($r=\sqrt{6}/(1+\sqrt{3})$, $\theta=5\pi/12$, $V_2=0.66$) exhibits systematically higher conductivity values, particularly in the $\hat{\kappa}_{22}/\kappa_1$ components, which peak at 6.69367, reflecting the influence of both geometric arrangement and slightly higher fiber volume fraction. The maximum observed deviation between AHM and \cite{Yan2016} remains below 0.00006 across all components, with the off-diagonal terms $\hat{\kappa}_{12}/\kappa_1$ showing perfect agreement with six decimal places in most cases. This remarkable precision underscores the reliability of the method for anisotropic conductivity analysis. The conductivity values of the rhombic configuration exceed those of the rectangular case by approximately 23.1\% for $\hat{\kappa}_{11}$, 0.19\% for $\hat{\kappa}_{22}$, and 30.5\% for $\hat{\kappa}_{12}$, revealing significant orientation-dependent effects that cannot be attributed solely to the modest 6. 5\% difference in the fiber volume fraction. These results collectively establish a rigorous validation of the AHM methodology while quantifying the subtle but important configuration-dependent variations in periodic fibrous composites.

The effective conductivity properties of periodic fibrous composites with imperfect thermal barriers have been investigated for different fiber arrangements and high fiber volume fractions \( V_2 \) in Table \ref{tabla2YanghighVF}. The results indicate that the Hex/Sq arrangement with \( \theta = \pi/3 \) yields the highest effective conductivity values, reaching \( 31.2597 \) for \( O_{10} \) and slightly increasing to \( 31.3126 \) for higher-order approximations \( O_{21} \) and \( O_{25} \). This suggests that the approximation converges rapidly for this configuration. The rectangular case shows anisotropic behavior, with \( \hat{\kappa}_{22}/\kappa_1 \) being significantly higher than \( \hat{\kappa}_{11}/\kappa_1 \), reflecting the influence of fiber alignment on thermal conductivity. The rhombic arrangement exhibits intermediate values, with a notable non-zero off-diagonal component \( \hat{\kappa}_{12}/\kappa_1 \), indicating coupling effects due to the non-orthogonal fiber orientation. The consistency of results across higher-order approximations (\( O_{21} \) and \( O_{25} \)) confirms the robustness of the computational method. Overall, the data highlight the strong dependence of effective conductivity on fiber arrangement, volume fraction, and approximation order, providing valuable insights for the design of fibrous composites with tailored thermal properties.

\subsection{Coated thermal layer model} 

In this subsection, we investigate the effective thermal conductivity of composites featuring a coated thermal barrier configuration. This model consists of circular fibers surrounded by a concentric mesophase layer, which acts as a thermal coating between the fibers and the matrix. The mesophase introduces an additional degree of thermal resistance or enhancement, depending on its conductivity relative to the matrix and fiber. The effective properties are evaluated as functions of the mesophase thickness-to-fiber radius ratio (\( t/R_2 \)), the total fiber volume fraction (\( V_f = V_2 + V_3 \)), and the conductivity contrast between phases (\( \rho_1 = \kappa_2/\kappa_1 \)). We compare the predictions obtained from the sixth-order approximation of the Asymptotic Homogenization Method (AHM-\( O_6 \)) with reference data from Lu’s model \cite{Lu1995} for hexagonal arrays of coated circular cylinders (\( \theta = 60^\circ \)), aiming to assess model accuracy across a wide range of geometric and material parameters.

Comparative results between the AHM-\( O_6 \) approximation and Lu’s 1995 model for different values of \( t/R_2 \), total fiber volume fraction (\( V_f = V_2 + V_3 \)), and conductivity contrast (\( \rho_1 \)) are presented in Tables~\ref{3fLurho990}--\ref{3fLurho10}. This analysis examines how variations in \( \rho_1 \) (990.5, 99.498, and 10.4195) influence the predicted effective conductivity (\( \kappa_{\text{eff}}/\kappa_1 \)) in coated fiber composites across different mesophase thicknesses and volume fractions.

For \( \rho_1 = 990.5 \), corresponding to a highly conductive mesophase, both models predict elevated values of \( \kappa_{\text{eff}}/\kappa_1 \), particularly at high fiber volume fractions (\( V_f \geq 0.8 \)). However, Lu’s model exhibits extreme sensitivity to \( V_f \) increases: at \( V_f = 0.906899 \), AHM predicts a value of 79.75, while Lu’s model does not report a corresponding result, suggesting potential numerical instability. Divergences are most pronounced for thin mesophases (\( t/R_2 = 0.1 \)), where Lu’s model predicts effective conductivities approximately 50\% higher than AHM at \( V_f = 0.905 \). For very thin mesophases (\( t/R_2 \leq 0.001 \)), both models converge to \( \kappa_{\text{eff}}/\kappa_1 \approx 1 \), indicating that the influence of the coating layer becomes negligible.

In the regime of intermediate conductivity (\( \rho_1 = 99.498 \)), discrepancies between models persist but are less severe. For example, at \( t/R_2 = 1.0 \) and \( V_f = 0.9 \), Lu’s model predicts \( \kappa_{\text{eff}}/\kappa_1 = 26.55 \) compared to AHM’s 25.76 (a 3\% difference). The gap widens with thinner mesophases; at \( t/R_2 = 0.1 \) and \( V_f = 0.905 \), Lu’s prediction exceeds AHM’s by 19\%. Notably, unlike in the highly conductive case, Lu’s model consistently provides values across all \( V_f \), although systematic overestimations are observed. Convergence to \( \kappa_{\text{eff}}/\kappa_1 = 1 \) for \( t/R_2 = 0.001 \) remains consistent, reinforcing the physical validity of both models in the limit of negligible coating thickness.

\begin{center}
\begin{table*}
\caption{Comparison of AHM and \cite{Lu1995} for hexagonal cell with \( \rho_1 = 990.5 \).\label{3fLurho990}}
\small
\begin{tabular*}{\textwidth}{@{\extracolsep\fill}lrrrr@{}}
\toprule
& \multicolumn{2}{c}{\( t/R_2 = 1.0 \)} & \multicolumn{2}{c}{\( t/R_2 = 0.1 \)} \\
\cmidrule(lr){2-3} \cmidrule(lr){4-5}
\textbf{\( V_f \)} & \textbf{AHM-\( O_6 \)} & \textbf{\cite{Lu1995}} & \textbf{AHM-\( O_6 \)} & \textbf{\cite{Lu1995}} \\
\midrule
0.200000  & 1.497905 & 1.497900 & 1.486932 & 1.486930 \\
0.500000  & 2.991260 & 2.991260 & 2.921961 & 2.921950 \\
0.700000  & 5.755734 & 5.755730 & 5.479869 & 5.479520 \\
0.800000  & 9.790154 & 9.790170 & 8.992523 & 8.990040 \\
0.880000  & 24.213168 & 24.300400 & 19.700908 & 19.718300 \\
0.900000  & 46.849705 & 50.494600 & 32.202896 & 33.608700 \\
0.905000  & 66.072540 & 90.380000 & 40.085097 & 46.959000 \\
0.906899  & 79.753283 & --        & 44.645393 & --        \\
\midrule
& \multicolumn{2}{c}{\( t/R_2 = 0.01 \)} & \multicolumn{2}{c}{\( t/R_2 = 0.001 \)} \\
\cmidrule(lr){2-3} \cmidrule(lr){4-5}
\textbf{\( V_f \)} & \textbf{AHM-\( O_6 \)} & \textbf{\cite{Lu1995}} & \textbf{AHM-\( O_6 \)} & \textbf{\cite{Lu1995}} \\
\midrule
0.200000  & 1.390020 & 1.390020 & 1.000000 & 1.000000 \\
0.500000  & 2.380334 & 2.380280 & 1.000000 & 1.000000 \\
0.700000  & 3.708175 & 3.706940 & 1.000000 & 1.000000 \\
0.800000  & 4.949063 & 4.943210 & 1.000000 & 1.000000 \\
0.880000  & 6.848813 & 6.819060 & 1.000000 & 1.000000 \\
0.900000  & 7.756607 & 7.732650 & 1.000000 & 1.000000 \\
0.905000  & 8.071261 & 8.103900 & 1.000000 & 1.000000 \\
0.906899  & 8.207049 & 8.340000 & 1.000000 & 1.000000 \\
\bottomrule
\end{tabular*}
\end{table*}
\end{center}

For low relative conductivity (\( \rho_1 = 10.4195 \)), the AHM-\( O_6 \) and Lu’s models demonstrate near-perfect agreement. At \( t/R_2 = 1.0 \) and \( V_f = 0.9 \), both predict an effective conductivity of approximately \( \kappa_{\text{eff}}/\kappa_1 \approx 5.12 \), differing by less than 0.1\%. Even at the highest fiber volume fraction considered (\( V_f = 0.906899 \)), the deviation remains negligible (5.257 for AHM-\( O_6 \) versus 5.262 for Lu’s model). This close correspondence reflects the mesophase’s marginal contribution to the overall thermal conductivity when the conductivity contrast \( \rho_1 \) is small, rendering theoretical differences between the models practically irrelevant. A minor exception occurs at \( t/R_2 = 0.01 \), where Lu’s model slightly underestimates \( \kappa_{\text{eff}}/\kappa_1 \) compared to AHM-\( O_6 \) (e.g., 0.152 versus 0.156 for \( V_f = 0.9 \)); however, such discrepancies are too small to have practical significance.

\begin{center}
\begin{table*}
\caption{Comparison of AHM and \cite{Lu1995} for hexagonal cell with \( \rho_1 = 10.4195 \).\label{3fLurho10}}
\small
\begin{tabular*}{\textwidth}{@{\extracolsep\fill}lrrrr@{}}
\toprule
& \multicolumn{2}{c}{\( t/R_2 = 1.0 \)} & \multicolumn{2}{c}{\( t/R_2 = 0.1 \)} \\
\cmidrule(lr){2-3} \cmidrule(lr){4-5}
\textbf{\( V_f \)} & \textbf{AHM-\( O_6 \)} & \textbf{\cite{Lu1995}} & \textbf{AHM-\( O_6 \)} & \textbf{\cite{Lu1995}} \\
\midrule
0.200000  & 1.338880 & 1.338880 & 1.000000 & 1.000000 \\
0.500000  & 2.137149 & 2.137140 & 1.000000 & 1.000000 \\
0.700000  & 3.080323 & 3.080190 & 1.000000 & 1.000000 \\
0.800000  & 3.841034 & 3.840510 & 1.000000 & 1.000000 \\
0.880000  & 4.781939 & 4.780500 & 1.000000 & 1.000000 \\
0.900000  & 5.120607 & 5.121110 & 1.000000 & 1.000000 \\
0.905000  & 5.218448 & 5.221470 & 1.000000 & 1.000000 \\
0.906899  & 5.257486 & 5.262220 & 1.000000 & 1.000000 \\
\midrule
& \multicolumn{2}{c}{\( t/R_2 = 0.01 \)} & \multicolumn{2}{c}{\( t/R_2 = 0.001 \)} \\
\cmidrule(lr){2-3} \cmidrule(lr){4-5}
\textbf{\( V_f \)} & \textbf{AHM-\( O_6 \)} & \textbf{\cite{Lu1995}} & \textbf{AHM-\( O_6 \)} & \textbf{\cite{Lu1995}} \\
\midrule
0.200000  & 0.725370 & 0.725369 & 0.677855 & 0.677855 \\
0.500000  & 0.430569 & 0.430491 & 0.350918 & 0.350900 \\
0.700000  & 0.283433 & 0.282762 & 0.192504 & 0.192343 \\
0.800000  & 0.219655 & 0.218070 & 0.122410 & 0.121979 \\
0.880000  & 0.171986 & 0.168952 & 0.065566 & 0.064235 \\
0.900000  & 0.160428 & 0.156849 & 0.049538 & 0.047241 \\
0.905000  & 0.157557 & 0.153825 & 0.045188 & 0.042386 \\
0.906899  & 0.156468 & 0.152676 & 0.043482 & 0.040424 \\
\bottomrule
\end{tabular*}
\end{table*}
\end{center}

\begin{figure}
	\centering
	\includegraphics[width=1.\textwidth]{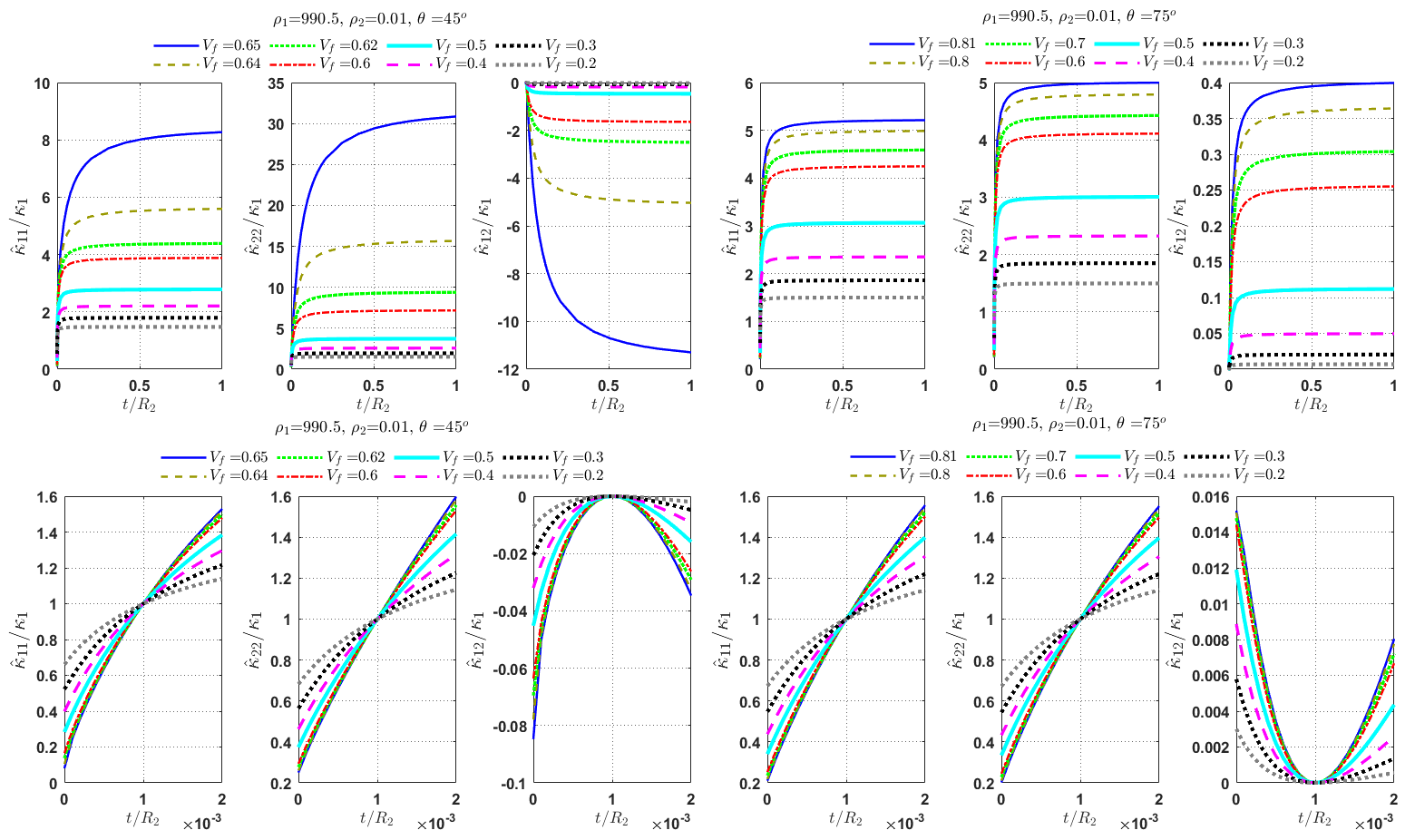} 
	\caption{Effective conductivity properties of periodic composite with rhombic cell of angles $\theta=45^0$ and $\theta=75^0$ with coated circular cylinders fibers vs the thickness of coated ring $t/R_2$. $V_f=V_2+V_3$ and $\rho_1=\kappa_2/\kappa_1= 990.5$ indicate a high conductive coated. In $t/R_2=0.001$, the composite has the critical interfacial layer.}  
	\label{fig:11}
\end{figure}

Figure~\ref{fig:11} presents the effective thermal conductivity of fibrous composite materials featuring a thermal barrier layer and a parallelogram periodic cell (\( \theta = 45^\circ \) and \( \theta = 75^\circ \)). The matrix has a thermal conductivity contrast of \( \rho_2 = \kappa_3/\kappa_1 = 1/100 \) relative to the inner fiber, while the mesophase/matrix contrast is significantly higher at \( \rho_1 = \kappa_2/\kappa_1 = 990.5 \). The study explores variations in interface thickness (\( t/R_2 \)) ranging from 0 to 1, and total fiber area fraction \( V_f \) (0.65 to 0.2 for \( \theta = 45^\circ \) and 0.81 to 0.2 for \( \theta = 75^\circ \)).

The results include the normalized axial conductivity (\( \hat{\kappa}_{11}/\kappa_1 \)), transverse conductivity (\( \hat{\kappa}_{22}/\kappa_1 \)), and the cross-term conductivity (\( \hat{\kappa}_{12}/\kappa_1 \)), plotted as functions of \( t/R_2 \). In this analysis, \( V_f \) denotes the combined volume fractions of the inner fiber (\( V_3 \)) and the coated fiber (\( V_2 \)), i.e., \( V_f = V_2 + V_3 \), following the notation used in Lu 1995~\cite{Lu1995}.

The figure is organized into two rows of six subfigures, displaying the evolution of the orthotropic effective conductivities versus \( t/R_2 \) for rhombic cells with \( \theta = 45^\circ \) (left panels) and \( \theta = 75^\circ \) (right panels). The top row spans the full range \( t/R_2 \in [0, 1] \), while the bottom row zooms on the initial interval \( t/R_2 \in [0, 0.02] \) to highlight the critical behavior near small interface thicknesses. The critical relative interfacial layer thickness \( \lambda \), defined in equation~(\ref{lam}), is located near \( t/R_2 = 0.01 \).

For small values of \( t/R_2 \) (below 0.01), both \( \hat{\kappa}_{11}/\kappa_1 \) and \( \hat{\kappa}_{22}/\kappa_1 \) increase sharply, particularly at high fiber volume fractions. This behavior suggests that the mesophase acts as a "thermal bridge," enhancing conductivity even with minimal thickness. Beyond \( t/R_2 = 0.01 \), the growth becomes nonlinear and eventually plateaus for \( t/R_2 \approx 0.3-1 \), indicating a saturation point where further increases in interfacial thickness yield diminishing returns.

The fiber volume fraction critically influences the effective properties. Higher fiber fractions lead to greater conductivity enhancements in both axial and transverse directions, as the percolating network of fibers and interface layers creates preferential pathways for heat conduction. In contrast, composites with lower fiber fractions (e.g., \( V_f = 0.3 \)) exhibit weaker dependence on \( t/R_2 \), as the matrix phase dominates thermal transport.

Anisotropy is clearly observed: for \( \theta = 75^\circ \), \( \hat{\kappa}_{11}/\kappa_1 \) consistently exceeds \( \hat{\kappa}_{22}/\kappa_1 \), indicating enhanced conductivity along the fiber axis. Conversely, for \( \theta = 45^\circ \), the transverse conductivity \( \hat{\kappa}_{22}/\kappa_1 \) becomes larger, reflecting the geometric alignment of fibers relative to the principal axes. The cross-term conductivity \( \hat{\kappa}_{12}/\kappa_1 \) captures the complex interplay between rhombic geometry and interface thickness, influencing the directional coupling of heat flow.

Symmetry effects are also evident: composites with \( \theta = 45^\circ \) exhibit higher effective conductivity than those with \( \theta = 75^\circ \), particularly at comparable fiber fractions. In general, lower fiber volume fractions achieve proportionally higher conductivity gains when a high-conductivity interface is present.

The interfacial layer plays a decisive role. Tiny increases in \( t/R_2 \) (e.g., from 0 to 0.001) trigger dramatic conductivity jumps, especially at high fiber loadings, as the high-conductivity mesophase offsets the relatively poor thermal performance of the inner fibers. However, for \( t/R_2 > 0.001 \), the curves flatten, signaling a physical limit where the additional thickness of the interface no longer substantially improves thermal performance.

In general, the presence of a critical relative interfacial thickness is crucial to improve the thermal properties of the composite relative to the matrix, particularly under conditions of high fiber load. Although inherent anisotropy (\( \hat{\kappa}_{11}/\hat{\kappa}_{22} > 1 \)) persists, it can be modulated through interface parameters, with the observed saturation at large \( t/R_2 \) emphasizing that the mesophase's influence eventually plateaus.

\begin{table*}
	\centering
	\caption{Effective conductivity properties of periodic fibrous composites with coating interface and periodic cell defined by \( \omega_1 = 1 \) and \( \omega_2 = re^{i\theta} \). Variations in effective conductivity are shown for different approximation orders \( O_k \) and fiber array geometries. Comparison of AHM results with numerical benchmarks from \cite{Yan2016} and \cite{ZEMLYANOVA2023}. Material parameters from Table 3 in \cite{Yan2016}:\( \rho_1 = 990.5 \), \( \rho_2 = 1/100 \), \( t/R_1 = 0.1 \),\( V_f = V_2+V_3\).\label{tabla3Yang1}}
	\footnotesize
	\begin{tabular*}{\textwidth}{@{\extracolsep{\fill}}lcccccccccc@{}}
		\toprule
		& \multicolumn{4}{c}{\textbf{Rhombic} ($r=\frac{\sqrt{5}}{2}$, $\theta=\arctan{2}$, $V_f=0.7$)} & \multicolumn{6}{c}{\textbf{Rhombic} ($r=\frac{2\sqrt{2}}{1+\sqrt{3}}$, $\theta=\frac{5\pi}{12}$, $V_2=0.7$)} \\
		\cmidrule(r){2-5} \cmidrule(r){6-11}
		& 
		\multicolumn{2}{c}{$\hat{\kappa}_{11}/\kappa_1$} & 
		\multicolumn{2}{c}{$\hat{\kappa}_{22}/\kappa_1$} & 
		\multicolumn{2}{c}{$\hat{\kappa}_{11}/\kappa_1$} & 
		\multicolumn{2}{c}{$\hat{\kappa}_{22}/\kappa_1$} & 
		\multicolumn{2}{c}{$\hat{\kappa}_{12}/\kappa_1$} \\
		\cmidrule(r){2-3} \cmidrule(r){4-5} \cmidrule(r){6-7} \cmidrule(r){8-9} \cmidrule(r){10-11}
		\textbf{$k$} & \cite{Yan2016} & AHM $O_{k-1}$ & \cite{Yan2016} & AHM $O_{k-1}$ & \cite{Yan2016} & AHM $O_{k-1}$ & \cite{Yan2016} & AHM $O_{k-1}$ & \cite{Yan2016} & AHM $O_{k-1}$\\
		\midrule
		1&	6.47373&	6.47373&	4.61585&	4.61585&	5.97169&	5.97169&	4.94750&	4.94750&	0.452284&	0.452284\\
		2&	6.90262&	6.90262&	4.68798&	4.68798&	6.80287&	6.80287&	5.44000&	5.43999&	0.342860&	0.342858\\
		3&	7.22368&	7.22368&	4.76057&	4.76057&	7.05683&	7.05683&	5.50599&	5.50599&	0.454821&	0.45482\\
		4&	7.25461&	7.25461&	4.76175&	4.76175&	7.08825&	7.08826&	5.51496&	5.51496&	0.438090&	0.438089\\
		6&	7.26463&	7.26463&	4.76176&	4.76176&	7.09915&	7.09915&	5.51601&	5.51600&	0.440132&	0.440131\\
		8&	7.26515&	7.26514&	4.76177&	4.76177&	7.09963&	7.09963&	5.51602&	5.51602&	0.440131&	0.44013\\
		10&	7.26517&	7.26517&	4.76177&	4.76177&	7.09965&	7.09966&	5.51602&	5.51602&	0.440128&	0.440127\\
             & \cite{ZEMLYANOVA2023}7.3080&    7.26517&    \cite{ZEMLYANOVA2023}4.7771 &   4.76177&    \cite{ZEMLYANOVA2023}7.1407 &   7.09966&    \cite{ZEMLYANOVA2023}5.5385 &   5.51602&    \cite{ZEMLYANOVA2023}0.4449&  0.440127\\     
		\bottomrule
	\end{tabular*}
\end{table*}
 
 \begin{center}
 	\begin{table*}
 		\caption{Effective conductivity properties of periodic fibrous composites with coating interface and periodic cell defined by \( \omega_1 = 1 \) and \( \omega_2 = re^{i\theta} \). Variations in effective conductivity are shown for different approximation orders \( O_k \) and fiber array geometries.  Material parameters: \( \rho_2 = 1/100 \), \( \rho_1 = 990.5 \), \( t/R_1 = 0.1 \), \( V_f = V_2+V_3 \).\label{tabla3Yang2}}
 		\footnotesize
 		\begin{tabular*}{\textwidth}{@{\extracolsep\fill}lccccccc@{}}
 			\toprule
 			\textbf{Model} & 
 			\multicolumn{2}{c}{Hex./Sqr (\( r = 1 \))} & 
 			\multicolumn{2}{c}{Rhom. (\( \theta=\arctan{2}, r = \frac{\sqrt{5}}{2}, V_f=0.785398 \))} & 
 			\multicolumn{3}{c}{Rhom. \(\left(\theta=\frac{5\pi}{12}, r = \frac{2\sqrt{2}}{1+\sqrt{3}} , V_f=0.785398\right)\)} \\
 			\cmidrule(lr){2-3} \cmidrule(lr){4-5} \cmidrule(lr){6-8}
 			& \( \theta = \pi/3,\ V_f = 0.906899 \) & \( \theta = \pi/2,\ V_f = 0.785398 \) &
 			\( \hat{\kappa}_{11}/\kappa_1 \) & \( \hat{\kappa}_{22}/\kappa_1 \) &
 			\( \hat{\kappa}_{11}/\kappa_1 \) & \( \hat{\kappa}_{22}/\kappa_1 \) & \( \hat{\kappa}_{12}/\kappa_1 \) \\
 			\midrule
 			\( O_{22} \)   &  61.8900  &  45.3454&    45.5679&      6.50777&      45.5231&       9.3779&     0.667772\\
 			\( O_{32} \)   & 65.8306  &    48.7244&   48.9107&      6.50777&      48.8819&      9.37863&     0.618489\\
 			\( O_{42} \)   &  68.0901  &   50.6967&   50.8636&      6.50777&      50.8466&      9.37905&     0.589661\\
 			\bottomrule
 		\end{tabular*}
        \label{table3YanhighVF3f}
 	\end{table*}
    
 \end{center}
The Table~\ref{tabla3Yang1} compares the effective conductivity properties of periodic fibrous composites with a coating interface, calculated using the Asymptotic Homogenization Method (AHM) at different approximation orders ($O_0$ to $O_{9}$), against numerical results from \cite{Yan2016} and \cite{ZEMLYANOVA2023}. The material parameters refer to Table 3 in \cite{Yan2016}, with $\rho_2 = 1/100$, $\rho_1 = 990.5$, and $t/R_1 = 0.1$. The complex periods of the cells are $\omega_1 = 1$ and $\omega_2 = re^{i\theta}$; all parameter values are provided in Table~\ref{tabla3Yang2}. 

The comparison between the AHM and the numerical benchmarks from \cite{Yan2016} reveals excellent agreement in predicting the effective conductivity of periodic rhombic fibrous composites. For both geometries analyzed—characterized by different lattice parameters ($r$, $\theta$) but equal fiber volume fraction ($V_f = 0.7$)—the AHM results converge precisely to the reference values as the approximation order $k$ increases. Notably, the convergence is rapid, with differences beyond the sixth decimal place vanishing by $k=4$ for most components of the effective conductivity tensor $\hat{\kappa}_{ij}/\kappa_1$.  

In the first rhombic configuration ($r=\sqrt{5}/2$, $\theta=\arctan{2}$), the diagonal components $\hat{\kappa}_{11}$ and $\hat{\kappa}_{22}$ exhibit anisotropy, with $\hat{\kappa}_{11}$ being consistently higher due to the lattice orientation. The AHM replicates the reference values exactly, even at low approximation orders ($k=1$), suggesting that the homogenized solution captures the dominant physics early in the expansion. For the second rhombic case ($r=2\sqrt{2}/(1+\sqrt{3})$, $\theta=5\pi/12$), the off-diagonal term $\hat{\kappa}_{12}$ reflects a slight shear coupling, which AHM reproduces with remarkable fidelity. The convergence here is marginally slower, requiring $k=3$ to stabilize within five decimal places, likely due to the more complex unit cell geometry.  This validates AHM as a robust alternative to numerical simulations for such microstructures, offering analytical insight without sacrificing accuracy.  

In Table \ref{table3YanhighVF3f} the effective conductivity properties of periodic fibrous composites with a coating interface have been investigated for different fiber arrangements at percolation-threshold volume fractions.The results demonstrate significant enhancement in effective conductivity due to the presence of the coating interface and fiber volume fraction. For the Hex./Sqr configuration with $\theta = \pi/3$, the conductivity increases from 61.8900 ($O_{22}$) to 68.0901 ($O_{42}$), showing a 10\% improvement with higher-order approximations. The rhombic arrangements exhibit anisotropic behavior, with the first case showing particularly strong directional dependence ($\hat{\kappa}_{11}/\kappa_1 \approx 50.86$ vs $\hat{\kappa}_{22}/\kappa_1 \approx 6.51$ at $O_{42}$). The second rhombic configuration reveals intermediate anisotropy with non-zero off-diagonal components ($\hat{\kappa}_{12}/\kappa_1 \approx 0.59$ at $O_{42}$), indicating coupling effects from the non-orthogonal fiber orientation. The convergence of results with increasing approximation order ($O_{22}$ to $O_{42}$) confirms the robustness of the method for these three-phase systems. Notably, the coating interface appears to amplify the anisotropic effects compared to two-phase composites, particularly in the rhombic arrangements where the $\hat{\kappa}_{11}/\kappa_1$ components are nearly an order of magnitude larger than $\hat{\kappa}_{22}/\kappa_1$ values. These findings highlight the critical role of interfacial phases in modifying the thermal transport properties of fibrous composites, especially near percolation thresholds.


\begin{figure}
	\centering
	\includegraphics[width=1.\textwidth]{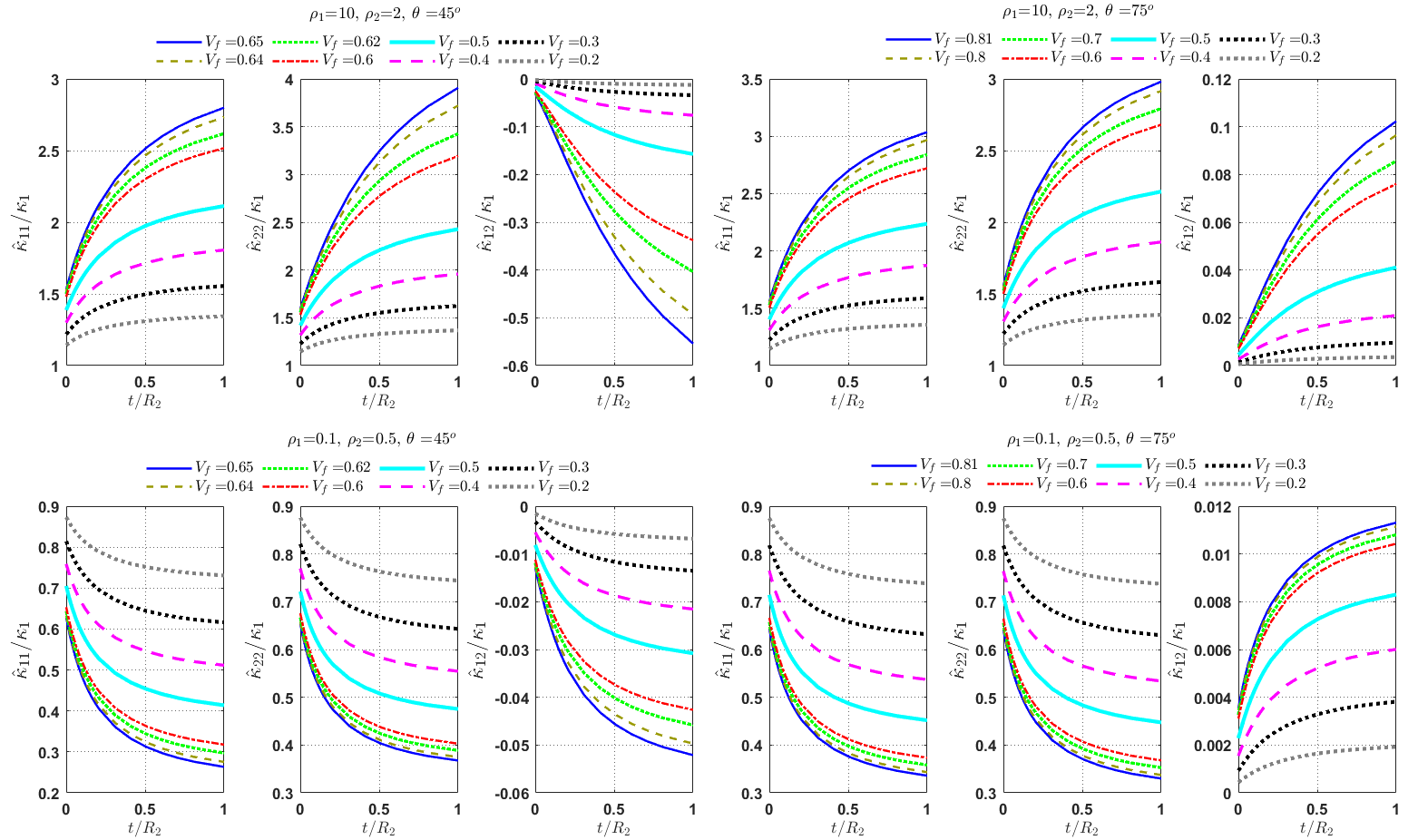} 
	\caption{Increase in the effective thermal conductivity properties of a periodic composite with a rhombic cell (\( \theta = 45^\circ \) and \( \theta = 75^\circ \)) featuring coated circular cylinder fibers, as a function of the coated ring thickness \( t/R_2 \). The total fiber volume fraction is \( V_f = V_2 + V_3 \), with conductivity contrasts \( \rho_1 = 10 \) and \( \rho_2 = 5 \). In this configuration, a critical coated layer thickness is not defined.
}  
	\label{fig:12}
\end{figure}

Figure~\ref{fig:12} presents the case where it is not possible to define a critical thickness of the interfacial layer \( \lambda \). This situation arises because the properties of the selected material satisfy the condition \( 1 < \rho_2 < \rho_1 \), which prevents the emergence of a distinct conductivity threshold. In this configuration, the effective thermal conductivities in both the axial (\( \hat{\kappa}_{11}/\kappa_1 \)) and transverse (\( \hat{\kappa}_{22}/\kappa_1 \)) directions exhibit a strong dependence on the total fiber volume fraction \( V_f \) and the symmetry of the periodic cell of the composite.

For high fiber volume fractions, the effective conductivity increases continuously throughout the full range of interface thicknesses \( t/R_2 \in [0, 1] \), without the sharp transition observed in previous cases. The influence of symmetry remains important: the periodic structure and fiber alignment continue to modulate the relative enhancement of conductivity along different directions.

The effect of microstructure on effective thermal conductivities was previously investigated by Lu and Lin (1995, 1996) for isotropic arrays, including square, hexagonal, and random configurations. In Lu’s model, the composite is assumed to exhibit globally isotropic behavior, in contrast to the Asymptotic Homogenization Method (AHM), which reveals that periodic composites generally display orthotropic characteristics. Nevertheless, it is insightful to compare the AHM results across different periodic unit cells to assess the influence of microstructural geometry.

\begin{figure}
	\centering
	\includegraphics[width=1.\textwidth]{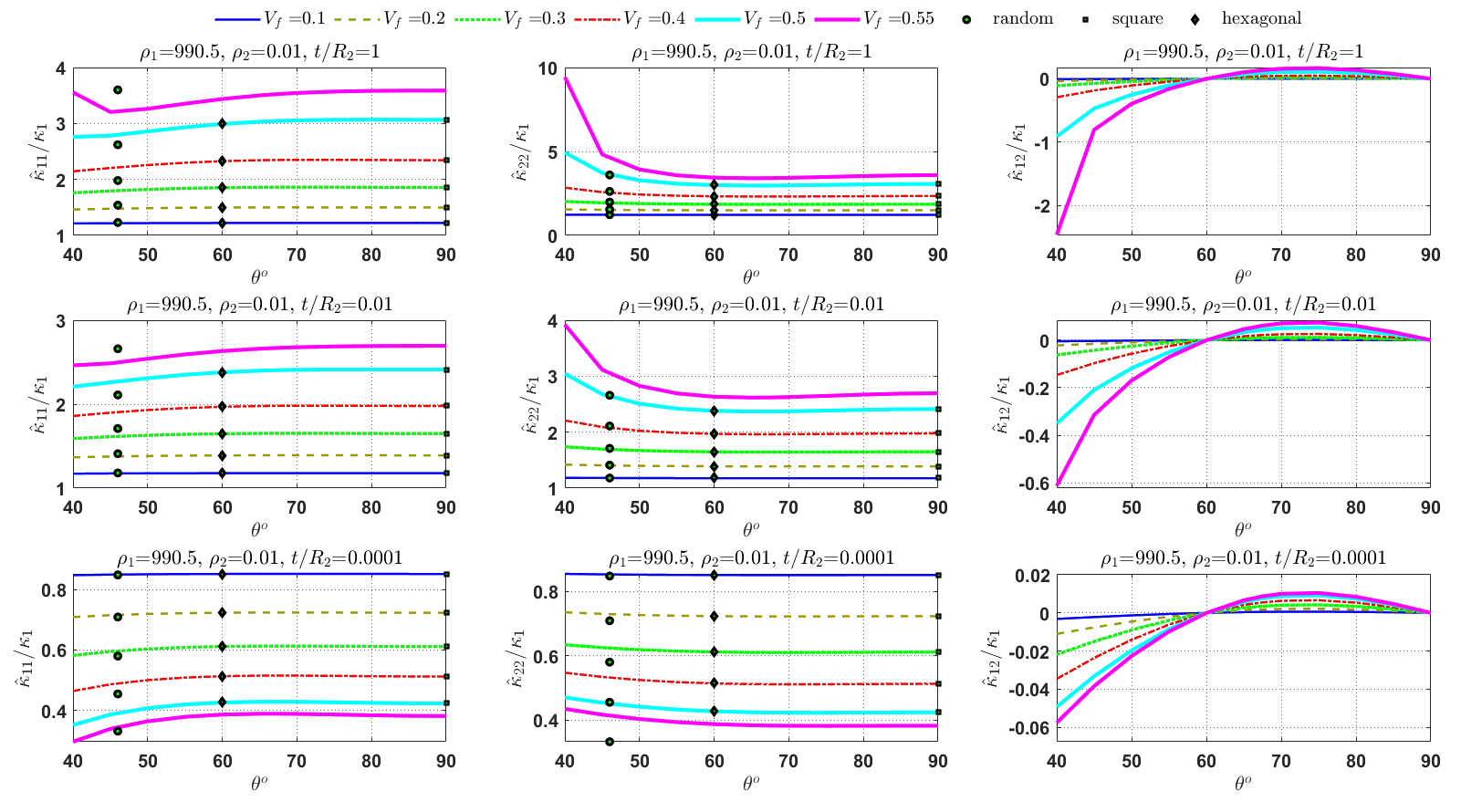}
	\caption{Angular dependence of the effective properties of the fiber composite with a thermal barrier. Comparison with Lu’s (1996) model shows excellent agreement for square and hexagonal arrangements ($\theta = 90^\circ$ and $\theta = 60^\circ$). The random arrangement aligns best with the coefficient $\kappa_{22}/\kappa_1$ at $\theta = 46^\circ$.}
	\label{angulo_varia}
\end{figure}

Figure~\ref{angulo_varia} presents the angular dependence of effective conductivity using the material parameters from Table~3 in Lu (1996), specifically with $\rho_1 = 990.5$, $\rho_2 = 0.01$, and total volume fraction $V_f = V_2 + V_3$ varying between 0.1 and 0.55. Various angles for the periodic cell orientation $\theta$ are considered. For the properties $\kappa_{11}/\kappa_1$ and $\kappa_{22}/\kappa_1$, the figure includes data reported in Lu’s work for square, hexagonal, and random arrays. Notably, AHM results align well with Lu’s model for square ($\theta = 90^\circ$) and hexagonal ($\theta = 60^\circ$) arrangements. The random array data reported by Lu aligns most closely with the AHM results for $\kappa_{22}/\kappa_1$ at $\theta = 46^\circ$.

The figure reveals a complex interplay between microstructure and effective thermal conductivity in fiber-reinforced composites with parallelogram periodicity. In the case of a thick mesophase ($t/R_2 = 1$), both $\kappa_{11}$ and $\kappa_{22}$ increase significantly with fiber volume fraction, especially for $V_f \geq 0.3$. For example, at $V_f = 0.5$, $\kappa_{11}/\kappa_1$ reaches approximately 2.76 for $\theta = 40^\circ$, exceeding values observed for thinner coatings. This strong dependence on mesophase thickness underscores the dominance of the highly conductive interface when it occupies a large portion of the composite volume.

Angular dependence is particularly pronounced in the $\kappa_{22}$ component. At $V_f = 0.5$ with $t/R_2 = 1$, $\kappa_{22}/\kappa_1$ drops from 4.93 at $\theta = 40^\circ$ to 3.07 at $\theta = \pi/2$, while $\kappa_{11}$ remains relatively stable. This behavior indicates that anisotropy is largely governed by the geometric orientation of the conductive interface, and it diminishes as the mesophase becomes thinner.

The off-diagonal component $\kappa_{12}$ exhibits a distinctive angular and volumetric response. It peaks at intermediate volume fractions ($V_f = 0.4$–0.55) and changes sign near $\theta = \pi/3$. The maximum negative value of $-0.912$ occurs at $\theta = 40^\circ$ for $t/R_2 = 1$ and $V_f = 0.5$, whereas $\kappa_{12}$ vanishes at $\theta = \pi/3$ and $\pi/2$. This suggests that certain orientations effectively decouple thermal transport directions, a behavior that can be exploited in applications requiring directional thermal control. Overall, the close agreement between AHM predictions and Lu’s (1996) model further validates the present methodology and confirms its robustness in capturing microstructure-driven anisotropic behavior.

\subsection {Disperse fiber composite with thermal barrier.} 

\begin{figure}
	\centering
	\includegraphics[width=.4\textwidth]{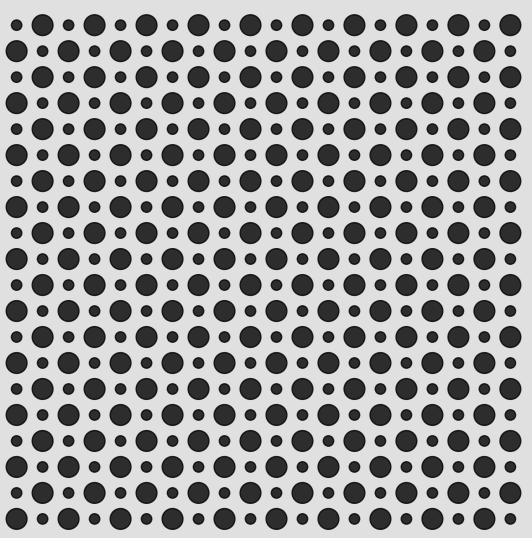}
	\caption{Schematic representation of fiber aggregation in a composite material. Large circles denote clustered inclusions, while smaller circles represent dispersed particles.}
	\label{nano_composite}
\end{figure}

 In this subsection, we apply the reiterated homogenization method to determine the effective thermal properties of fibrous composites featuring dispersed fibers and a thermal barrier, as illustrated in Figure~\ref{nano_composite}. Following the approach developed in ~\cite{iglesias2023conductivity, Nascimento2017}, the procedure consists of two sequential homogenization steps to account for the aggregation process and the influence of interfacial thermal resistance on the effective conductivity in two-dimensional square arrays of circular cylinders.

The reiterated homogenization method proceeds in two stages. First, an intermediate effective conductivity is calculated by accounting for the interaction between individual nanoinclusions and the surrounding matrix. Then, fiber clusters are treated as inclusions embedded within this intermediate medium, leading to the final effective conductivity (\( \hat{\kappa}_{RH} \)).

The improvement due to the dispersion process and the thermal barrier is quantified by the conductivity gain (\( \hat{\kappa}_{gain} \)), defined as
\begin{equation}
	\hat{\kappa}_{gain} = \frac{\hat{\kappa}_{RH}}{\hat{\kappa}_{CH}},
\end{equation}
where \( \hat{\kappa}_{CH} \) is the effective conductivity computed for a single-scale microstructure without dispersed fibers, but with the same overall inclusion volume fraction. This parameter measures the enhancement in thermal transport provided by the multiscale architecture and interfacial effects.

To apply the analytical formulae developed in the previous sections to a two-phase composite with dispersed inclusions of different sizes, the fiber population is divided into two groups: clustered fibers and non-clustered fibers, with corresponding volume fractions \( \phi_c \) and \( \phi_{nc} \), respectively, such that the total fiber volume fraction is \( \phi = \phi_c + \phi_{nc} \). The degree of aggregation is characterized by the parameter \( \alpha = \phi_c / \phi \), where \( \alpha \in [0,1] \) represents the fraction of fibers forming clusters.

In the first step of the homogenization procedure, the partial effective conductivity \( \kappa_p \) is computed by considering the interaction between the matrix (\( \kappa_1 \)) and the non-clustered inclusions (\( \kappa_2 \)). The corresponding fiber volume fraction for this step is given by

\begin{equation}
V_2 = \frac{(1-\alpha)\phi}{1-\phi_c},
\label{eq:V2_first_step}
\end{equation}
and the effective properties are evaluated using equations~(\ref{kappa11spring})--(\ref{kappa22spring}).

In the second step, the final effective conductivity \( \hat{\kappa}_{RH} \) is obtained by treating the fiber clusters as inclusions embedded within the intermediate medium. In this case, the matrix properties are set as \( \kappa_1 = \kappa_p \), the fiber properties remain \( \kappa_2 \), and the volume fraction of fibers is \( V_2 = \alpha \phi \).
 
\begin{figure}
	\centering
	\includegraphics[width=1\textwidth]{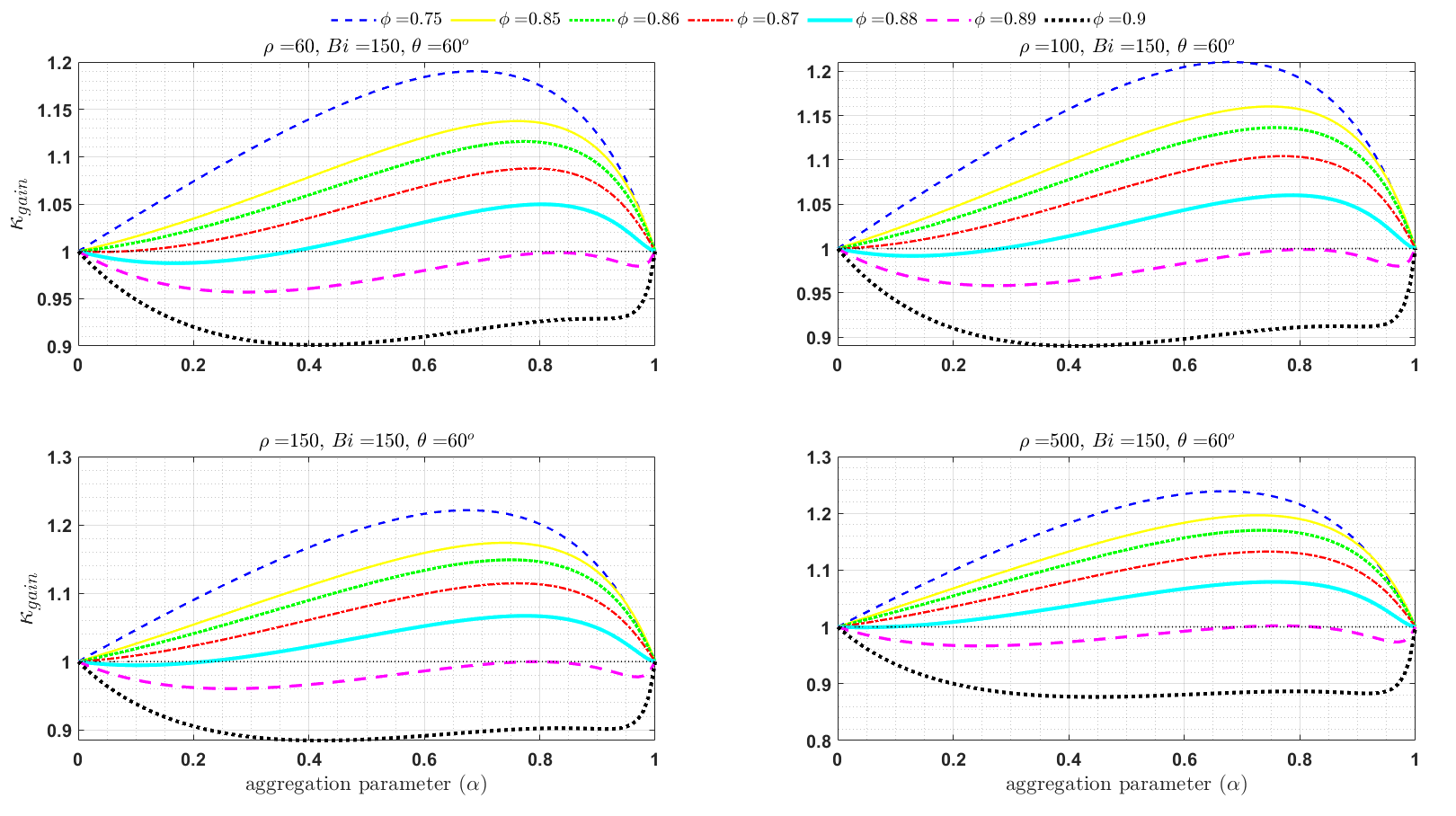}
	\caption{Gain in effective thermal conductivity (\( \hat{\kappa}_{gain} \)) for a composite with dispersed fibers compared to a composite without dispersed fibers, considering a thermal spring barrier and a rhombic periodic cell with \( \theta = 60^\circ \).}
	\label{Kgain_spring_60}
\end{figure}

\begin{figure}
	\centering
	\includegraphics[width=1\textwidth]{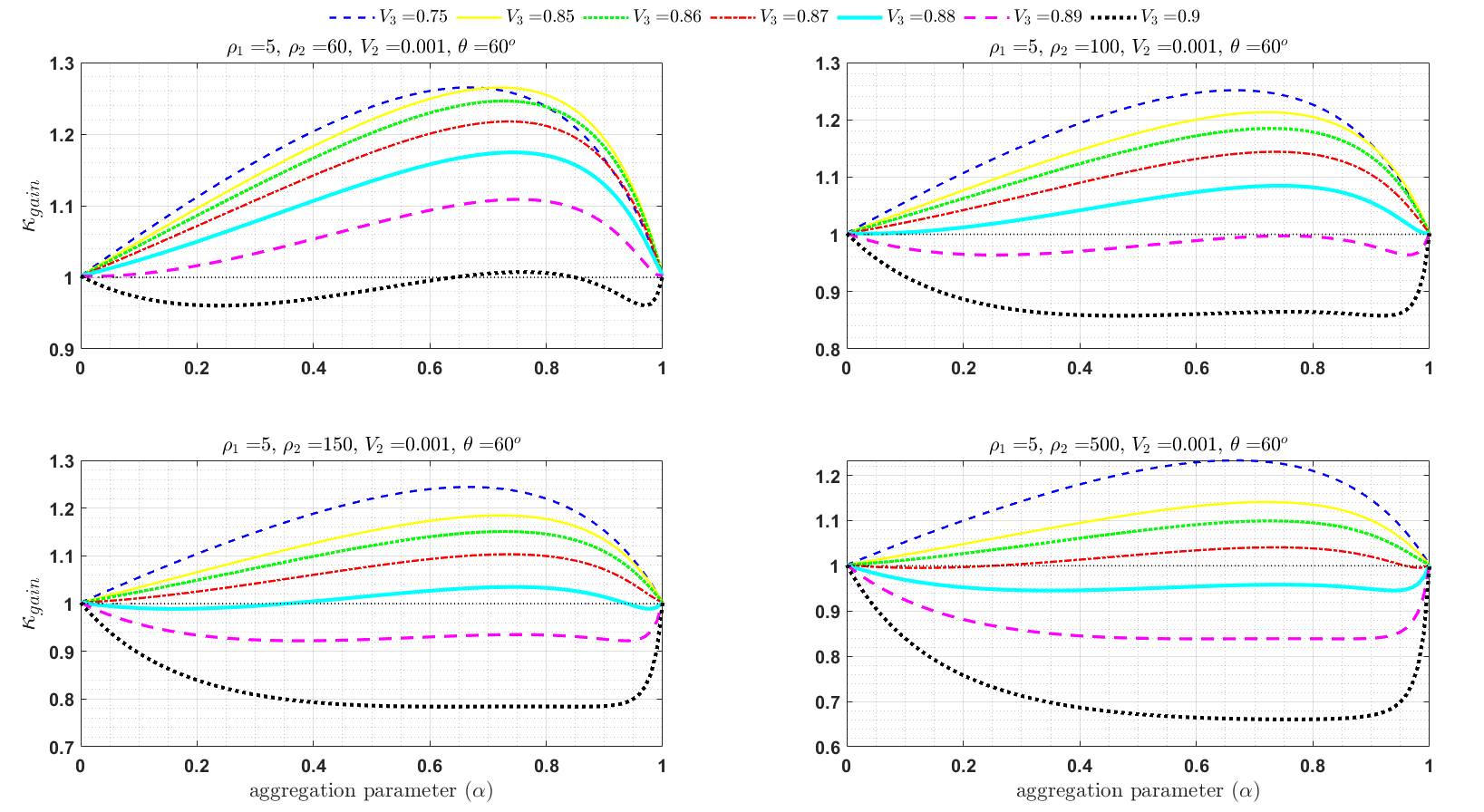}
	\caption{Gain in effective thermal conductivity (\( \hat{\kappa}_{gain} \)) for a composite with dispersed fibers compared to a composite without dispersed fibers, considering a thermal mesophase barrier (three-phase composite) and a rhombic periodic cell with \( \theta = 60^\circ \).}
	\label{Kgain_3f_60}
\end{figure}

The effective properties for the spring model, representing imperfect contact or a thermal spring barrier, are given by equations~(\ref{kappa11spring})--(\ref{kappa22spring}) for different periodic distributions. The spring model was also used by Iglesias et al.~\cite{iglesias2023conductivity} to calculate the effective properties and the conductivity gain parameter (\( \hat{\kappa}_{gain} \)) for composites with square symmetry.

Figure~\ref{Kgain_spring_60} illustrates the influence of fiber aggregation on the conductivity gain in the presence of a spring thermal barrier, for a rhombic periodic cell with \( \theta = 60^\circ \) and high inclusion volume fractions near the percolation threshold (\( \phi \in [0.8, 0.9] \)). Different conductivity contrasts (\( \rho_2 \in \{60, 100, 150, 500\} \)) are considered to analyze their impact on \( \hat{\kappa}_{gain} \). It is observed that increasing the contrast \( \rho_2 \) extends the range of aggregation parameters \( \alpha \) for which conductivity enhancement occurs. For example, for \( \phi \in [0.87, 0.89] \), a higher \( \rho_2 \) increases the interval \( (\alpha_1, \alpha_2) \) where \( \hat{\kappa}_{gain} \geq 1 \). In contrast, for \( \phi = 0.75 \), conductivity gain is achieved across all \( \alpha \in [0, 1] \), while for \( \phi = 0.90 \), \( \hat{\kappa}_{gain} \leq 1 \) for all \( \alpha \). These alternating gain and loss regions were also reported for high volume fractions in the perfect contact case by Berlyand and Mityushev~\cite{BerlyandMityushev}, and in the presence of a thermal barrier by Iglesias et al.~\cite{iglesias2023conductivity}.

The thermal barrier can also be modeled by introducing an interfacial mesophase layer between the fiber and the matrix, as depicted schematically in Figure~\ref{fig:Fig_1}. In this case, we apply the three-phase model given by equations~(\ref{c1313masC2313})--(\ref{c1323masC2323}) to determine the effective properties. When combined with the reiterated homogenization method, this model successfully captures the oscillatory behavior of the conductivity gain.

Figure~\ref{Kgain_3f_60} presents analogous results for a fibrous composite with rhombic periodicity (\( \theta = 60^\circ \)) incorporating a mesophase layer. Here, the behavior of the conductivity gain curves for \( V_3=\phi \in [0.87, 0.90] \) and different contrasts \( \rho_2 \in \{60, 100, 150, 500\} \) (with \( \rho_1 = 5 \)) shows that increasing \( \rho_2 \) leads to a reduction in the interval \( (\alpha_1, \alpha_2) \) where \( \hat{\kappa}_{gain} \geq 1 \). For \( \phi = 0.75 \), conductivity gain is maintained for all \( \alpha \in [0, 1] \), whereas for \( \phi = 0.90 \), alternating gain and loss regions are again observed, particularly for lower contrasts like \( \rho_2 = 60 \).

Overall, the occurrence of conductivity gain or loss is influenced by the symmetry of the composite, the contrast between the fiber and the matrix, and the fiber volume fraction. These factors jointly determine whether the incorporation of a thermal barrier and fiber aggregation enhances or diminishes the composite’s effective thermal conductivity.

\section{Conclusions}\label{sec5}

The effective thermal behavior of composites made of three-phase unidirectional periodic fibers with isotropic constituents is studied. The fibers are concentric, have a circular cross-section, and are distributed along a parallelogram cell. Different types of periodic cells and three contact scenarios were examined: perfect contact and two types of imperfect contact (with a thin-layer mesophase model and a spring-type interface model). The main results are: (i) unified formulas for the calculation of all effective thermal conductivity tensor coefficients were obtained for any period 1 parallelogram; (ii) these formulas were analytically and numerically validated, demonstrating their accuracy for different truncation orders, even for high fiber volume fractions very close to percolation; (iii) the formulas have simple expressions, and a calculation procedure is provided for any truncation order and any parallelogram cell. All programs created for numerical validation appear in an open-access repository, and (iv) the application of the unified formulas to study the gain in effective thermal conductivity in tubular fiber composites, as well as in the case of composites with dispersed fibers where they were combined using the reiterated homogenization method. The results show how these models effectively capture the oscillatory behavior of thermal conductivity at different fiber concentrations. These findings highlight the importance of choosing the appropriate model and periodic cell configuration to accurately predict the effective thermal behavior.


\bmsection*{Author contributions}

This is an author contribution text. This is an author contribution text. This is an author contribution text. This is an author contribution text. This is an author contribution text.

\bmsection*{Acknowledgments}
The authors acknowledge the financial support of the PREI-DGAPA-UNAM program (DGAP/DFAJ27 2OI2O23) and the PAPIIT-DGAPA-UNAM grant IN101822. This research was conducted at the Unidad Académica del Instituto de Investigaciones en Matemáticas Aplicadas y en Sistemas (IIMAS) en el Estado de Yucatán,Universidad Nacional Autónoma de México (UNAM). We also appreciate the valuable assistance of the technical and administrative staff of the IIMAS-Yucatán unit throughout this work. Furthermore, FJS, RGD, and JBC acknowledge the support of the CONAHCYT project CF-2023G-1458, Mexico, and the Coordinating of Superior-Level Staff Improvement (CAPES), PRAPG Program No. 23038.003836/2023-39.

\bmsection*{Financial disclosure}

This work was supported by the PREI-DGAPA-UNAM program (DGAP/DFAJ27 2OI2O23) and the PAPIIT-DGAPA-UNAM grant IN101822. Raúl Guinovart-Díaz and Julián Bravo-Castillero acknowledge this support through the Unidad Académica del Instituto de Investigaciones en Matemáticas Aplicadas y en Sistemas (IIMAS) en el Estado de Yucatán, Universidad Nacional Autónoma de México (UNAM).

\bmsection*{Conflict of interest}

The authors declare no potential conflict of interests.

\bibliography{Bibliography}

\bmsection*{Supporting information}

Additional supporting information may be found in the
online version of the article at the publisher’s website.

\end{document}